\documentclass[intlimits,twoside,a4paper]{article}

\usepackage{amsmath,amssymb,amsbsy}
\usepackage{graphicx}

\usepackage[T2A]{fontenc}
\usepackage[cp1251]{inputenc}

\usepackage[eqsecnum]{cmpj2}

\def\bm{\boldsymbol}
\issue{2014}{17}{4}{43704}
\doinumber{10.5488/CMP.17.43704}

\title[Effect of correlated hopping on thermoelectric properties]
{Effect of correlated hopping on thermoelectric properties: Exact solutions for the Falicov-Kimball model}
\author{A.M. Shvaika}
\address{Institute for Condensed Matter Physics of the National Academy of Sciences of Ukraine, \\ 1 Svientsitskii St., 79011 Lviv, Ukraine}

\date{Received September 30, 2014, in final form November 18, 2014}

\begin{document}

\maketitle

\begin{abstract}
The effect of correlated hopping on the charge and heat transport is investigated for the Falicov-Kimball model. Exact solutions for the electrical and thermal conductivities and thermoelectric power are obtained within the dynamical mean field theory. The temperature dependences of the transport coefficients are analysed for particular values of correlated hopping which correspond to the significant reconstruction of the density of states and transport function. The cases with strong enhancement of thermoelectric properties are elucidated.
\pacs 72.15.Jf, 72.20.Pa, 71.27.+a, 71.10.Fd
\end{abstract}

\section{Introduction}

Direct transformation of the heat flow into the electric current and vice versa attracts much attention in science and engineering~\cite{mahan:42}, but its applications are limited by the low thermoelectric \textit{figure of merit} of the traditional bulk materials --- metals and semiconductors --- good electric conductors are also good heat conductors. On the other hand, bad metals or compounds with strong electron correlations, possessing large variability of their band spectrum and density of states depending on the chemical structure and doping, are considered as possible new candidates for thermoelectric materials~\cite{mahan:42,costi:2519}. In many cases, the anomalous properties of strongly correlated materials are ascribed to the on-site Coulomb or spin interactions, which are the leading contributions.

It was already noticed by Hubbard in his seminal article~\cite{hubbard:238} that besides the local Coulomb-type interaction $U\sum_i \hat{n}_{i\uparrow}\hat{n}_{i\downarrow}$ there should be other nonlocal contributions: the intersite Coulomb interaction $\sum_{ij} V_{ij} \hat{n}_{i}\hat{n}_{j}$ and the so-called correlated hopping
\begin{equation}
\sum_{ij\sigma} t_{ij}^{(2)} (\hat{n}_{i\bar{\sigma}}+\hat{n}_{j\bar{\sigma}})c_{i\sigma}^{\dagger}c_{j\sigma} \quad\text{and}\quad \sum_{ij\sigma} t_{ij}^{(3)} \hat{n}_{i\bar{\sigma}}c_{i\sigma}^{\dagger}c_{j\sigma}\hat{n}_{j\bar{\sigma}},
\end{equation}
which reflects the fact that different many-body states can overlap in different extent and, as a result,
the value of intersite hopping depends on the occupation of these states. The origin of the correlated
hopping can be either a direct intersite interaction or an indirect effective one~\cite{foglio:4554,simon:7471}.
The effect of local Coulomb interaction is the subject of the famous Hubbard model which is widely investigated
in the theory of strongly correlated electron systems.

The correlated hopping is less popular. Even the term used is not well established.
Apart from the term ``correlated hopping'', many other terms circulate, such as ``assisted hopping'',
``bond-charge interaction (repulsion)'', ``occupation-dependent hopping'', ``correlated hybridization'', etc.
Similar contributions in the theory of disordered systems are also known as an ``off-diagonal disorder''~\cite{blackman:2412}.
Correlated hopping was considered in connection with the new mechanisms of high temperature superconductivity~\cite{hirsch:326,bulka:10303}, electron-hole asymmetry~\cite{didukh:7893}, and enhancement of magnetic properties~\cite{kollar:045107}. Recently, the correlated hopping has been examined in relation to the quantum dots~\cite{meir:196802,hubsch:196401,tooski:055001} and optical lattices~\cite{eckholt:093028,jurgensen:043623,liberto:013624}.

In this article we shall consider the effect of correlated hopping on the charge and heat transport for the Falicov-Kimball model~\cite{falicov:997}, which possesses exact solutions in the dynamical mean field theory (DMFT)~\cite{freericks:1333,schiller:15660,shvaika:075101,shvaika:119901}. In section~\ref{sec:Kubo}, we recall the main arguments of the linear response Kubo theory for the charge and heat transport which will be important in further considerations. Section~\ref{sec:ch-FKM} provides the DMFT solutions for the Falicov-Kimball model with correlated hopping and derivations
of the charge and energy current operators and transport coefficients in the homogeneous phase. It should be noted that at low temperatures, the phase transitions to the ordered phases could occur, e.g., the ground state phase diagrams for the one-dimensional $(D=1)$ and two-dimensional $(D=2)$ models display a variety of the modulated phases~\cite{gajek:3473,wojtkiewicz:233103,wojtkiewicz:3467,cencarikova:42701}, but phase diagrams for the $D=\infty$ Falicov-Kimball model with correlated hopping are unknown so far and we assume that a homogeneous solution is valid down to $T=0$. In section~\ref{sec:numerics} we consider peculiarities of the charge and heat transport with the change of correlated hopping value and doping and we summarize in section~\ref{sec:concl}.

\section{Macroscopic and microscopic levels in the description of thermoelectric effect}\label{sec:Kubo}

In the linear response Kubo theory, the charge $\bm{j}_c(\bm{r})$ and energy (heat) $\bm{j}_Q(\bm{r})$ currents are
caused by the electro-chemical potential gradient, including electrical field and charge distribution
inhomogeneities, and temperature gradient and can be obtained from the following equations~\cite{mahan:book}
\begin{align}
\bm{j}_c(\bm{r}) &= - e\int \mathbf{L}_{11}(\bm{r},\bm{r}') \bm{\nabla} \tilde{\mu}(\bm{r}') \rd \bm{r}' - e \int \mathbf{L}_{12}(\bm{r},\bm{r}') \frac{\boldsymbol{\nabla} T(\bm{r}')}{T(\bm{r}')} \rd \bm{r}',
\\
\bm{j}_Q(\bm{r}) &= - \int \mathbf{L}_{21}(\bm{r},\bm{r}') \bm{\nabla} \tilde{\mu}(\bm{r}') \rd \bm{r}' -  \int \mathbf{L}_{22}(\bm{r},\bm{r}') \frac{\bm{\nabla} T(\bm{r}')}{T(\bm{r}')} \rd \bm{r}',
\end{align}
where $\tilde{\mu}(\bm{r})=\mu - eV(\bm{r})$ is an electro-chemical potential.

For the uniform (steady) dc charge and energy currents, we can define the dc electric conductivity
\begin{equation}
\bm{\sigma}_{\textrm{dc}} = e^2 \mathbf{L}_{11}\,,
\end{equation}
Seebeck coefficient (thermoelectric power $\bm{E}=\mathbf{S}\nabla T$)
\begin{equation}
\mathbf{S} = \frac{1}{eT} \mathbf{L}_{11}^{-1} \mathbf{L}_{12}\,,
\end{equation}
and electronic contribution in thermal conductivity
\begin{equation}\label{thermcond}
\bm{\kappa}_{\textrm{e}} = \frac1T \left[\mathbf{L}_{22} - \mathbf{L}_{21} \mathbf{L}_{11}^{-1} \mathbf{L}_{12}\right]
\end{equation}
in terms of the generalized transport integrals $\mathbf{L}_{lm}$. The efficiency of the thermoelectric material is characterized by the dimensionless \textit{figure of merit}
\begin{equation}\label{ZT}
ZT = T\frac{{\sigma}_{\textrm{dc}} S^2}{\kappa_{\textrm{e}} + \kappa_{\textrm{ph}}}\,.
\end{equation}

The standard route to introduce macroscopic currents in the microscopic lattice models is as follows~\cite{mahan:book}. First of all, one can define the charge polarization by
\begin{equation}
\bm{\hat{P}} = \sum_{i\alpha} \bm{R}_i z_{\alpha} \hat{n}_{i\alpha},
\end{equation}
where $\bm{R}_i$ is the lattice site vector and $\hat{n}_{i\alpha}$ is the particle number operator at site $i$ for particles of the kind $\alpha$ with charge $z_{\alpha}$.
Now, the charge current can be defined using the continuity equation as follows:
\begin{equation}
\bm{\hat{j}} = \frac{\rd \bm{\hat{P}}}{\rd t} = \frac{1}{\ri} \left[\bm{\hat{P}}, \hat H\right].
\end{equation}

In a similar way, one can define the energy polarization by
\begin{equation}
\bm{\hat{P}}_Q = \sum_{i} \bm{R}_i \hat{H}_{i}\,,
\end{equation}
where $\hat{H}_{i}$ are the single site (local) contributions in total energy $\hat{H} = \sum_i \hat{H}_{i}$, and the energy current by
\begin{equation}\label{jQ_def}
\bm{\hat{j}}_Q = \frac{\rd \bm{\hat{P}}_Q}{\rd t} = \frac{1}{\ri} \left[\bm{\hat{P}}_Q, \hat H\right].
\end{equation}

Let us recall how it works for the case of noninteracting electrons. The Hamiltonian for noninteracting electrons on the lattice can be written as follows:
\begin{equation}
\hat{H} %= \sum_{ij\sigma} t_{ij} c_{i\sigma}^{\dagger} c_{j\sigma} - \mu \sum_{i\sigma} c_{i\sigma}^{\dagger} c_{i\sigma}
 = \sum_i \hat{H}_i\,,\qquad
\hat{H}_i = \frac12 \sum_{j\sigma} \left(t_{ij} c_{i\sigma}^{\dagger} c_{j\sigma} + t_{ji} c_{j\sigma}^{\dagger} c_{i\sigma}\right) - \mu \sum_{\sigma} c_{i\sigma}^{\dagger} c_{i\sigma}\,,
\end{equation}
where $\hat{H}_i$
%\begin{equation}
%\end{equation}
is a ``local'' contribution for site $i$. Notice that the intersite hopping term contributes to different sites with a half weight.

The electrical charge polarization and current operators are defined by
\begin{equation}
\bm{\hat{P}} = e \sum_{i\sigma} \bm{R}_i c_{i\sigma}^{\dagger} c_{i\sigma}\,, \qquad
\bm{\hat{j}} = -\ri e  \sum_{ij\sigma} \left( \bm{R}_i - \bm{R}_j\right) t_{ij} c_{i\sigma}^{\dagger} c_{j\sigma}
\end{equation}
and the energy polarization and current operators now take the form
\begin{equation}
\bm{\hat{P}}_Q = \frac12 \sum_{ij\sigma} \left(\bm{R}_i + \bm{R}_j\right) t_{ij} c_{i\sigma}^{\dagger} c_{j\sigma}\,,
\qquad %\\
\bm{\hat{j}}_Q = -\ri  \sum_{ij\sigma} \left( \bm{R}_i - \bm{R}_j\right) \left(\sum_l t_{il} t_{lj}\right) c_{i\sigma}^{\dagger} c_{j\sigma}\,.
\end{equation}
On the other hand, the energy current operator can be rewritten as follows:
\begin{equation}
\bm{\hat{j}}_Q = \sum_{ij\sigma} \left( \bm{R}_i - \bm{R}_j\right) t_{ij} \left(c_{i\sigma}^{\dagger} \frac{\rd c_{j\sigma}}{\rd t} - \frac{\rd c_{i\sigma}^{\dagger}}{\rd t} c_{j\sigma}\right)
%\end{equation}
\qquad \text{or}\qquad
%\begin{equation}
\bm{\hat{j}}_Q \sim\lim\limits_{t'\to t} \frac{1}{2} \left(\frac{\rd}{\rd t} - \frac{\rd}{\rd t'}\right) \bm{\hat{j}}(t,t'),
\label{jQ_vs_jc}
\end{equation}
where we have introduced
\begin{equation}
\bm{\hat{j}}(t,t') = -\ri e  \sum_{ij\sigma} \left( \bm{R}_i - \bm{R}_j\right) t_{ij} c_{i\sigma}^{\dagger}(t) c_{j\sigma}(t').
\end{equation}

Relation (\ref{jQ_vs_jc}) is very important. In the cases when it holds, one can immediately write down an
expressions for the generalized transport integrals $\mathbf{L}_{lm}$~\cite{freericks:035133},
the so-called Boltzmann relations, also known as the Johnson-Mahan theorem in the
theory of metals and semiconductors~\cite{jonson:4223,jonson:9350}.
In the simplest form, when the dynamical screening effects can be ignored,
the Boltzmann relations state that the generalized transport integrals for electrons $\mathbf{L}_{11}$, $\mathbf{L}_{12}=\mathbf{L}_{21}$,
and $\mathbf{L}_{22}$ can be written in terms of one transport function (relaxation time) $\mathbf{I}(\omega)$:
\begin{align}
\mathbf{L}_{lm} &= \frac{\sigma_0}{e^2} \int_{-\infty}^{+\infty} \rd \omega \left[-\frac{\rd f(\omega)}{\rd \omega}\right] \mathbf{I}(\omega) \omega^{l+m-2},
\label{JM}
\end{align}
where
\(
f(\omega) = 1/(\re^{\beta\omega}+1)
\)
is the Fermi distribution function. Once the screening effects and inelastic scattering being taken into account,
 one have to replace these relations by the generalized one~\cite{freericks:035133}.

For the case of noninteracting electrons, the transport function has the form
\begin{equation}
I^{\alpha\beta}(\omega) = \frac{1}{N} \sum_{\bm{k}\sigma} \frac{\partial \varepsilon_{\bm{k}}}{\partial k_{\alpha}} \frac{\partial \varepsilon_{\bm{k}}}{\partial k_{\beta}} \delta(\omega + \mu - \varepsilon_{\bm{k}}),
\end{equation}
where $\varepsilon_{\bm{k}}$ is the band energy (Fourier transform of the hopping integral $t_{ij}$), and we get the known relations
\begin{align}
L_{lm}^{\alpha\beta} &= \frac{\sigma_0}{e^2} \frac{1}{N} \sum_{\bm{k}\sigma} \frac{\partial \varepsilon_{\bm{k}}}{\partial k_{\alpha}} \frac{\partial \varepsilon_{\bm{k}}}{\partial k_{\beta}} \left[-\frac{\rd f(\varepsilon_{\bm{k}}-\mu)}{\rd \varepsilon_{\bm{k}}}\right] (\varepsilon_{\bm{k}} - \mu)^{l+m-2}.
\end{align}

\begin{figure}[tbh]
\centering
\includegraphics[width=0.6\linewidth]{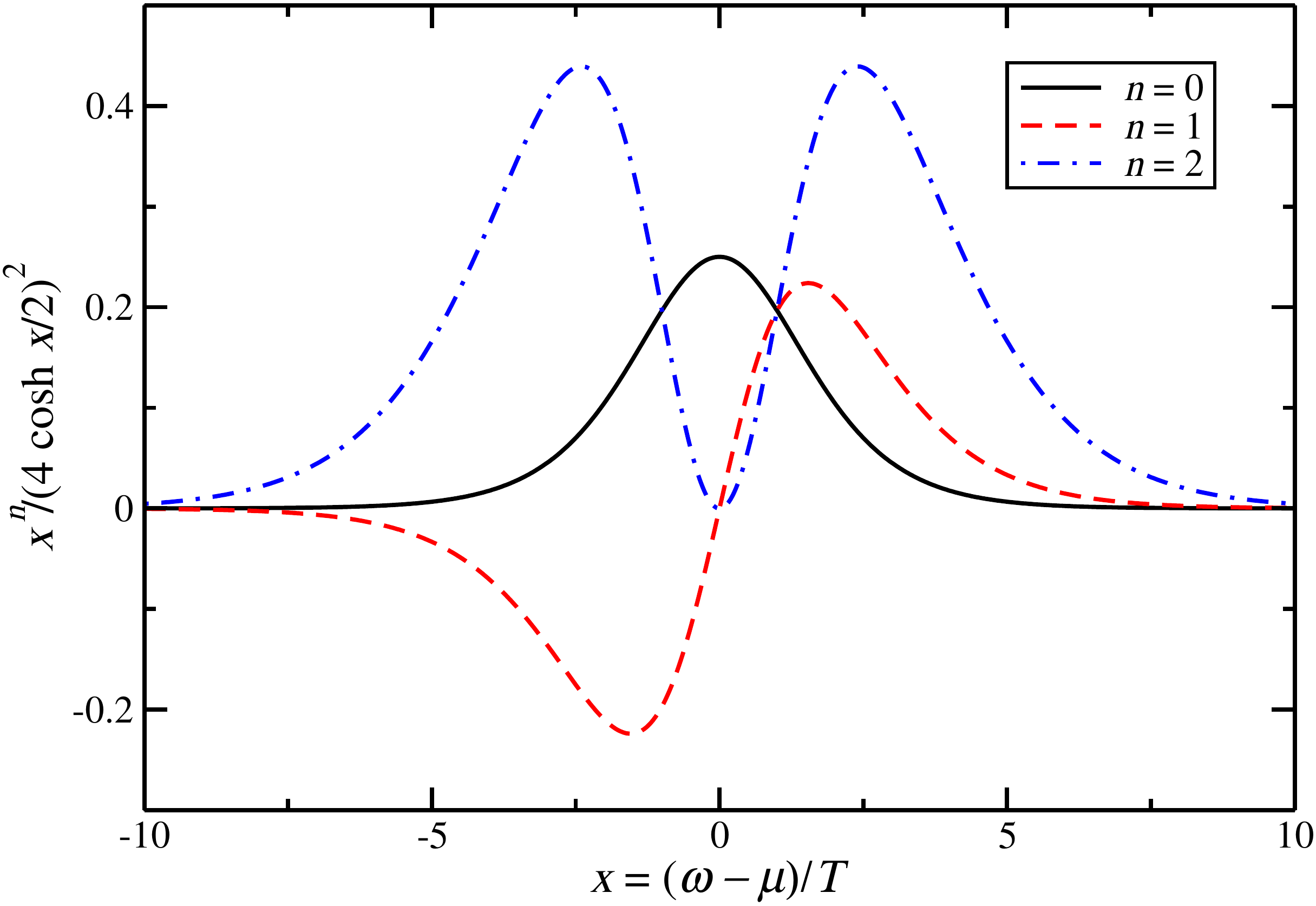}
\caption{(Color online) Dimensionless Fermi window ($n=0$) and its ``moments'' with extrema at points $x_0=0$, $x_1=\pm 1.543404638$, and $x_2=\pm 2.399357280$, respectively.}
\label{fig:Fw_n}
\end{figure}

The main consequences from equations (\ref{JM}) are as follows. The values of the transport integrals $\mathbf{L}_{lm}$ are determined by the features of the transport function $\mathbf{I}(\omega)$ only within the Fermi window ($n=0$) of width $\sim 4T$ and its ``moments''
\begin{equation}
(\omega - \mu)^n \left[-\frac{\rd f(\omega - \mu)}{\rd \omega}\right] = \frac{(\omega - \mu)^n}{4T\cosh^2 \dfrac{\omega - \mu}{2T}} = \dfrac{T^{n-1}}{4} \; \frac{x^n}{\cosh^2 x/2}\,, \qquad x=\dfrac{\omega - \mu}{T}
\end{equation}
which spread over a larger energy interval (see figure~\ref{fig:Fw_n}).
For the metals, the transport function $\mathbf{I}(\omega)$ is almost constant within the Fermi window which results in a weak temperature dependence of electrical dc conductivity, the absence of thermoelectric power, and the linear temperature dependence of thermal conductivity in agreement with the Wiedemann-Franz law. On the other hand, it is obvious from equation~(\ref{ZT}) that large values of the \textit{figure of merit} could be obtained when the thermal conductivity is strongly reduced, and it was shown by Mahan and Sofo~\cite{mahan:7436} that these could be achieved for the $\delta$-function like transport function $I(\omega)\sim \delta(\omega-\varepsilon_0)$, when the electronic contribution in thermal conductivity~(\ref{thermcond}) vanishes.

In order to get high values of the \textit{figure of merit} $ZT$ (\ref{ZT}) for thermoelectric properties in some temperature interval, one has to look for the systems with strongly asymmetric, within the Fermi window, transport function $\mathbf{I}(\omega)$ with sharp peak at $(\omega-\mu)/T=1.543404638$ or $-1.543404638$ to enhance the Seebeck coefficient and minima at $(\omega-\mu)/T=\pm 2.399357280$ to reduce the thermal conductivity. That is why, the materials with strong electron correlations as well as disordered systems attract much attention as prospective thermoelectric applications. This is due to the strong variability of their band structure and density of states in the Fermi window depending on the chemical structure and doping. In particular, it was already shown that the doping of Mott insulator can strongly improve its thermoelectric properties~\cite{freericks:195120,zlatic:155101}. In this case, the doping shifts the chemical potential into the lower or upper Hubbard band which produces a strongly asymmetric density of states (d.o.s.) and transport function. Besides the Hubbard type local Coulomb interaction, one can also consider the non-local contributions which can produce additional enhancements.

\section{Charge and energy current operators and transport coefficients for the Falicov-Kimball model with correlated hopping}\label{sec:ch-FKM}

A simplest model in the theory of strongly correlated electron systems is the Falicov-Kimball one~\cite{falicov:997}, which considers the local interaction between the itinerant $d$ electrons and localized $f$ electrons. It is a binary alloy type model and it has an exact solution in the dynamical mean field theory (DMFT)~\cite{freericks:1333}. Its enhancement by correlated hopping was also considered, and the DMFT solutions with a nonlocal self-energy were obtained~\cite{schiller:15660,shvaika:075101,shvaika:119901}.

The Hamiltonian of the Falicov-Kimball model with correlated hopping contains two contributions: the local correlations $H_{\textrm{loc}}$ and the nearest-neighbour intersite hopping $H_t$ on the $D$-dimensional hypercubic lattice
\begin{align}
H &= H_{\textrm{loc}} + H_t\,, \qquad
H_{\textrm{loc}} = \sum_i \left[Un_{id}n_{if} - \mu_f n_{if} - \mu_d n_{id}\right],
\nonumber\\
H_t&=\frac{1}{2\sqrt{D}}\sum_{\langle ij\rangle}\Bigl[  t_1d_i^{\dag}d_j +
  t_2d_i^{\dag}d_j\left(n_{if}+n_{jf}\right) + t_3d_i^{\dag}d_jn_{if}n_{jf}\Bigr].
\end{align}
It is convenient to project on the occupation of the local $f$-states by introducing operators $P_i^+=n_{if}$, $P_i^-=1-n_{if}$ and to rewrite the Hamiltonian in matrix notations
\begin{align}
  H_t&=\frac{1}{2\sqrt{D}}\sum_{\langle ij\rangle}\Bigl[
  t_{ij}^{++}P_i^+d_i^{\dag}d_jP_j^+ \!+ t_{ij}^{--}P_i^-d_i^{\dag}d_jP_j^-
  \!+t_{ij}^{+-}P_i^+d_i^{\dag}d_jP_j^- \!+
  t_{ij}^{-+}P_i^-d_i^{\dag}d_jP_j^+\Bigr]
  \nonumber\\
  &=\frac{1}{2\sqrt{D}}\sum_{\langle ij\rangle}\bm{d}_i^{\dag}\mathbf{t}_{ij}\bm{d}_j,
\qquad  \bm{d}_i=\begin{pmatrix} P_i^+ d_i \\ P_i^- d_i \end{pmatrix}, \qquad
  \mathbf{t}_{ij}=\begin{pmatrix}
                    t_{ij}^{++} & t_{ij}^{+-} \\
                    t_{ij}^{-+} & t_{ij}^{--}
                  \end{pmatrix}.
\end{align}
The connection between the elements of the hopping matrix $\mathbf{t}_{ij}$ and the initial hopping amplitudes is as follows
\begin{align}
  t^{--}&=t_1\,,             &t_1&=t^{--}\,,
  \nonumber\\
  t^{+-}&=t^{-+}=t_1+t_2\,,  &t_2&=t^{+-(-+)}-t^{--}\,,
  \nonumber\\
  t^{++}&=t_1+2t_2+t_3\,,    &t_3&=t^{++}+t^{--}-t^{+-}-t^{-+}\,.
\end{align}

In infinite dimensions $D\to\infty$, the self-energy is nonlocal but an irreducible over the hopping part $\bm{\Xi}(\omega)$ of matrix Green's function $\mathbf{G}_{\bm{k}}(\omega)$, when one performs an expansion around the atomic limit, is local which allows one to develop the DMFT approach for the systems with correlated hopping~\cite{shvaika:075101,shvaika:119901}. In terms of this irreducible part, the lattice Green's function can be written as follows:
\begin{equation}\label{G_k}
  \mathbf{G}_{\bm{k}}(\omega)=\left[\bm{\Xi}^{-1}(\omega) - \mathbf{t}_{\bm{k}}\right]^{-1},
\end{equation}
where for the hopping matrix, we neglect the possible chirality
\begin{equation}\label{t_mtrx}
  \mathbf{t}_{\bm{k}}=\begin{pmatrix}
                    t^{++} & t^{+-} \\
                    t^{-+} & t^{--}
                  \end{pmatrix}\varepsilon_{\bm{k}}\,,
\end{equation}
and assume that each term is proportional to the unperturbed band energy
\begin{equation}\label{band_energy}
 \varepsilon_{\bm{k}}=\frac{1}{\sqrt{D}} \sum_{\alpha=1}^{D} \cos k_{\alpha}\,.
\end{equation}
Below we shall use the hopping amplitude over empty states as an energy unit: $t^{--}=t_1=1$.

An irreducible part $\bm{\Xi}(\omega)$ as well as the matrix of the $\lambda$-fields
$\bm{\Lambda}(\omega)=\left\|\lambda^{\alpha\beta}(\omega)\right\|$  are solutions of the system of equations
\begin{equation}\label{SIAM}
  \frac1N\sum_{\bm{k}}
  \left[\bm{\Xi}^{-1}(\omega) - \mathbf{t}_{\bm{k}}\right]^{-1}
  =\left[\bm{\Xi}^{-1}(\omega) - \bm{\Lambda}(\omega)\right]^{-1} = \mathbf{G}_{\text{imp}}(\omega),
\end{equation}
where the expressions for the Green's function $\mathbf{G}_{\text{imp}}(\omega)$ of the local impurity problem for the Falicov-Kimball model are known
\begin{align}
  G_{\text{imp}}^{++}(\omega) & =
  \frac{\langle P^+\rangle}{\omega+\mu_d - U - \lambda^{++}(\omega)}\,,
  \\
  G_{\text{imp}}^{--}(\omega) & =
  \frac{\langle P^-\rangle}{\omega+\mu_d - \lambda^{--}(\omega)}\,,
  \\
  G_{\text{imp}}^{+-}(\omega) & = G_{\text{imp}}^{-+}(\omega) = 0.
\end{align}

\begin{figure}[tb]
\centering
\includegraphics[scale=0.19,clip]{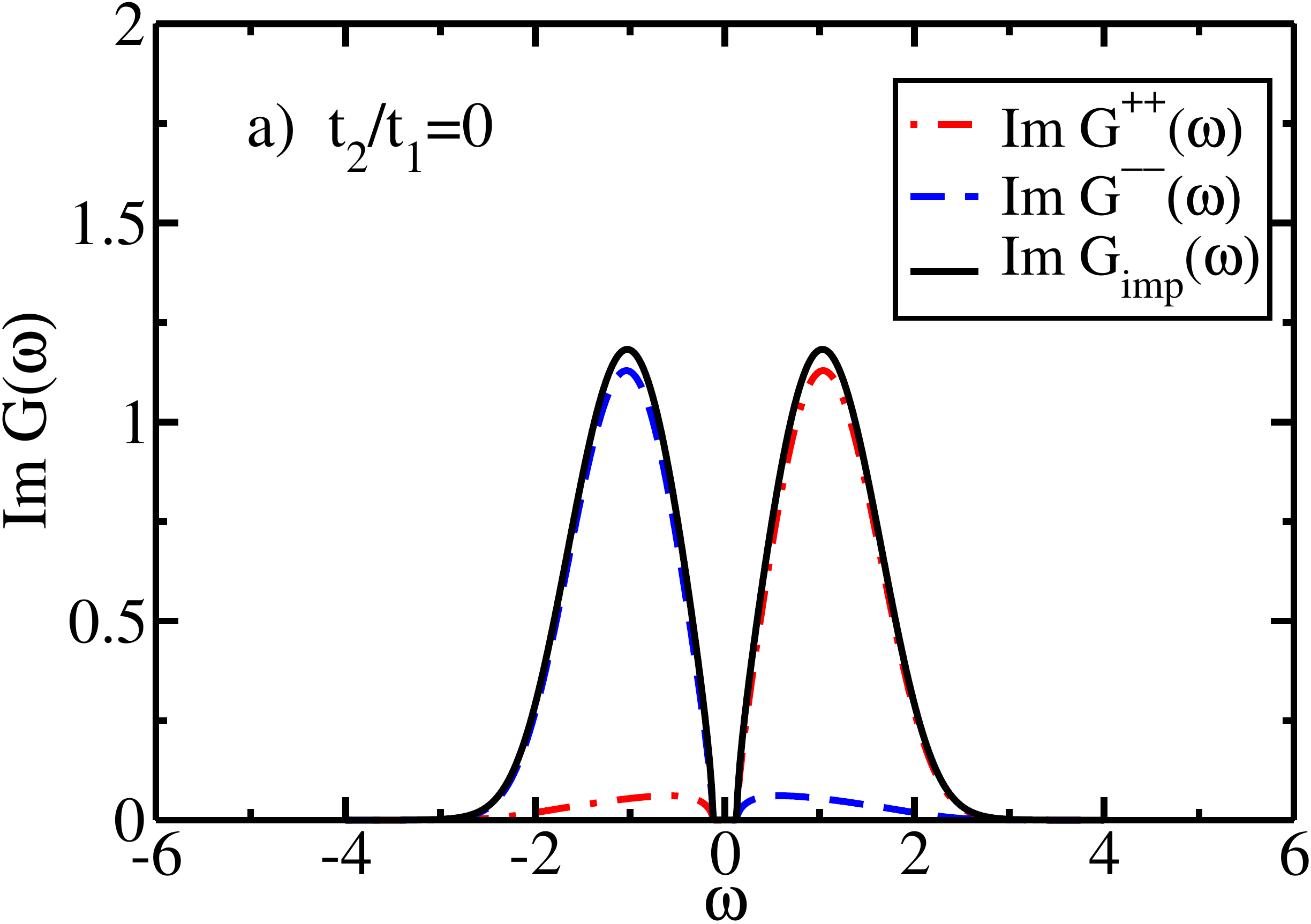}
\quad
\includegraphics[scale=0.19,clip]{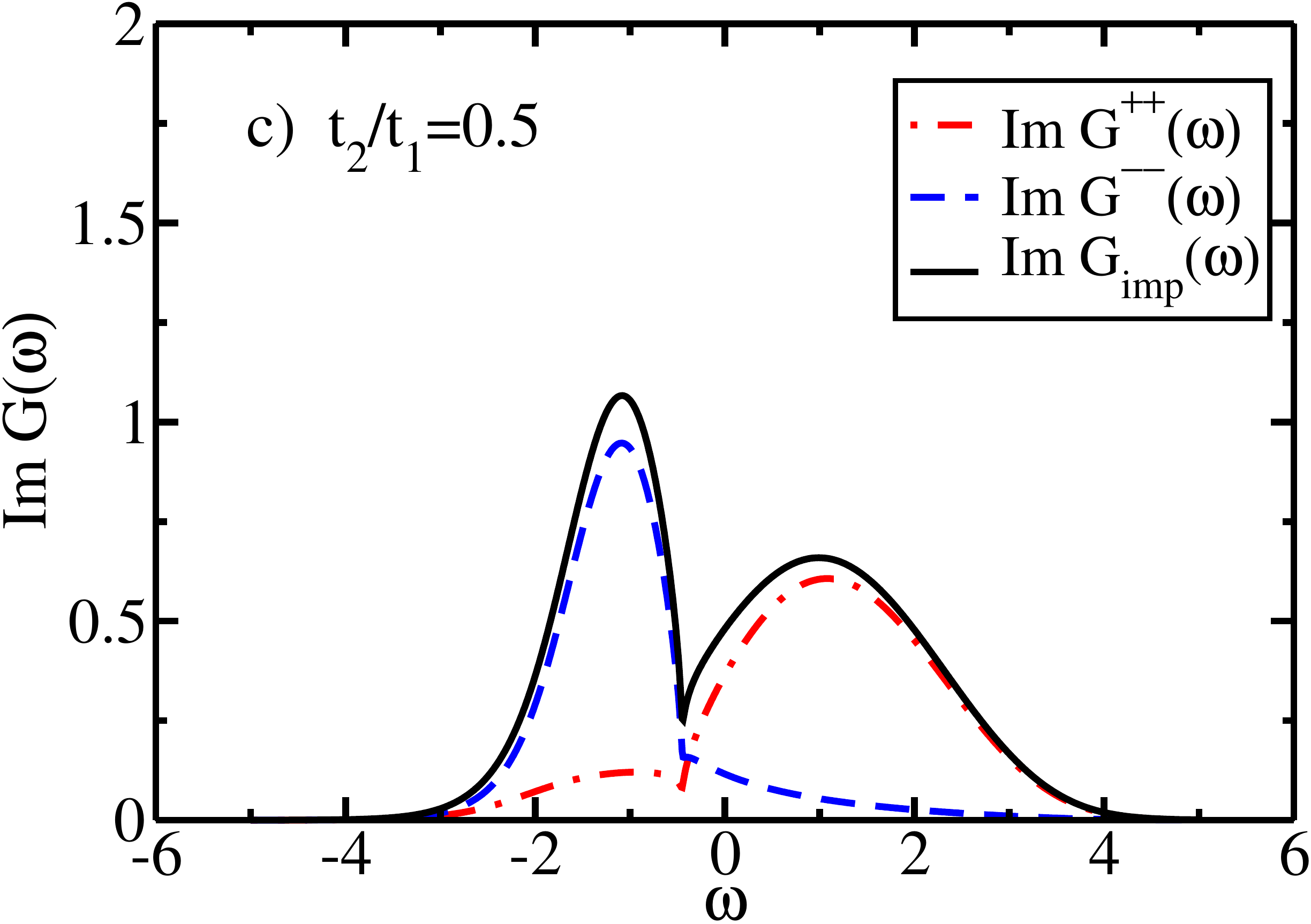}
\quad
\includegraphics[scale=0.19,clip]{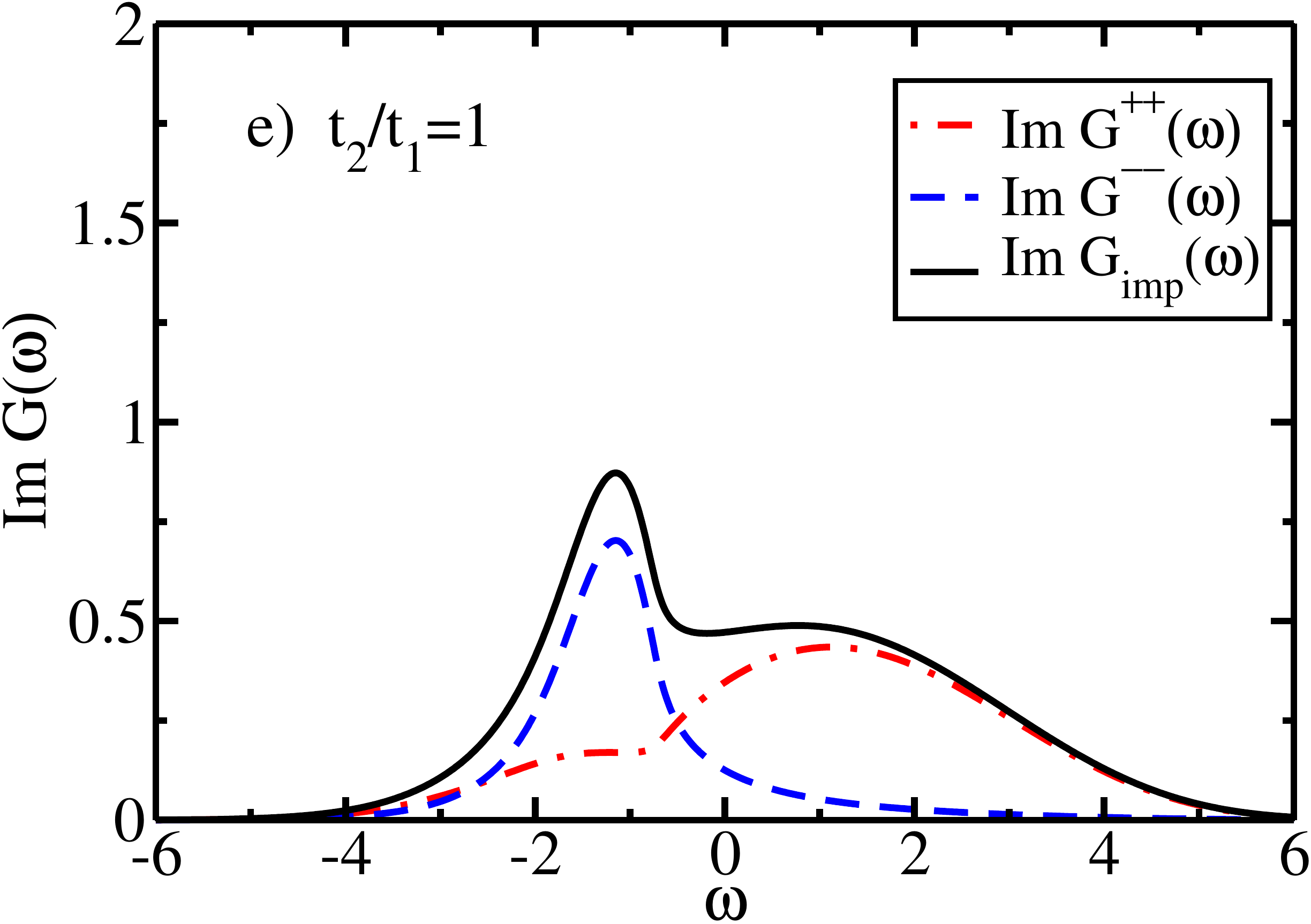}
\\ [1em]
\includegraphics[scale=0.19,clip]{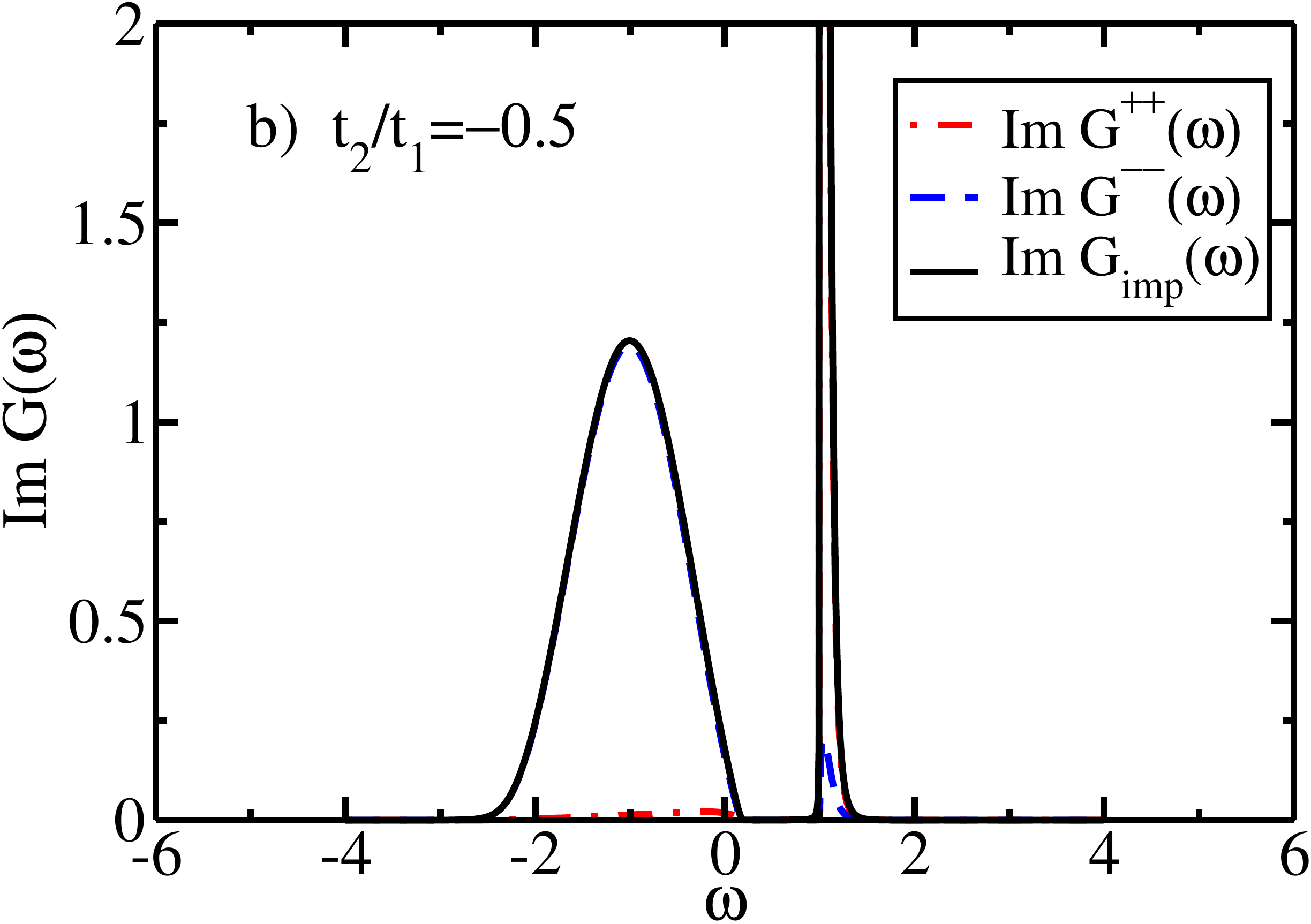}
\quad
\includegraphics[scale=0.19,clip]{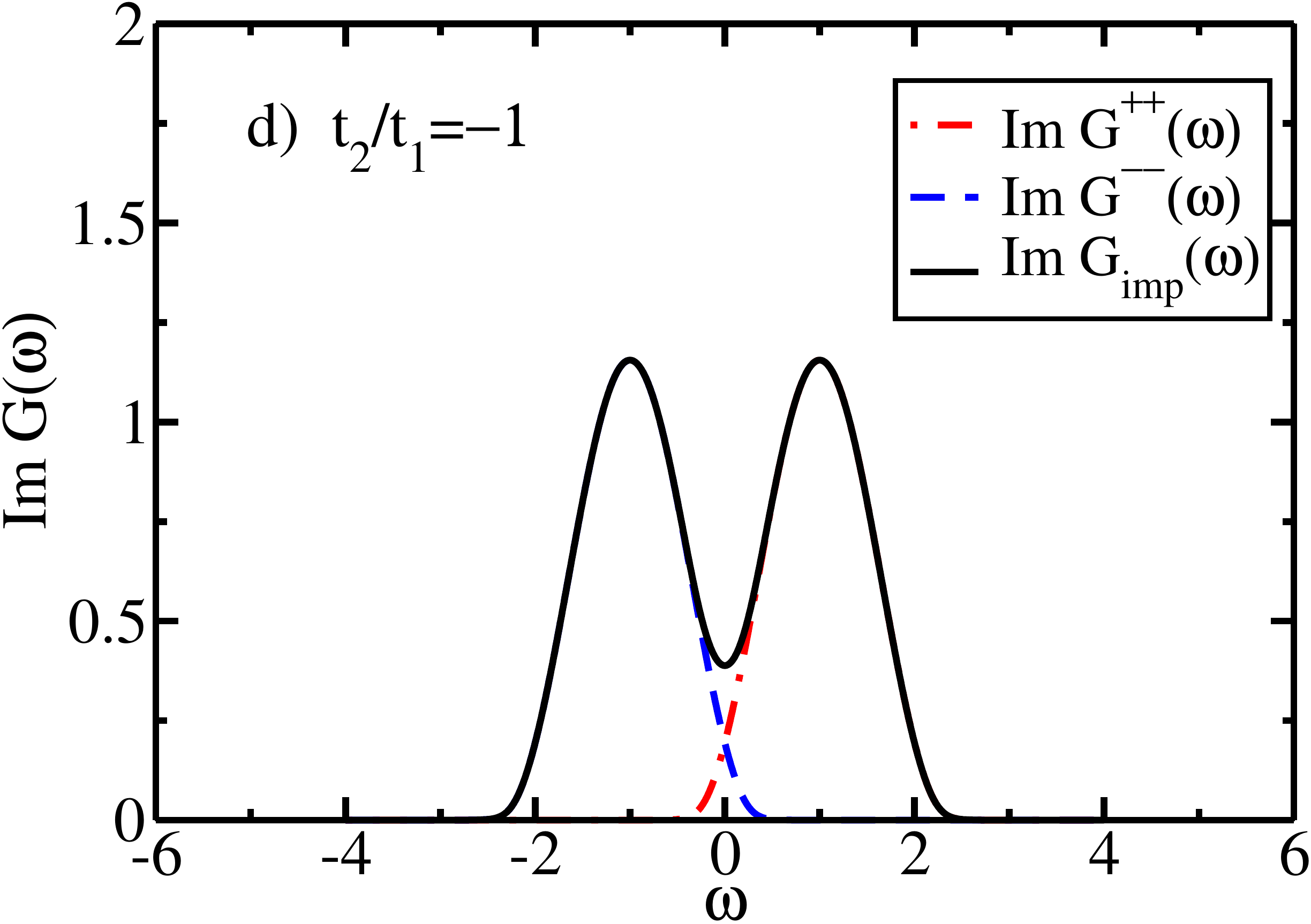}
\quad
\includegraphics[scale=0.19,clip]{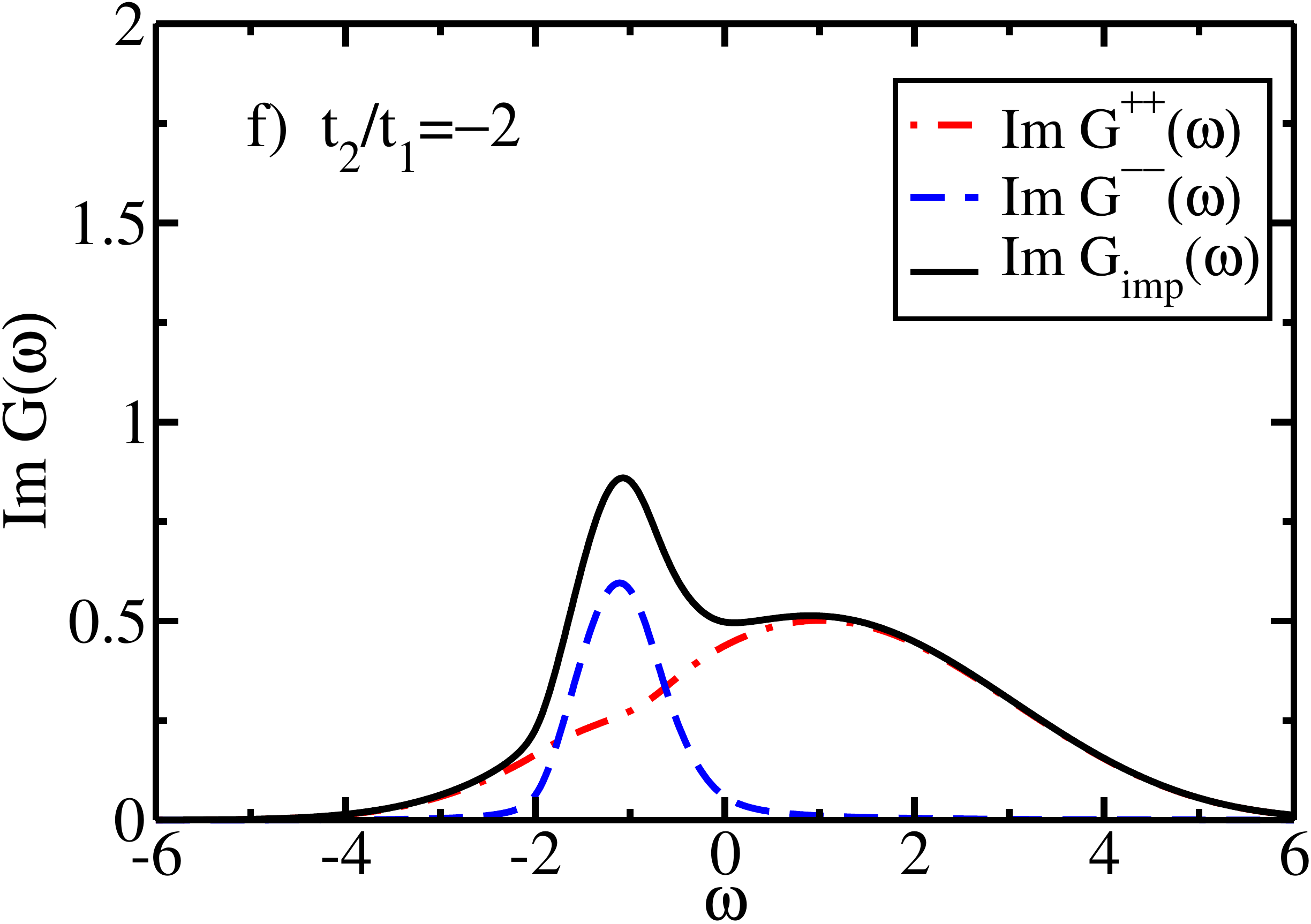}
\caption{(Color online) Evolution of the one-particle d.o.s. with a change of the correlated hopping amplitude $t_2$ ($t_3=0$, $t_1=1$, $U=2$) (from reference~\cite{shvaika:119901}).}\label{fig:dos}
\end{figure}

Previous investigations of the Falicov-Kimball model with correlated hopping~\cite{shvaika:075101,shvaika:119901} have elucidated several special cases with a strong reconstruction of the one-particle d.o.s., which in the considered case is equal to:
\begin{equation}
A_d(\omega) = -\frac{1}{\pi} \sum_{\alpha,\beta=\pm} \Im G_{\text{imp}}^{\alpha\beta}(\omega).
\end{equation}
These crossover points correspond to the special forms of the hopping matrix (\ref{t_mtrx}):
\begin{enumerate}\parskip=0pt\itemsep=3pt
\item The regular Falicov-Kimball model without correlated hopping corresponds to the case of $t_2=t_3=0$, when all components of the hopping matrix (\ref{t_mtrx}) are the same $t^{\alpha\beta}=t_1$ and its determinant is equal to zero $\det \mathbf{t}_{\bm{k}} = 0$ [figure~\ref{fig:dos}~(a)].
\item The case when one or both of the diagonal components of the hopping matrix (\ref{t_mtrx}) are equal to zero, e.g. $t^{++} = 0$ for $t_2=-\frac12 (t_1+t_3)$, with strong reduction of the band widths and an \textit{orthogonality singularity} at the edge of one or both bands [figure~\ref{fig:dos}~(b)].
\item The case of diagonal hopping matrix (\ref{t_mtrx}), when $t_2=-t_1$ and $t^{+-} = t^{-+} = 0$ and d.o.s. consists of independent bands [figure~\ref{fig:dos}~(d)].
\end{enumerate}
In other cases, the correlated hopping results in the break of the electron-hole symmetry with strong asymmetry of the d.o.s. and in the spread of the bands due to an effective increase of the hopping amplitude.
The similar behaviour could be also predicted for the transport function and will be considered below.

The Falicov-Kimball model with correlated hopping is similar to the binary alloy model with off-diagonal disorder, and the first calculations of the thermoelectric power for the last one were done by Hoshino and Niizeki~\cite{hoshino:1320} based on the Mott's relation and for the special case of the zero determinant $\det \mathbf{t}_{\bm{k}} = 0$ of hopping matrix~(\ref{t_mtrx}). Below we shall provide a more general approach.

First of all, one has to rederive the charge and energy current operators in the presence of correlated hopping. Starting from the expression for charge polarization which contains contributions both from the $d$ and $f$ particles
\(
\bm{\hat{P}} = \sum_i \bm{R}_i \left(z_d \hat{n}_{id} + z_f \hat{n}_{if}\right),
\)
one can get an expression for the charge current in the presence of correlated hopping
\begin{equation}
\bm{\hat{j}}_c = -\ri z_d \sum_{\substack{\langle ij\rangle \\ \alpha\beta=\pm}} \left( \bm{R}_i - \bm{R}_j\right) t_{ij}^{\alpha\beta} P_i^{\alpha} d_{i}^{\dagger} d_{j} P_j^{\beta}.
\end{equation}

The dc electric conductivity
\begin{equation}\label{dc_cond}
\sigma_{\text{dc}} = -z_d^2 \lim\limits_{\Omega\to0} \dfrac{1}{\Omega} \Img \chi(\Omega)
\end{equation}
is connected with the current-current Green's function, which in the considered case has the following form:
\begin{equation}\label{chi_jj}
\chi(\ri\nu) = \frac{T}{2}\sum_m \frac{1}{N} \sum_{\bm{k}} \sum_{\alpha\beta\alpha'\beta'=\pm} t^{\alpha\beta} G_{\bm{k}}^{\beta\alpha'} (\ri\omega_m) t^{\alpha'\beta'} G_{\bm{k}}^{\beta'\alpha}(\ri\omega_m+\ri\nu),
\end{equation}
where for the $D\to\infty$ hypercubic lattice with unperturbed band energy (\ref{band_energy}), a rigorous replacement $\left({\rd \varepsilon_{\bm{k}}}/{\rd \bm{k}}\right)^2 \to \frac12$ was used. After substituting the components of the lattice Green's function (\ref{G_k}) in (\ref{chi_jj}) and performing summation over the momentum $\bm{k}$, one can get an expression for the dc conductivity in a standard form through the transport function
\begin{equation}
\sigma_{\textrm{dc}} = z_d^2 L_{11}\,, \qquad
L_{11} = \int_{-\infty}^{+\infty} \rd \omega \left[-\frac{\rd f(\omega)}{\rd \omega}\right] I(\omega)\,.
\end{equation}
Analytic expressions for the transport function $I(\omega)$ in the homogeneous phase in terms of the solutions of the DMFT impurity problem (\ref{SIAM}) were derived and are presented in the appendix~\ref{sec:append}.

On the other hand, the energy polarization operator for the Falicov-Kimball model with correlated hopping is equal to
\begin{equation}
\bm{\hat{P}}_Q = \sum_i \bm{R}_i \left[Un_{id}n_{if} - \mu_f n_{if} - \mu_d n_{id}\right] + \frac12 \sum_{\substack{\langle ij\rangle \\ \alpha\beta=\pm}} \left(\bm{R}_i+\bm{R}_j\right) t_{ij}^{\alpha\beta}P_i^{\alpha}d_i^{\dag}d_jP_j^{\beta}.
\end{equation}
Now, from the continuity equation (\ref{jQ_def}) we get an expression for the energy current operator
\begin{equation}
\bm{\hat{j}}_Q = -\frac{\ri}{2} \sum_{\substack{\langle ij \rangle\\ \alpha\beta=\pm}} \left(\bm{R}_i-\bm{R}_j\right) P_i^{\alpha}d_i^{\dag}d_jP_j^{\beta} \left\{ t_{ij}^{\alpha\beta} \left[U(n_{if}+n_{jf}) - 2\mu_d \right] + \sum_{l\gamma} t_{il}^{\alpha\gamma} P_l^{\gamma} t_{lj}^{\gamma\beta}  \right\}
\end{equation}
and one can check that it is equal to
\begin{equation}
\bm{\hat{j}}_Q = \frac{1}{2} \sum_{\substack{\langle ij\rangle\\ \alpha\beta=\pm}} \left(\bm{R}_i-\bm{R}_j\right) t_{ij}^{\alpha\beta}  \left\{ P_i^{\alpha}d_i^{\dag}\frac{\rd (d_jP_j^{\beta})}{\rd t} - \frac{\rd (P_i^{\alpha}d_i^{\dag})}{\rd t}d_jP_j^{\beta} \right\}.
\end{equation}
This means that the connection (\ref{jQ_vs_jc}) between the energy and charge current operators holds in the case of correlated hopping, and the Boltzmann relations (\ref{JM}) can be also used in this case. The numerical results for the charge and heat transport in the systems with correlated hopping are presented in the next section.

\section{Charge and heat transport in presence of correlated hopping}\label{sec:numerics}

It was already mentioned above that for different values of correlated hopping, different shapes of the one-particle d.o.s. can be realized (figure~\ref{fig:dos}) and the crossover from one regime to another takes place at special forms of the hopping matrix~(\ref{t_mtrx}): either its determinant or some matrix elements are equal to zero. Let us check the behaviour of the charge and heat transport in the vicinity of these crossover points.

\begin{figure}[tbhp]
\centering
\includegraphics[scale=0.19]{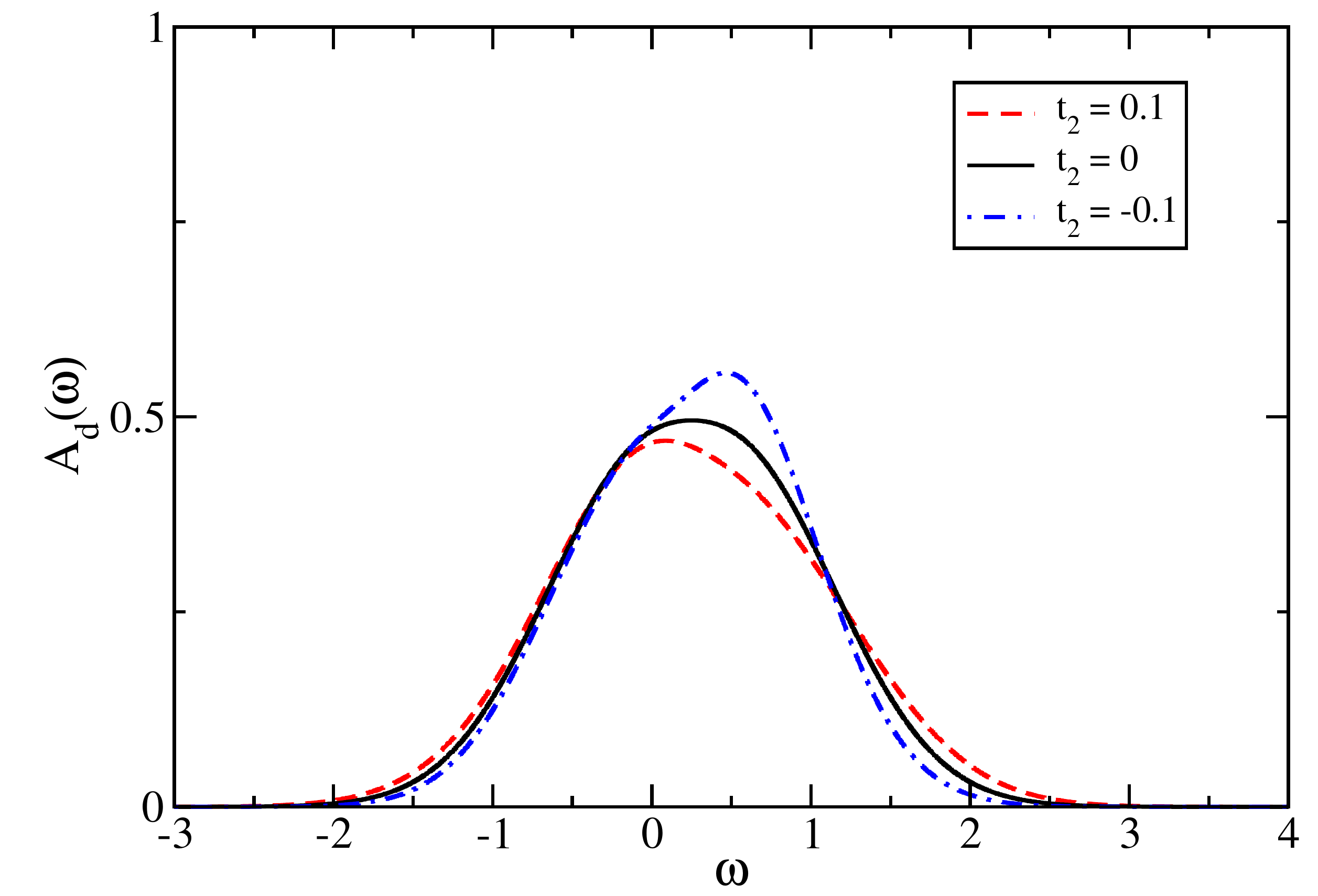}
\includegraphics[scale=0.19]{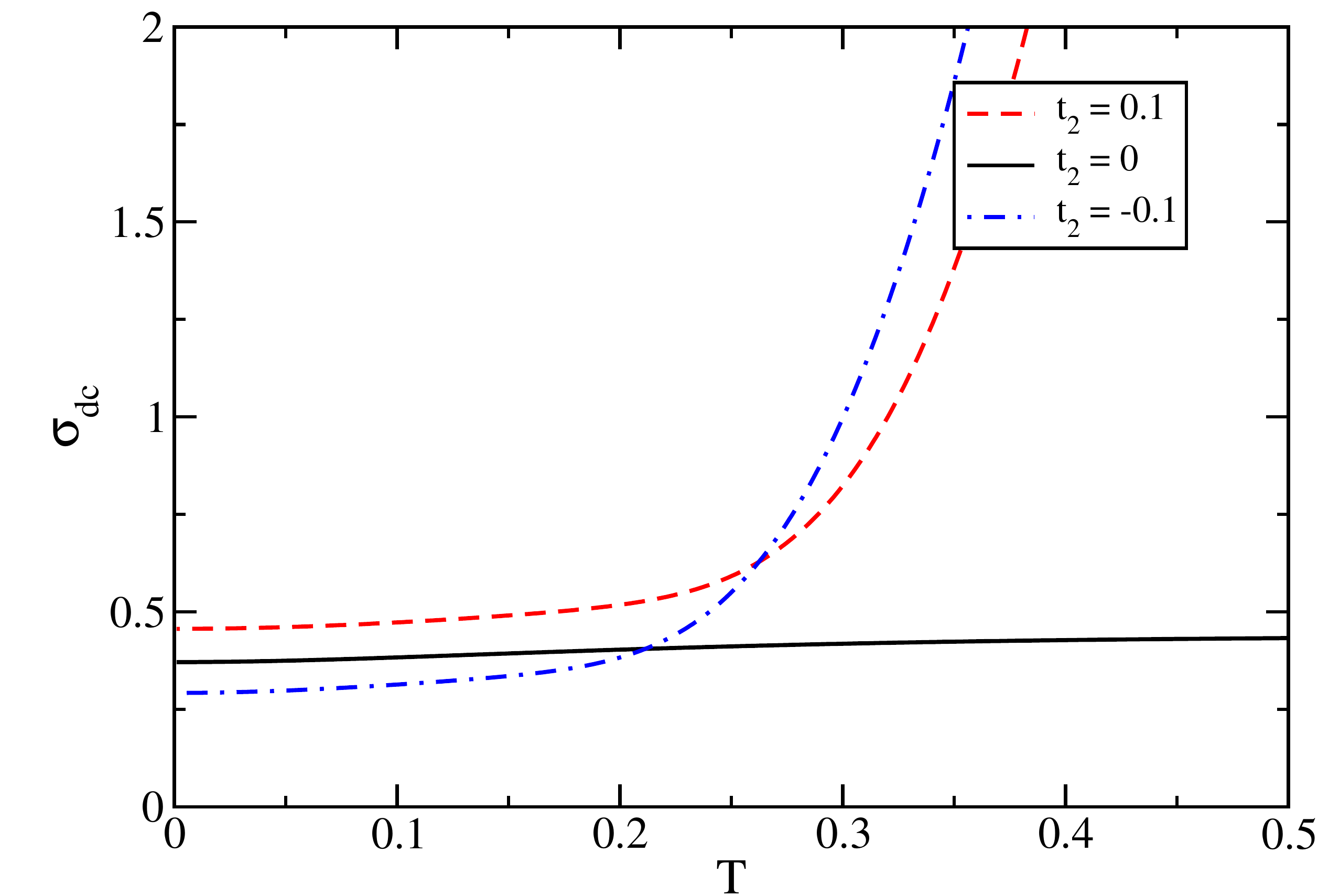}
\includegraphics[scale=0.19]{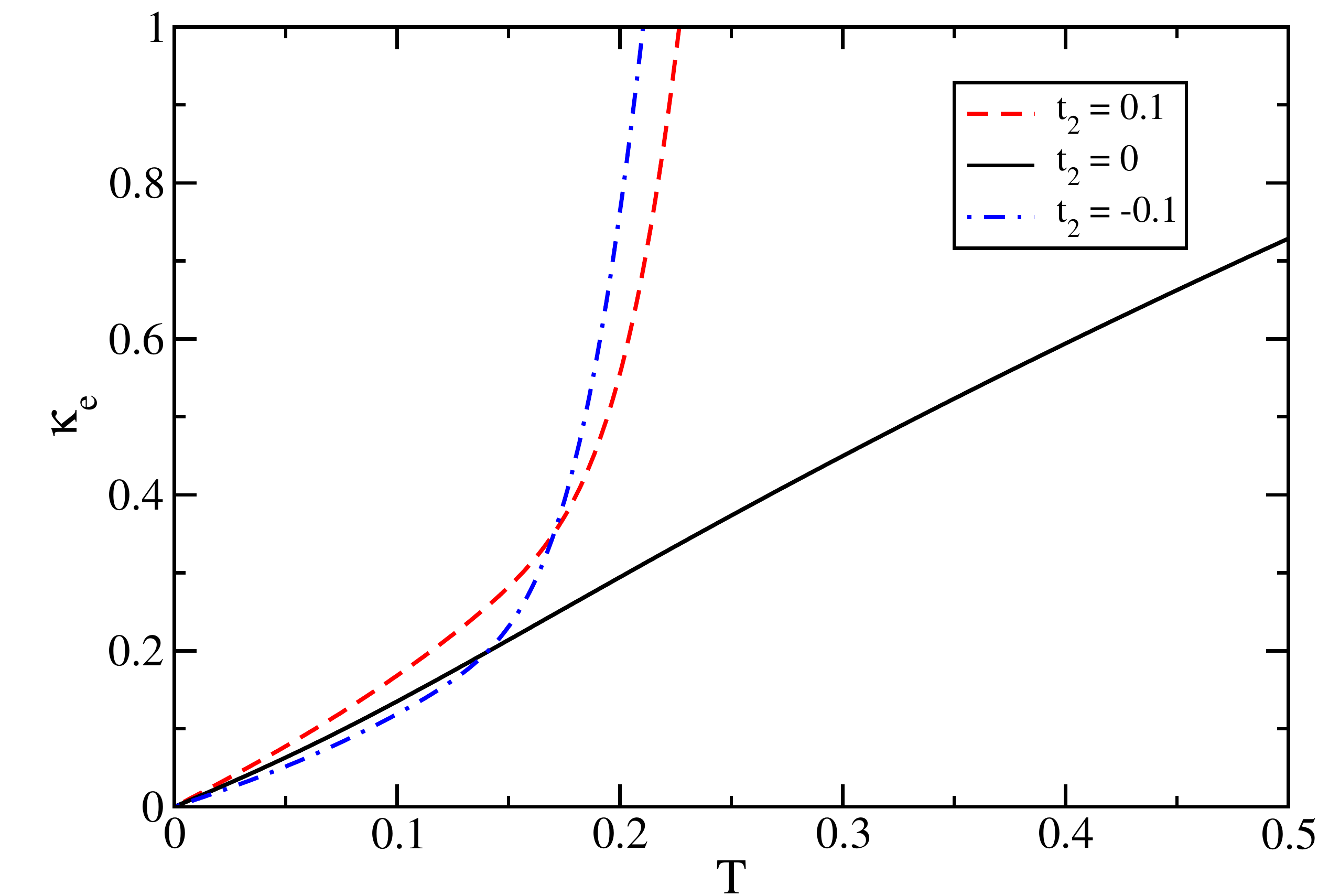}
\\
\includegraphics[scale=0.19]{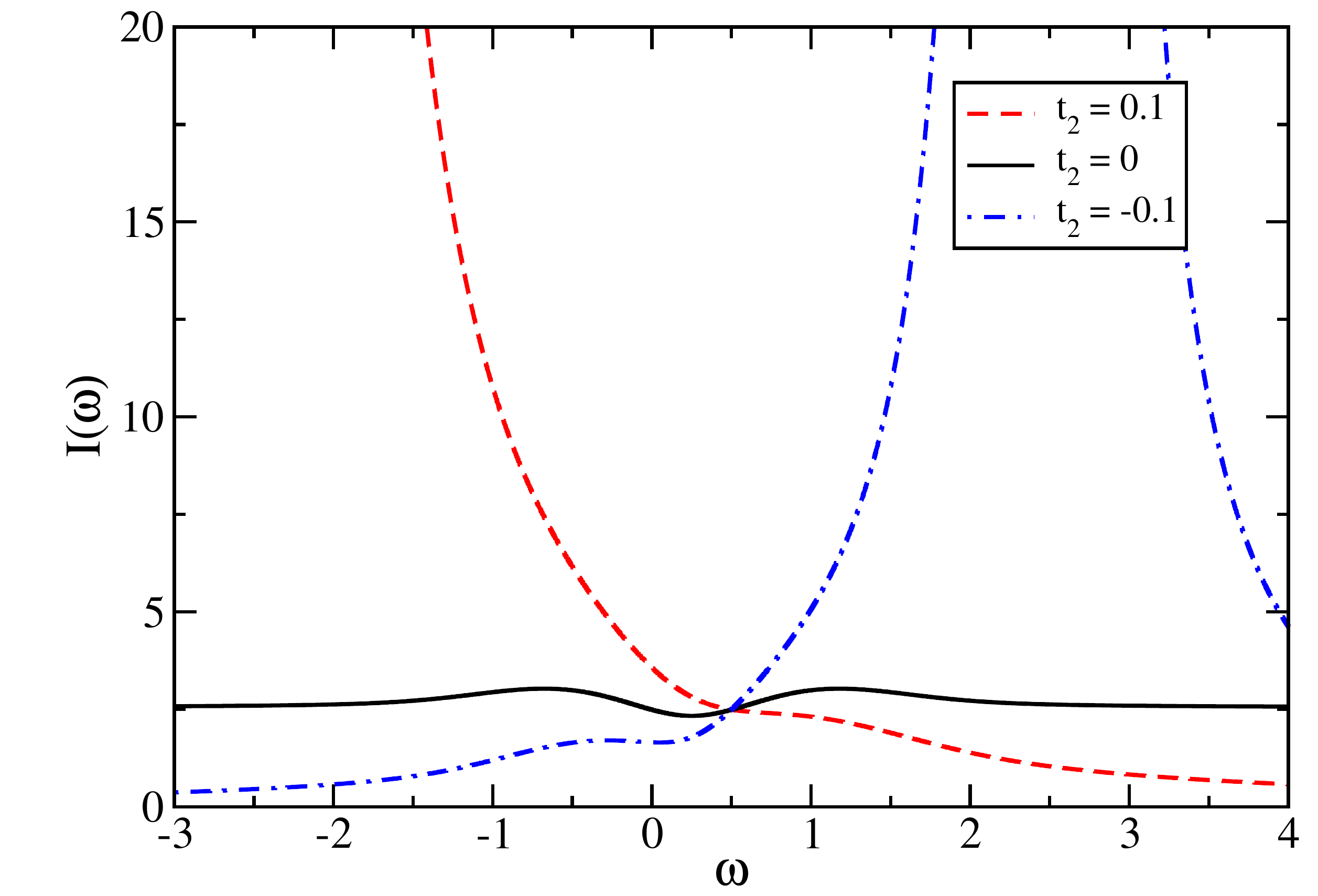}
\includegraphics[scale=0.19]{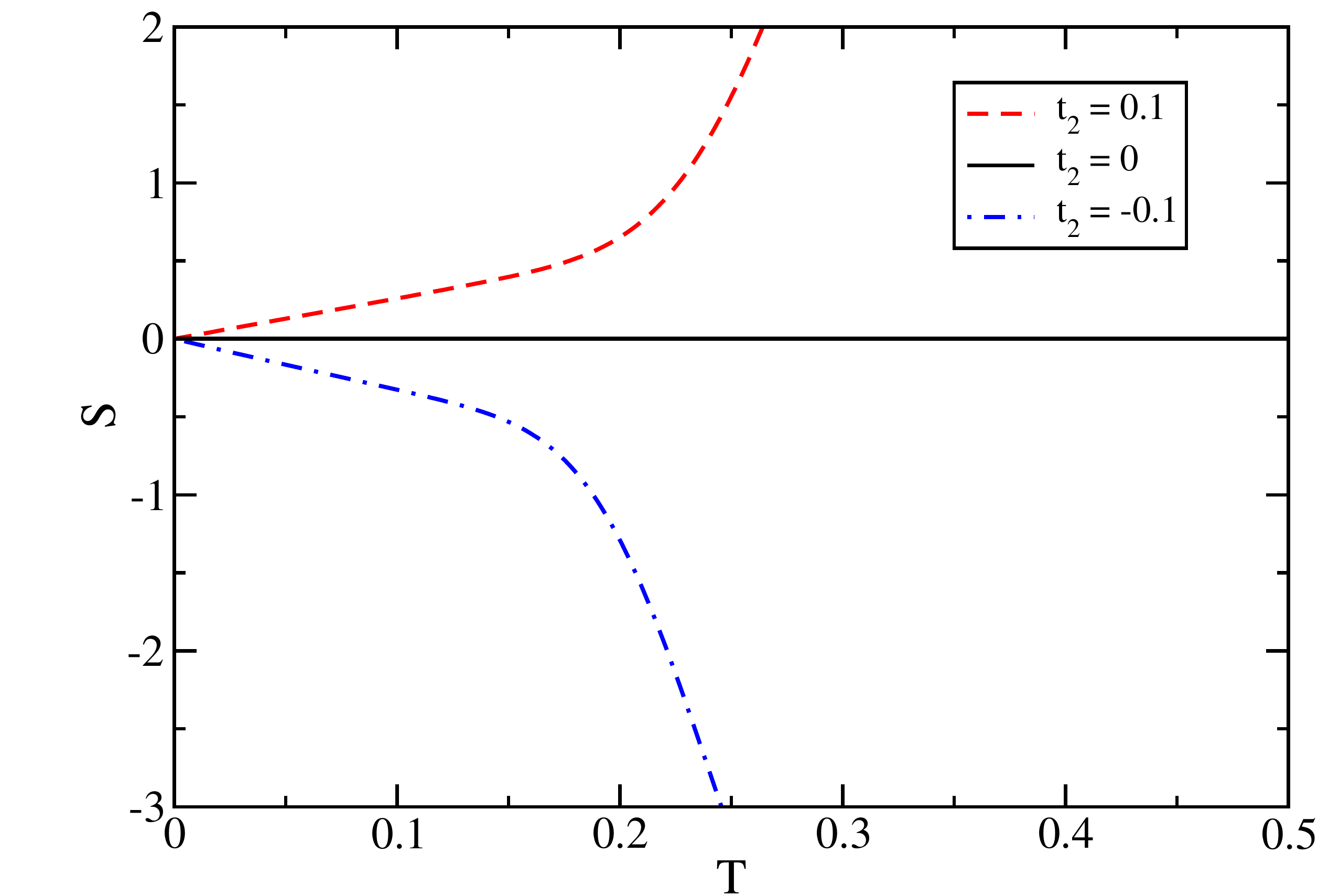}
\includegraphics[scale=0.19]{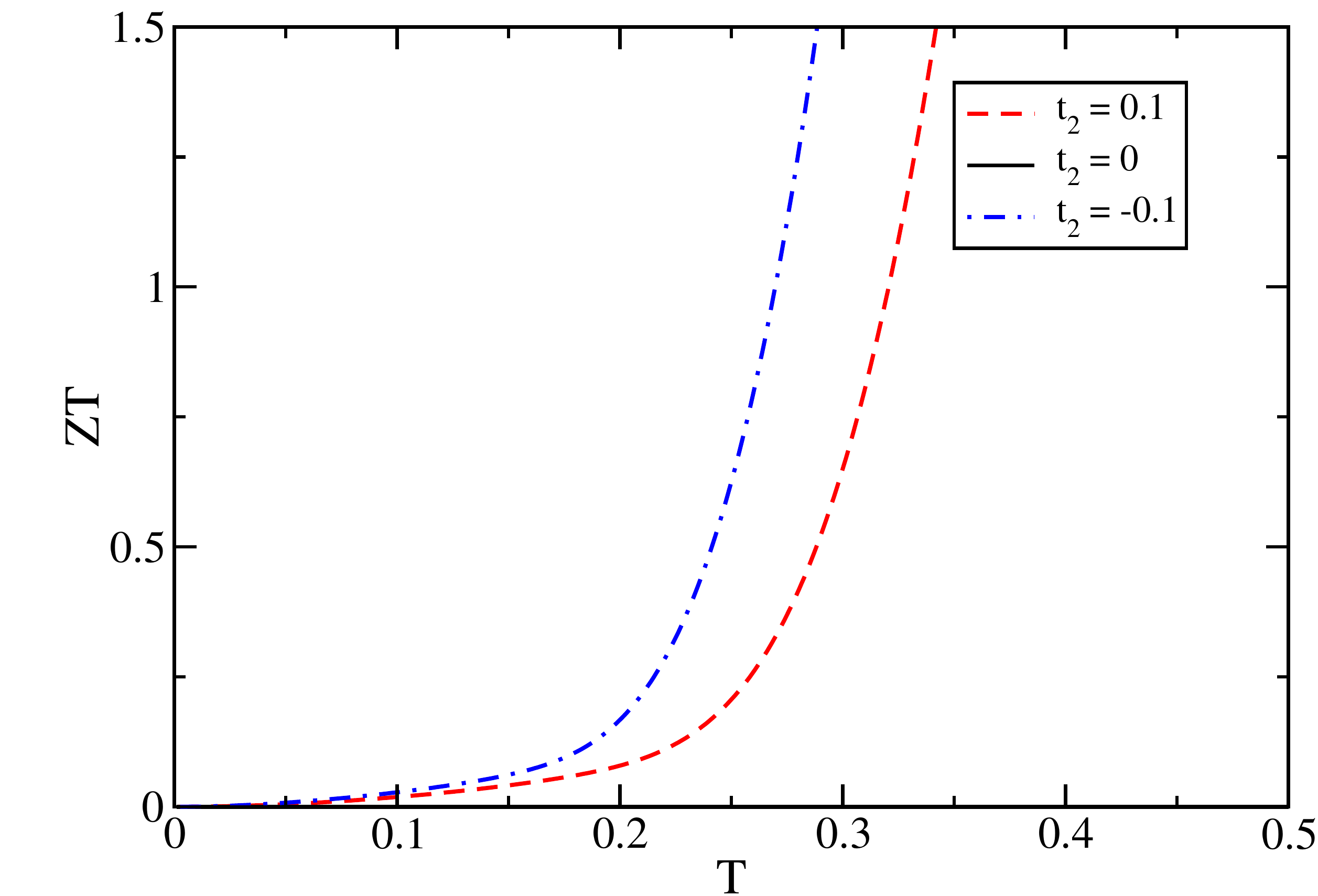}
\caption{(Color online) Density of states $A_d(\omega)$, transport function $I(\omega)$, and temperature dependences of the dc electric $\sigma_{\textrm{dc}}$ and thermal $\kappa_{\textrm{e}}$ conductivities, Seebeck coefficient $S$, and \textit{figure of merit} $ZT$ at half filling $n_f=n_d=1$ for $U=0.5$, $t_1=1$, and $t_2=0.1$, $0$, $-0.1$.}\label{fig:U05_t2_0}
\end{figure}

First of all, we consider the small values of correlated hopping $t_2$ (below we shall put $t_3=0$). In the absence of correlated hopping $t_2=0$ and for small values of the Coulomb interaction $U$ (figure~\ref{fig:U05_t2_0}),
the shape of the d.o.s. $A_d(\omega)$ deviates by a small amount from the unperturbed one (Gaussian for the $D\to\infty$ hypercubic lattice). In this case, the transport function $I(\omega)$ slightly varies in the vicinity of the chemical potential value $\mu_d=U/2$ and approaches the constant value at large frequencies. The last feature is caused by the Gaussian d.o.s.~\cite{freericks:195120}, when for the large energy values we still have an exponentially small density of states with a finite relaxation rate (see appendix~\ref{sec:append}), which cannot be observed in real systems. Thus, we consider only the small and moderate temperature values when the Fermi window and its moments (see figure~\ref{fig:Fw_n}) are inside the features of the d.o.s.

The correlated hopping $t_2$ being switched on leads to slight changes of the d.o.s., which become asymmetric, and to tremendous changes in the shape and values of the transport function $I(\omega)$ with a strong enhancement at the band edges. At low temperatures, the effect of correlated hopping on the electric $\sigma_{\textrm{dc}}$ and thermal $\kappa_{\textrm{e}}$ conductivities is minor and leads to the appearance of the Seeback effect due to electron-hole asymmetry. At higher temperature values, we observe an enhancement of all transport coefficients because the moments of the Fermi window (figure~\ref{fig:Fw_n}) now start to cover the sharp features of the transport function at the band edges and this behaviour
could be non-physical as a consequence of the Gaussian~d.o.s.

\begin{figure}[!b]
\centering
\includegraphics[scale=0.19]{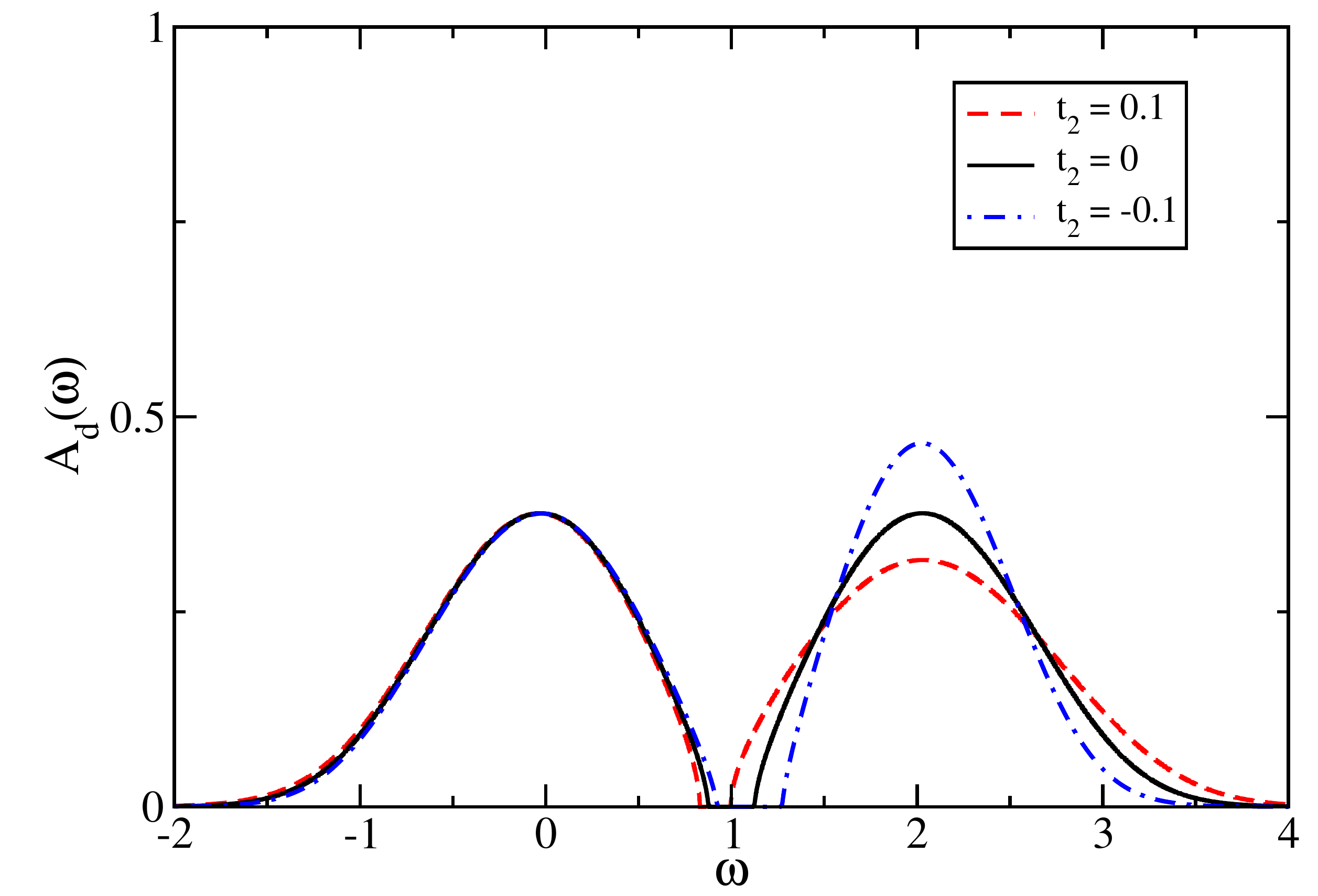}
\includegraphics[scale=0.19]{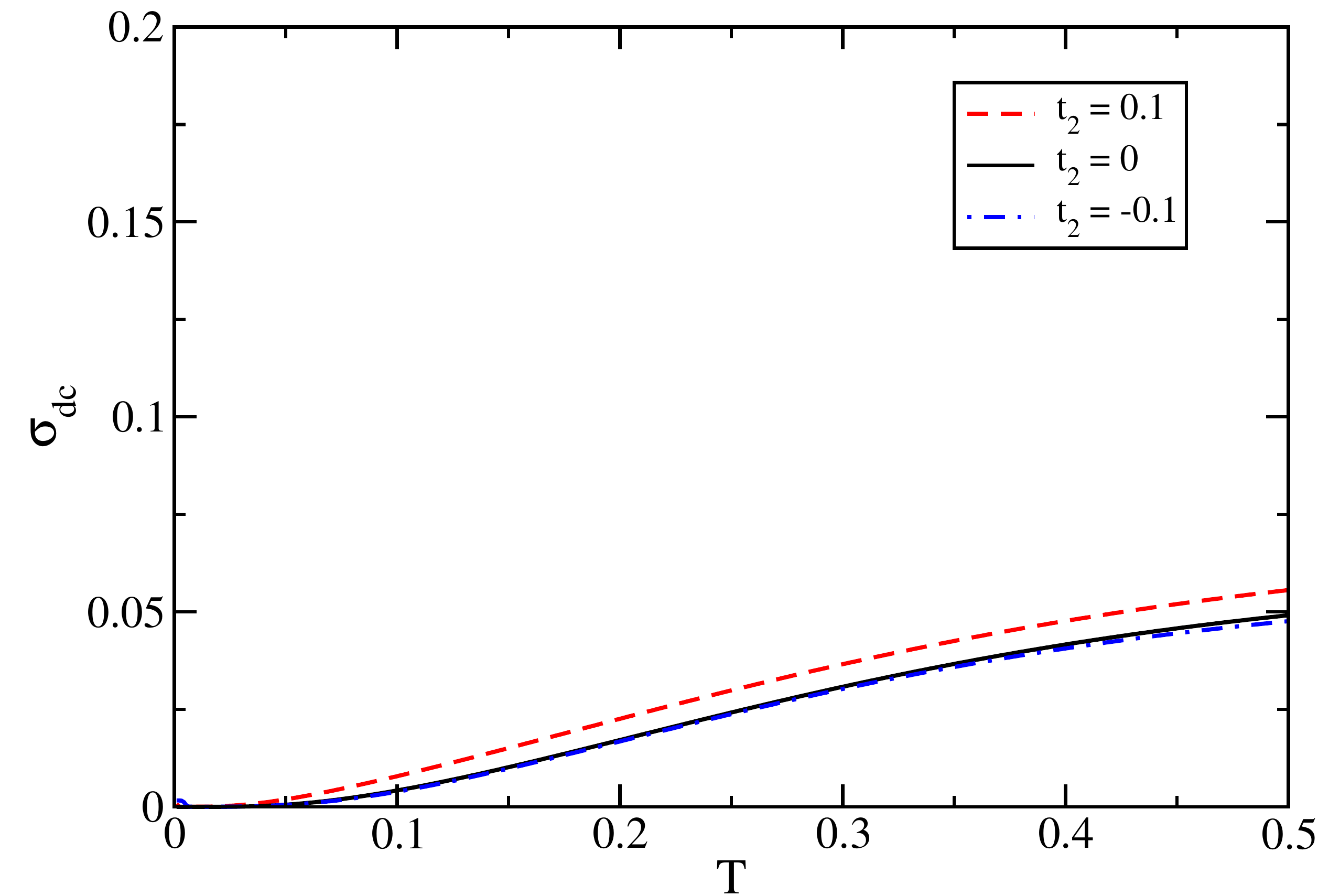}
\includegraphics[scale=0.19]{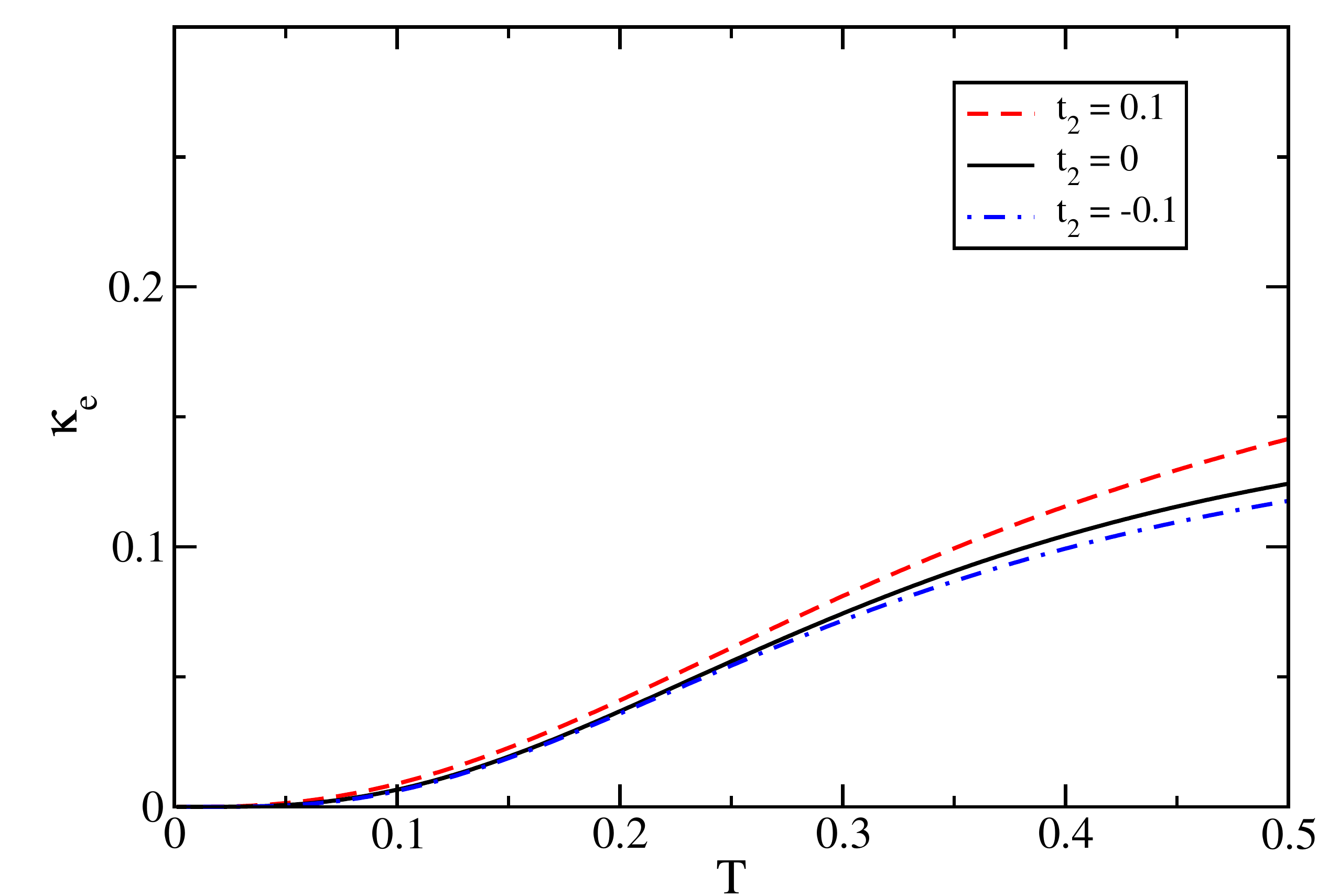}
\\
\includegraphics[scale=0.19]{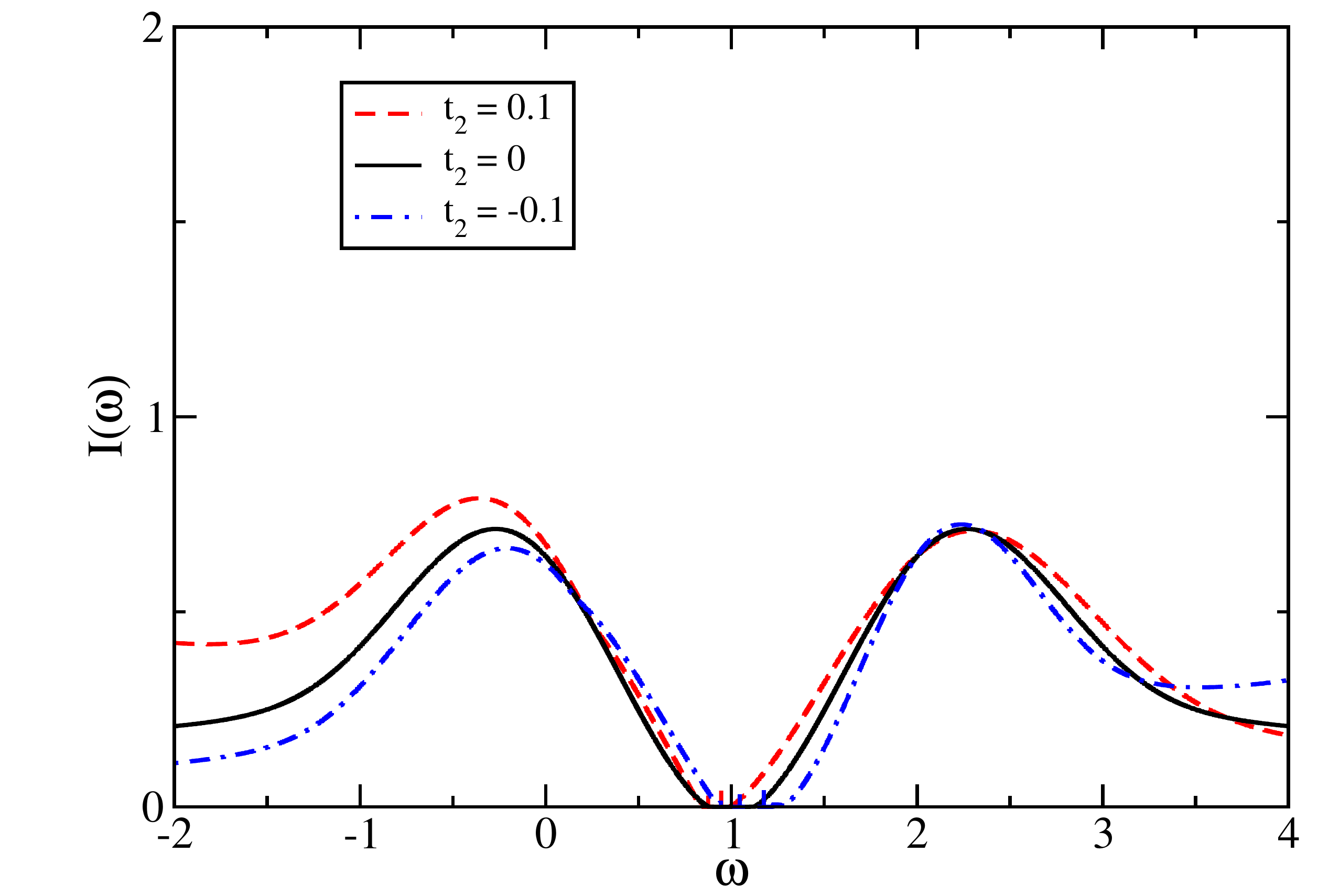}
\includegraphics[scale=0.19]{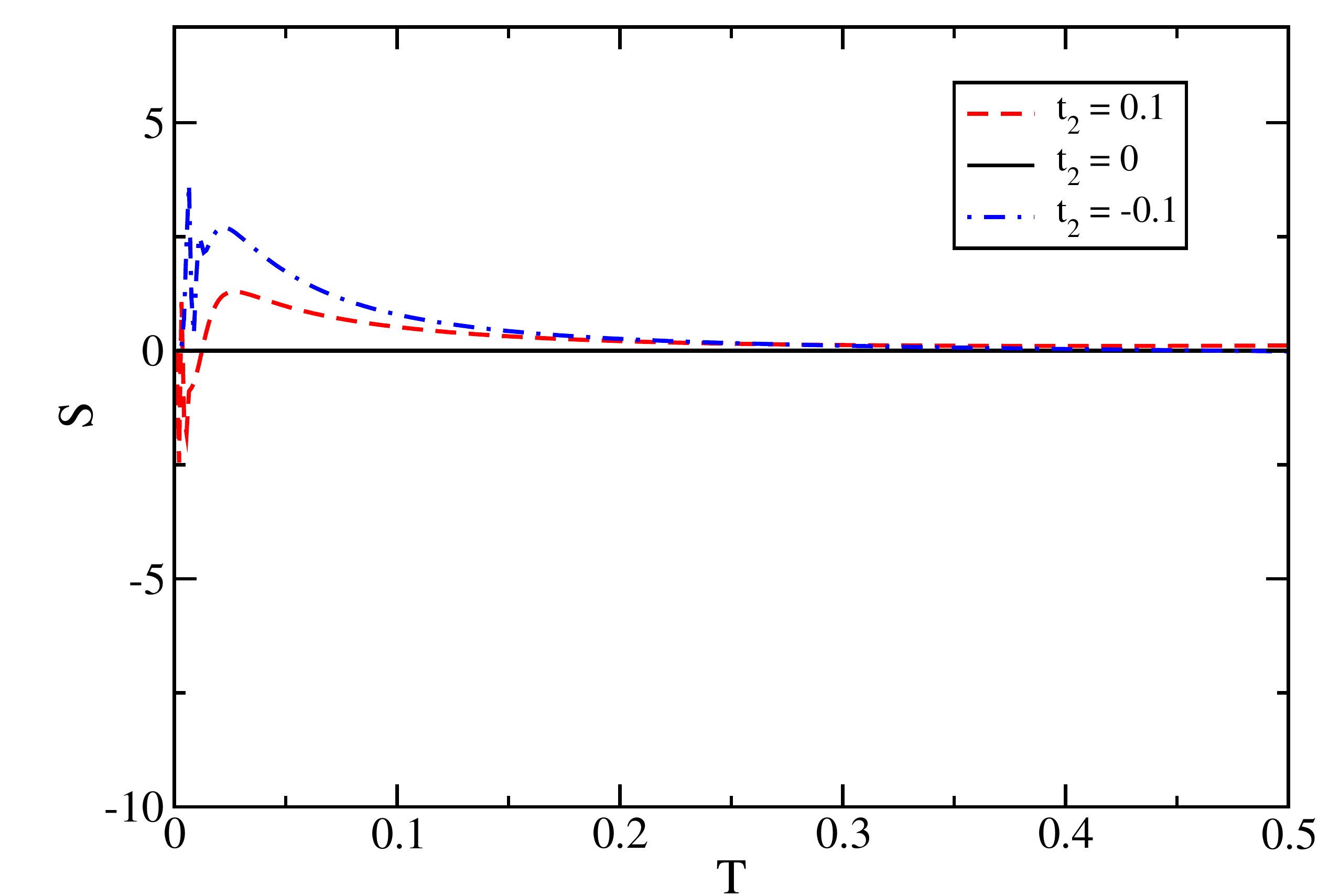}
\includegraphics[scale=0.19]{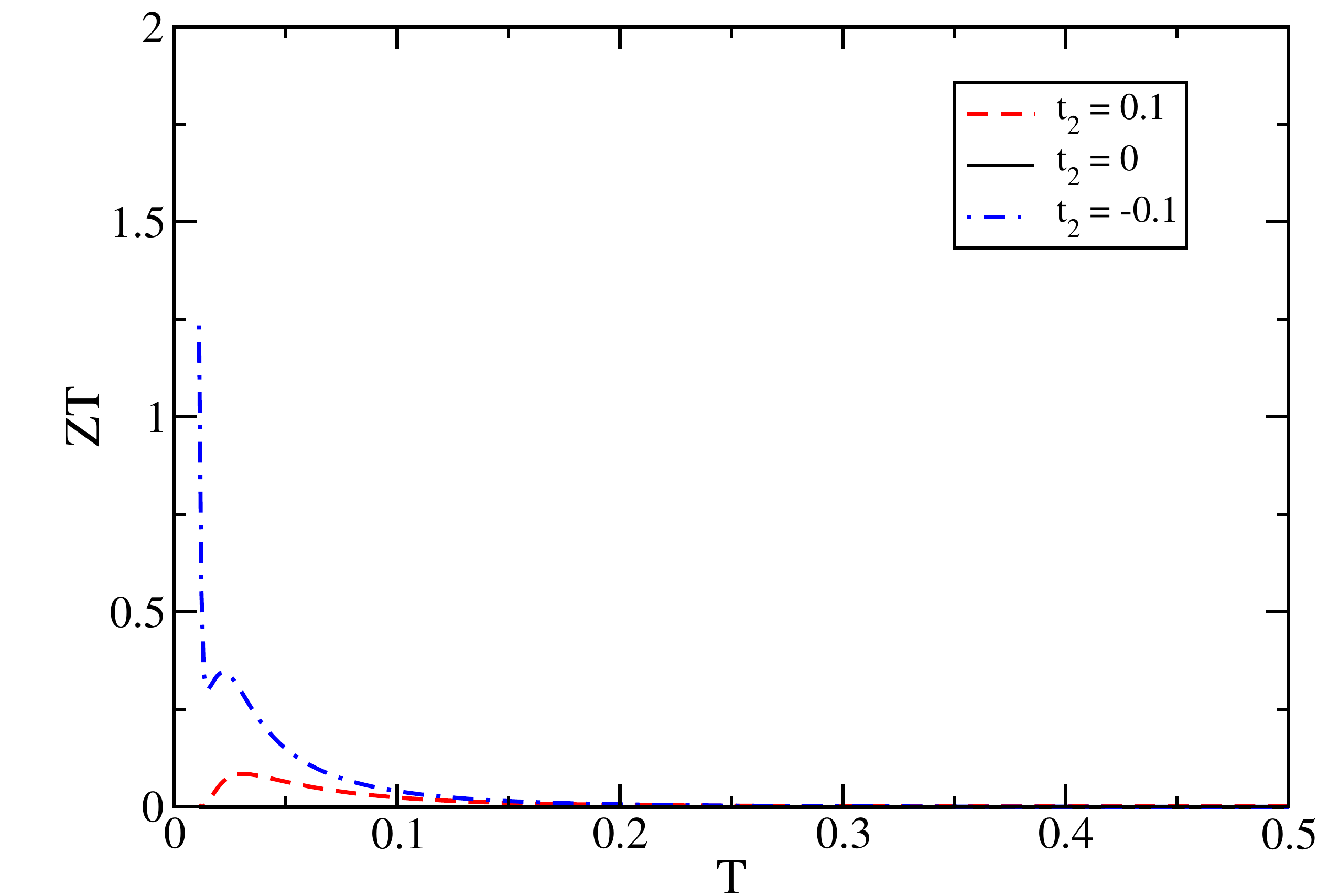}
\caption{(Color online) Same as in figure~\ref{fig:U05_t2_0} for $U=2$.}\label{fig:U20_t2_0}
\end{figure}

In the Mott insulator phase for $U=2$, the effect of small values of correlated hopping is much smaller (figure~\ref{fig:U20_t2_0}).
Both the d.o.s. $A_d(\omega)$ and the transport function $I(\omega)$ change only slightly with respect to the case without correlated hopping $t_2=0$. The temperature dependence of the electric $\sigma_{\textrm{dc}}$ and thermal $\kappa_{\textrm{e}}$ conductivities is almost the same and we observe only an increase of the Seeback effect at low temperatures similar to the one in the doped Mott insulators~\cite{freericks:195120} (non-zero correlated hopping breaks the electron-hole symmetry like doping in the case without correlated hopping). It is known that the temperature behaviour of thermoelectric properties in Mott insulator to a great extent is determined by the temperature dependence of the chemical potential~\cite{freericks:195120,zlatic:155101}, and due to the numerical issues we were not able to determine the chemical potential values in the gap at very low temperatures with a precision sufficient to get smooth dependences of the Seebeck coefficient.

\begin{figure}[!t]
\centering
\includegraphics[scale=0.19]{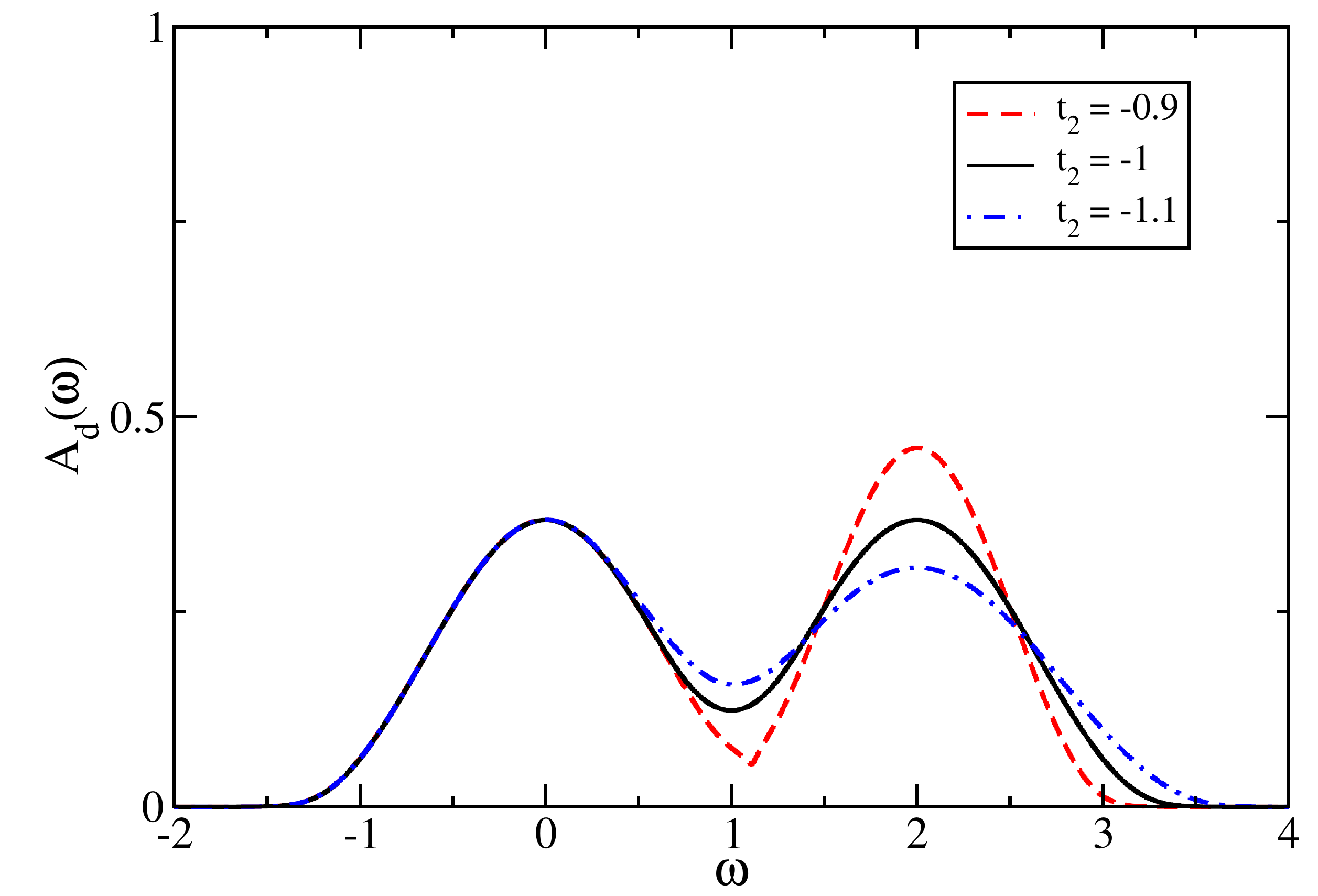}
\includegraphics[scale=0.19]{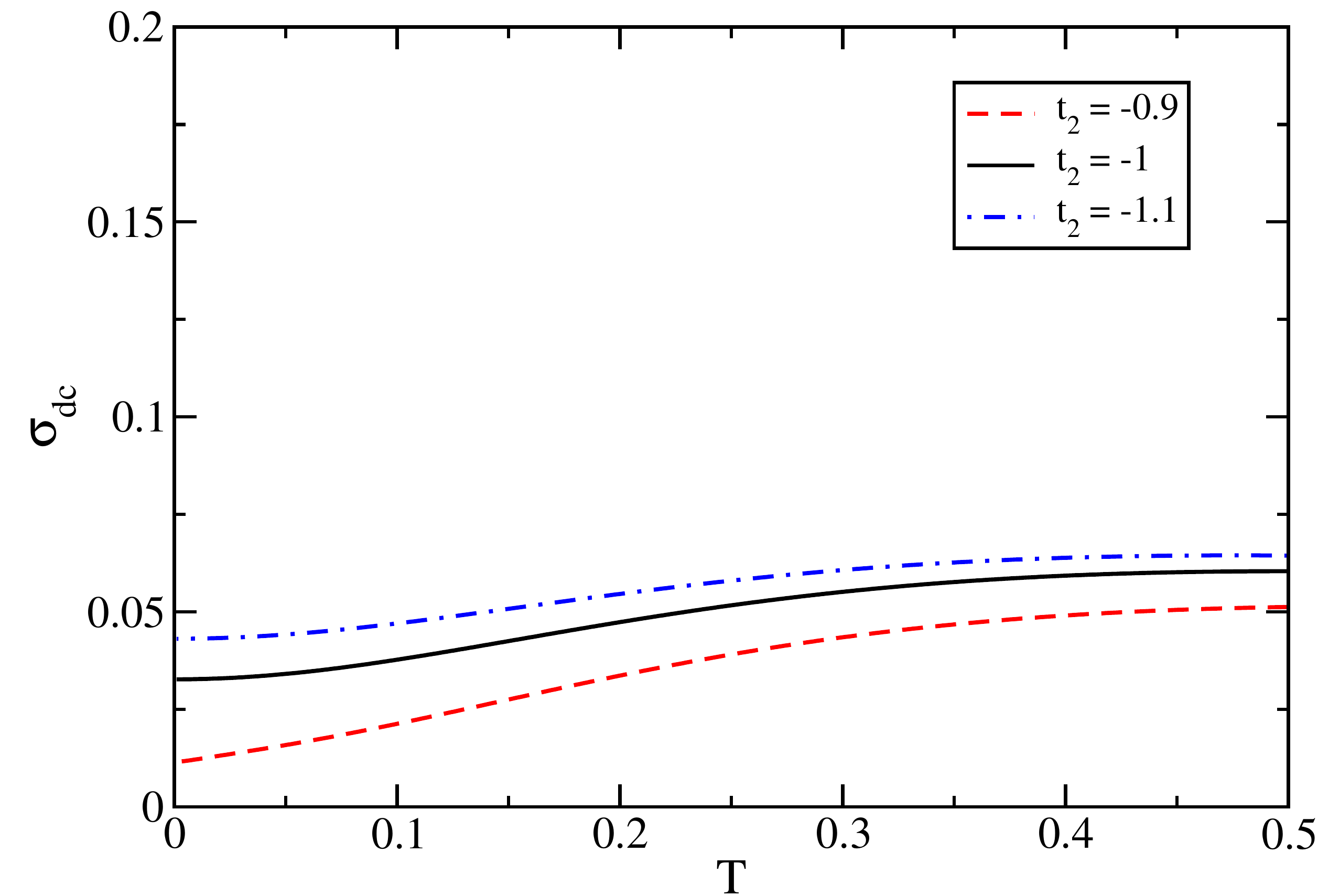}
\includegraphics[scale=0.19]{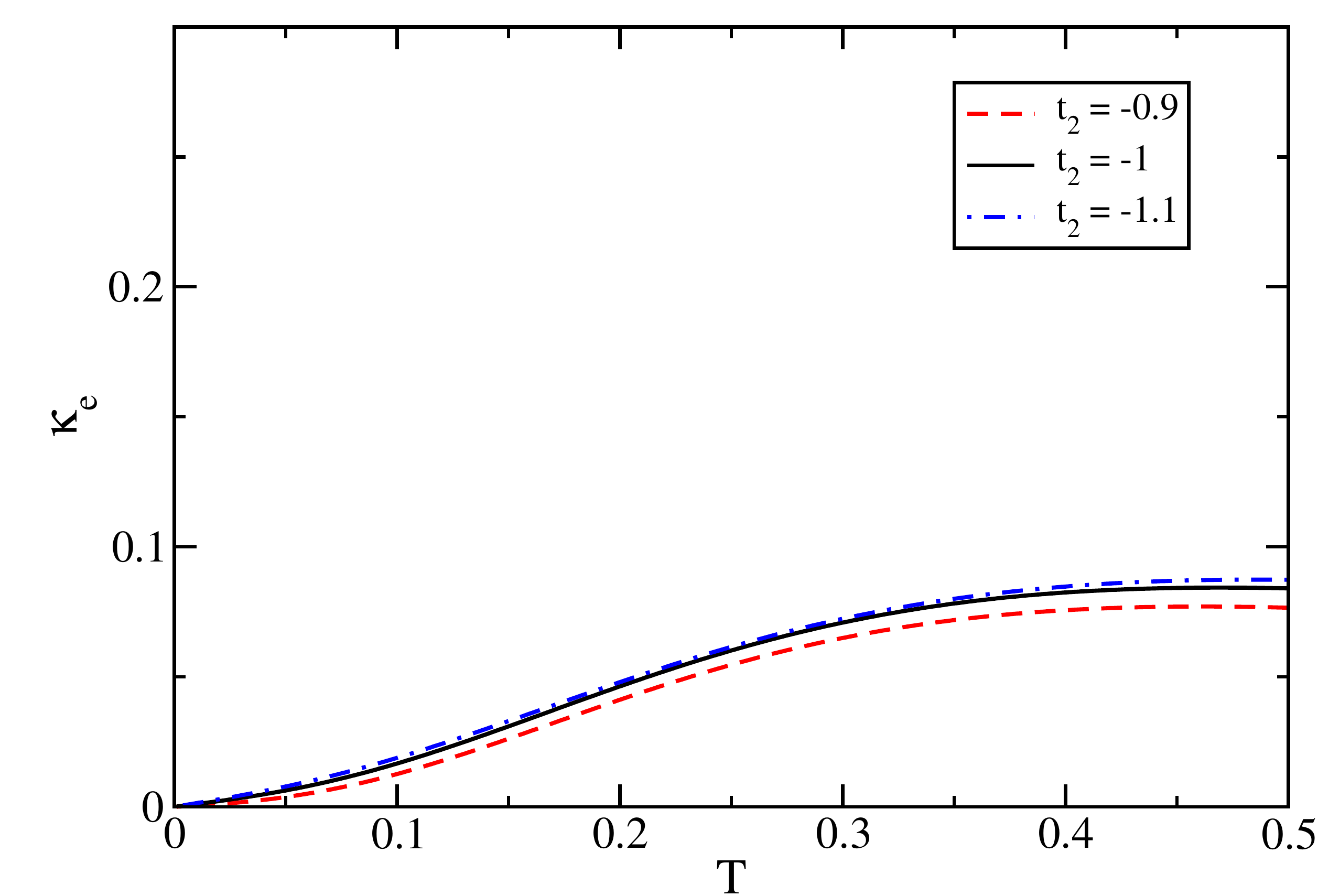}
\\
\includegraphics[scale=0.19]{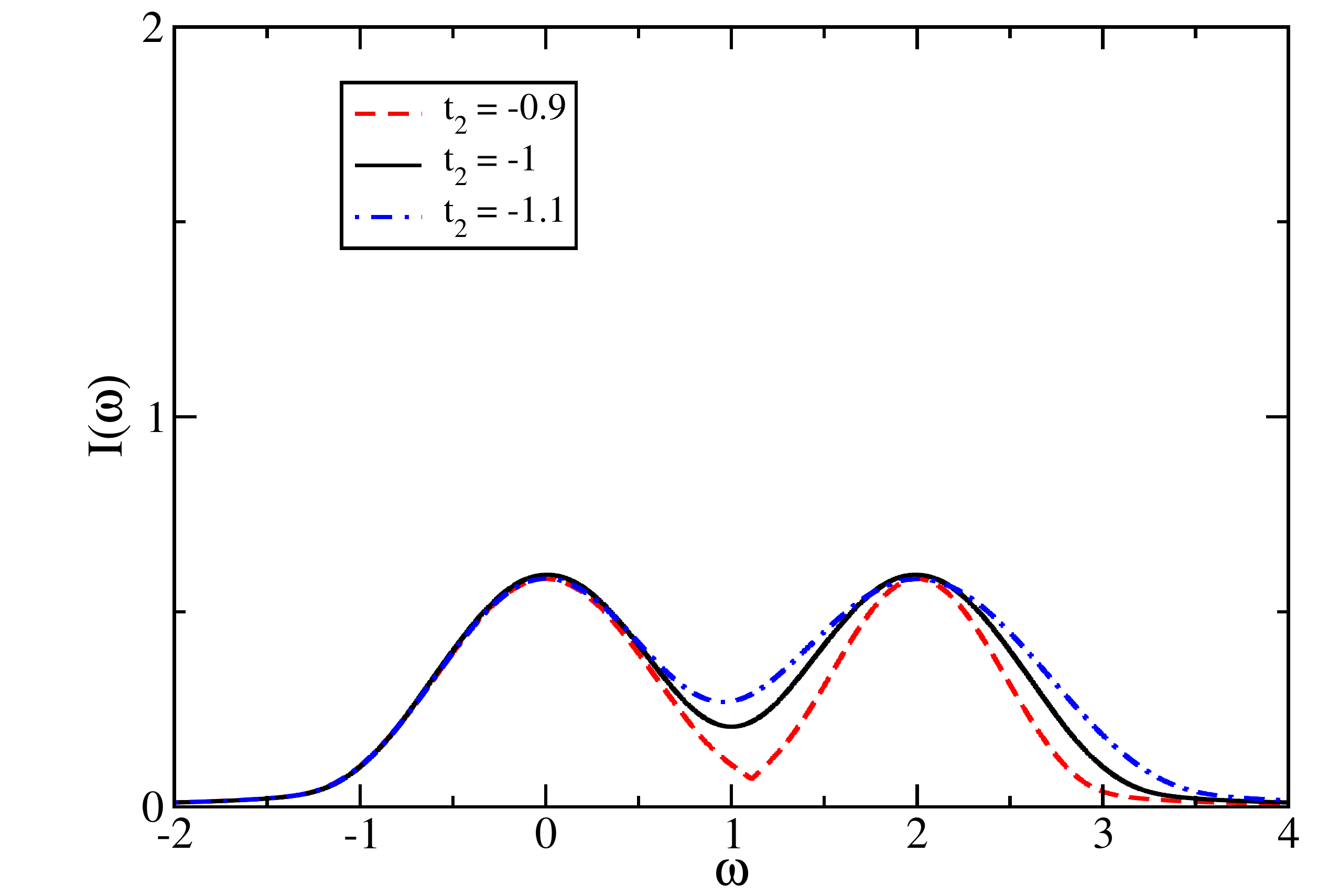}
\includegraphics[scale=0.19]{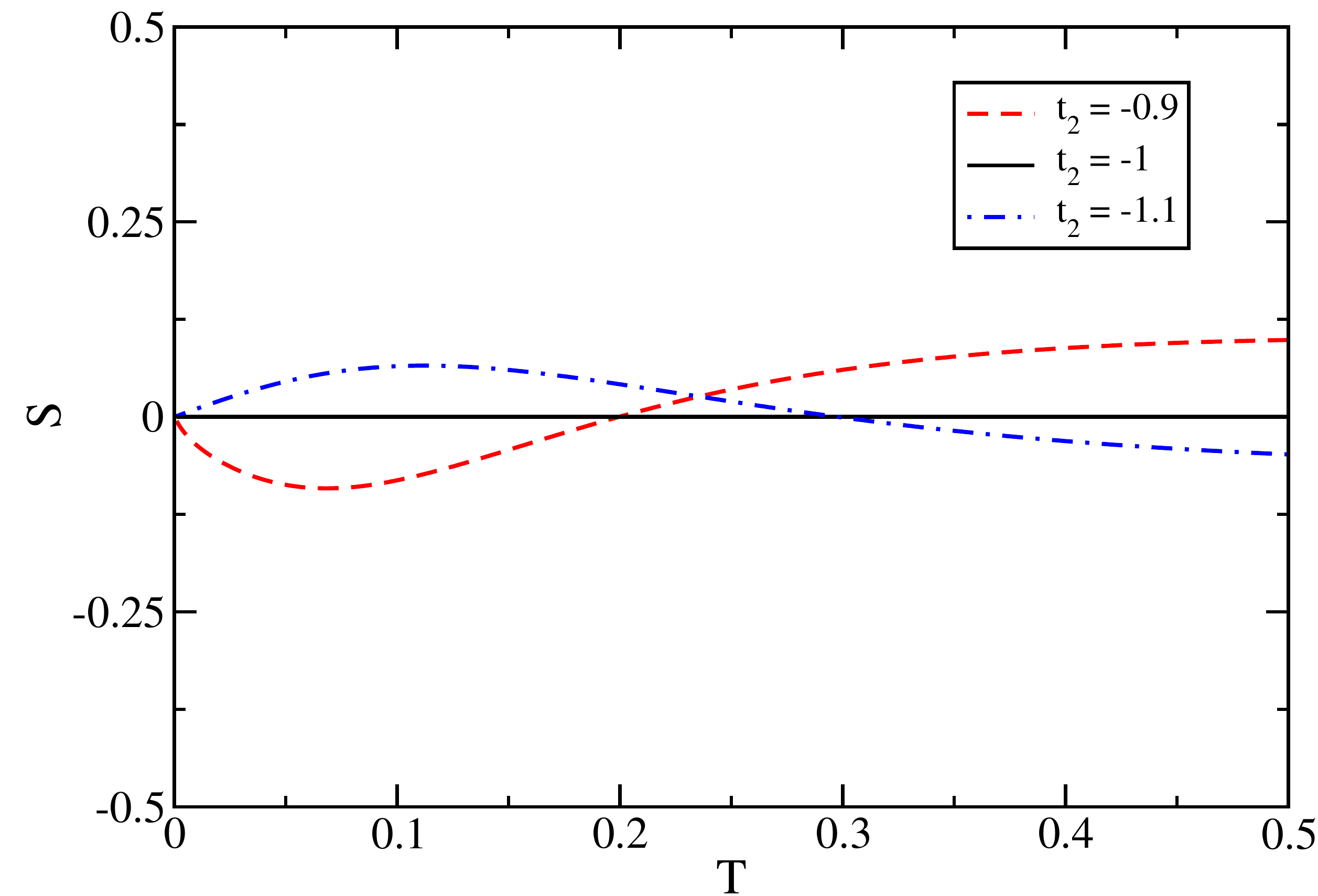}
\includegraphics[scale=0.19]{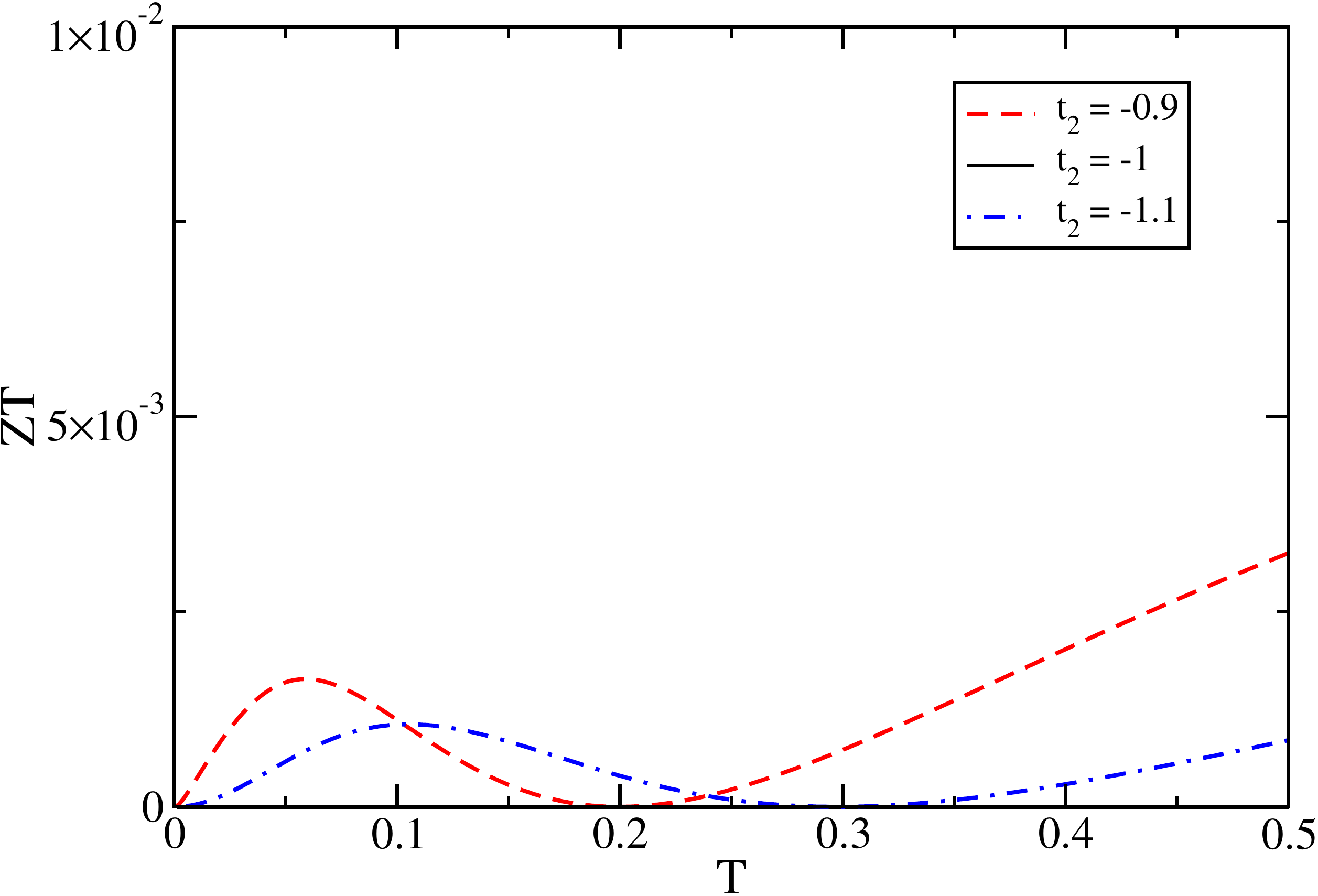}
\caption{(Color online) Density of states $A_d(\omega)$, transport function $I(\omega)$,
and temperature dependences of the dc electric $\sigma_{\textrm{dc}}$ and thermal $\kappa_{\textrm{e}}$
conductivities, Seebeck coefficient $S$, and \textit{figure of merit} $ZT$ at half filling $n_f=n_d=1$
for $U=2$, $t_1=1$, and $t_2=-0.9$, $-1$, $-1.1$.}
\label{fig:U20_t2_1}
\end{figure}

For the opposite case of almost independent bands, when $t_2\approx -1$ and off-diagonal elements of the hopping matrix~(\ref{t_mtrx}) are small and change the sign ($t^{+-}=t^{-+}\to0$), the d.o.s. and transport function are very smooth in the Fermi window, which results in the metallic behaviour of the electric and thermal conductivity and in weak thermoelectric properties (figure~\ref{fig:U20_t2_1}). The Coulomb interaction $U$ is less important in this case.
%%%

\begin{figure}[!b]
\centering
\includegraphics[scale=0.19]{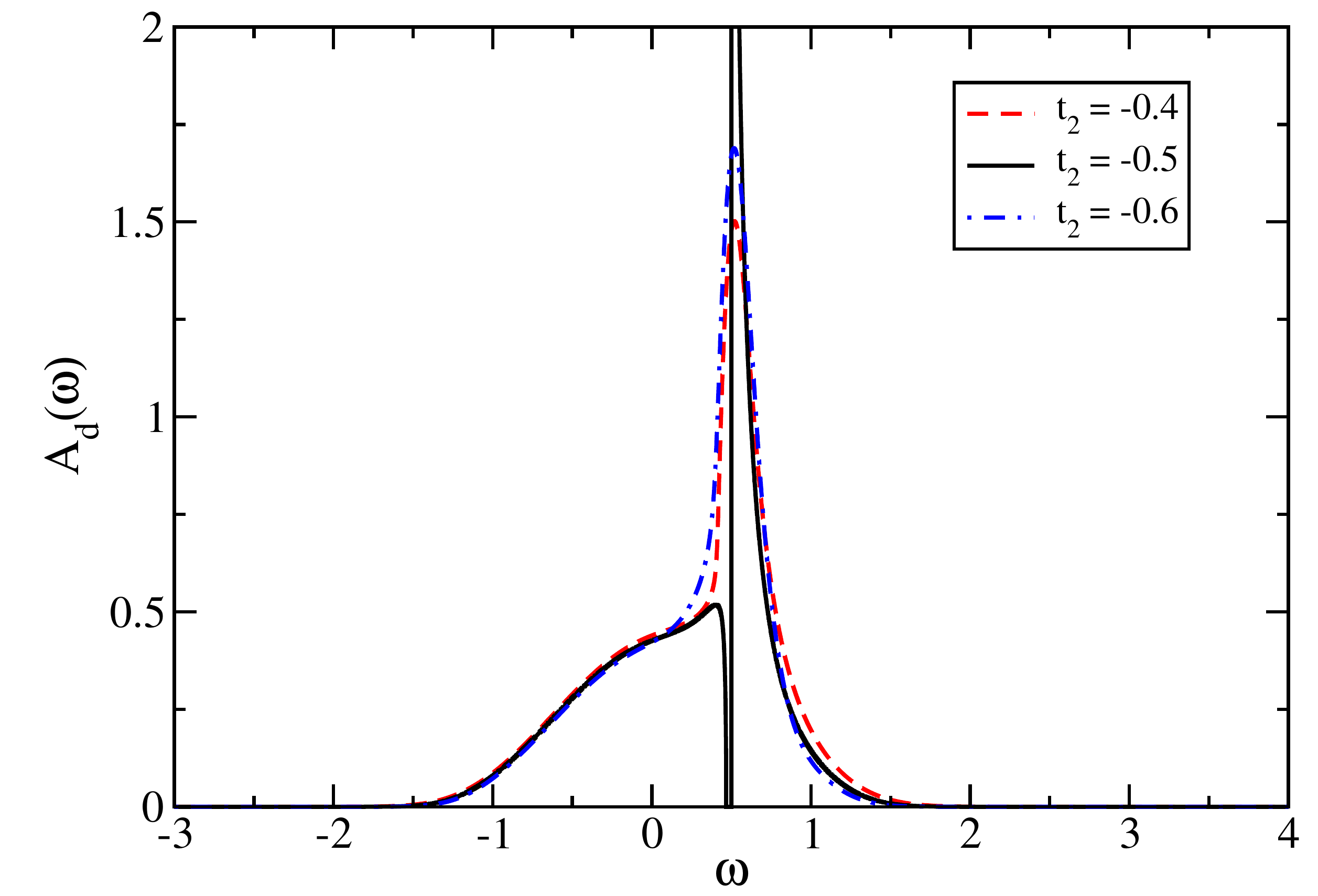}
\includegraphics[scale=0.19]{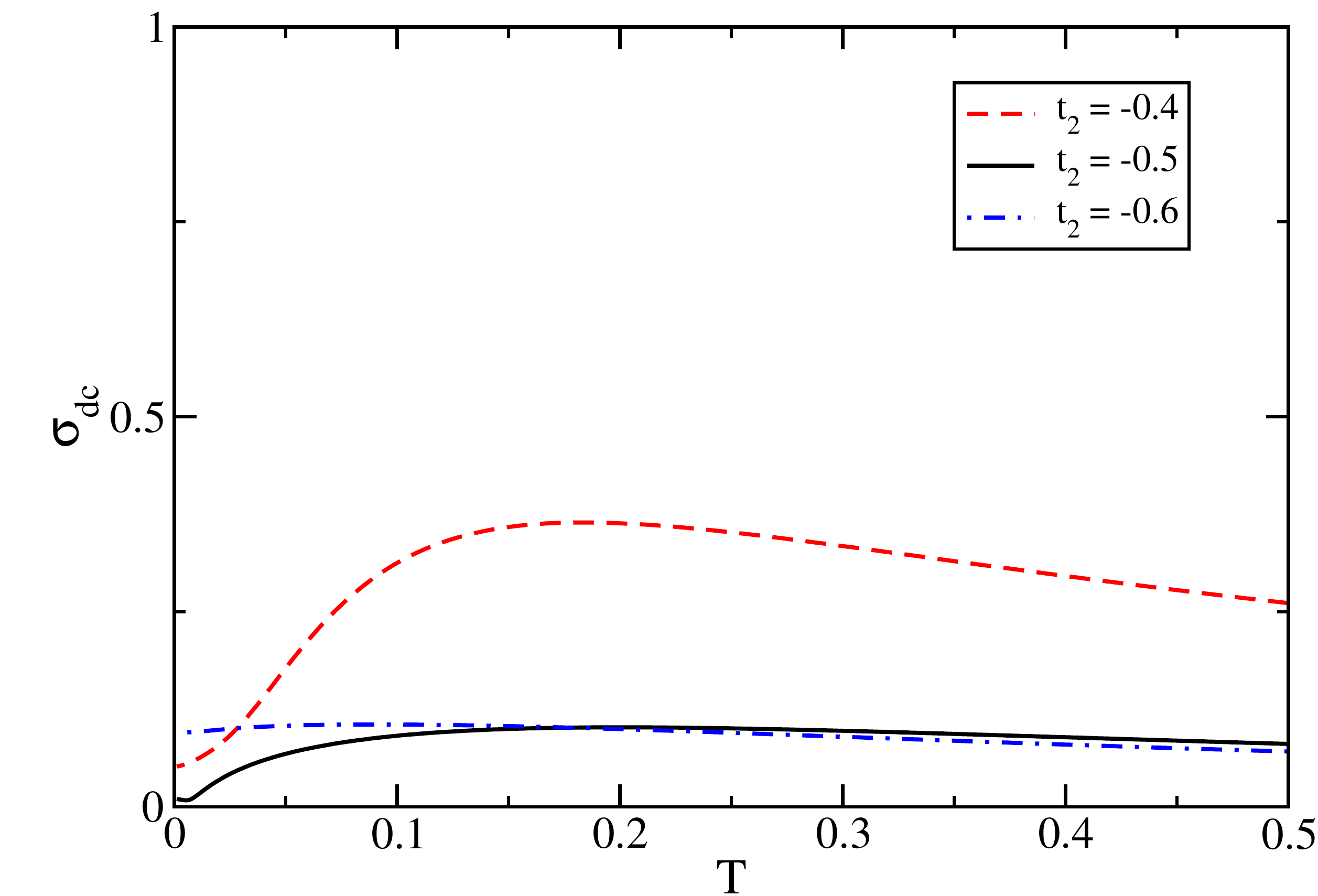}
\includegraphics[scale=0.19]{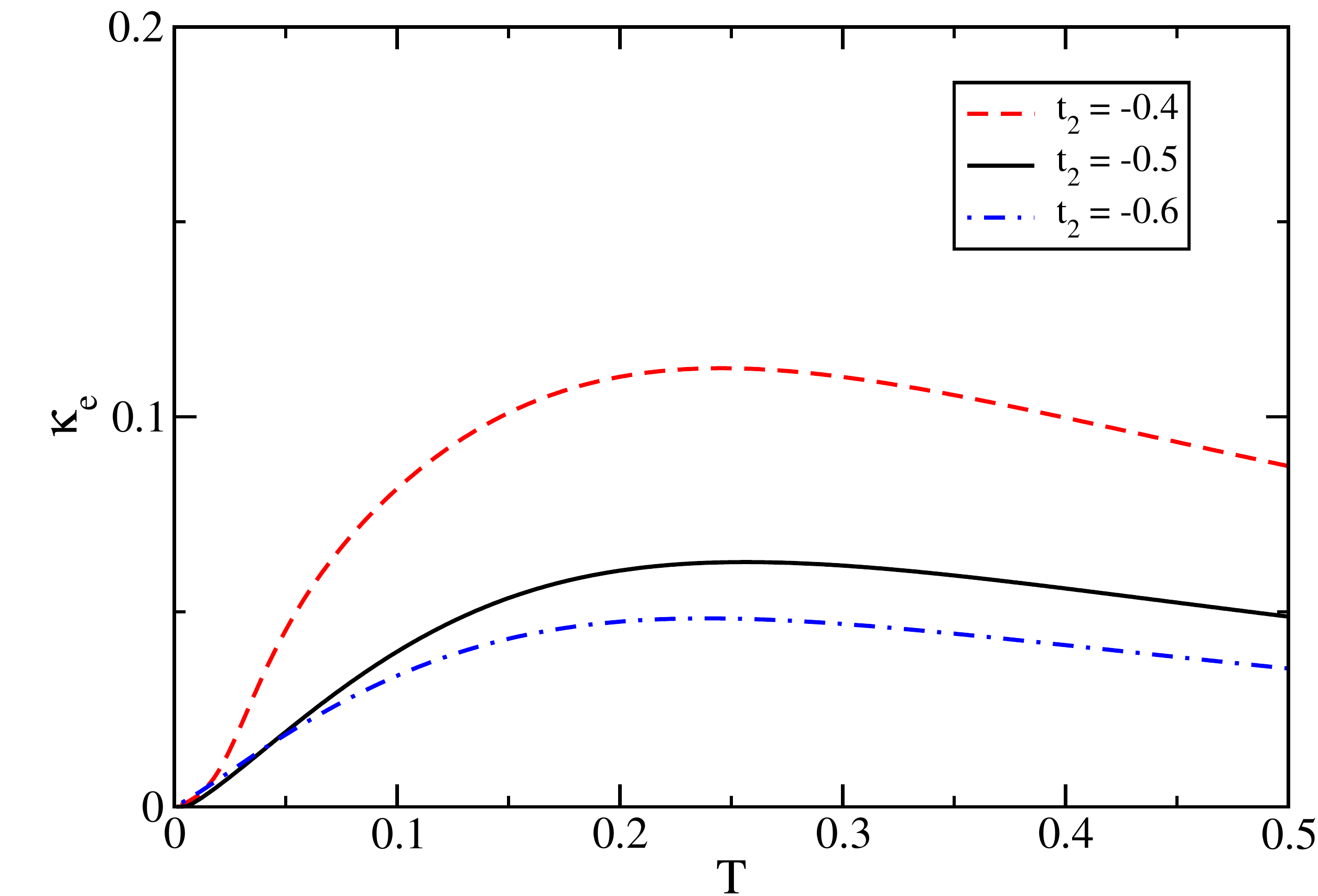}
\\
\includegraphics[scale=0.19]{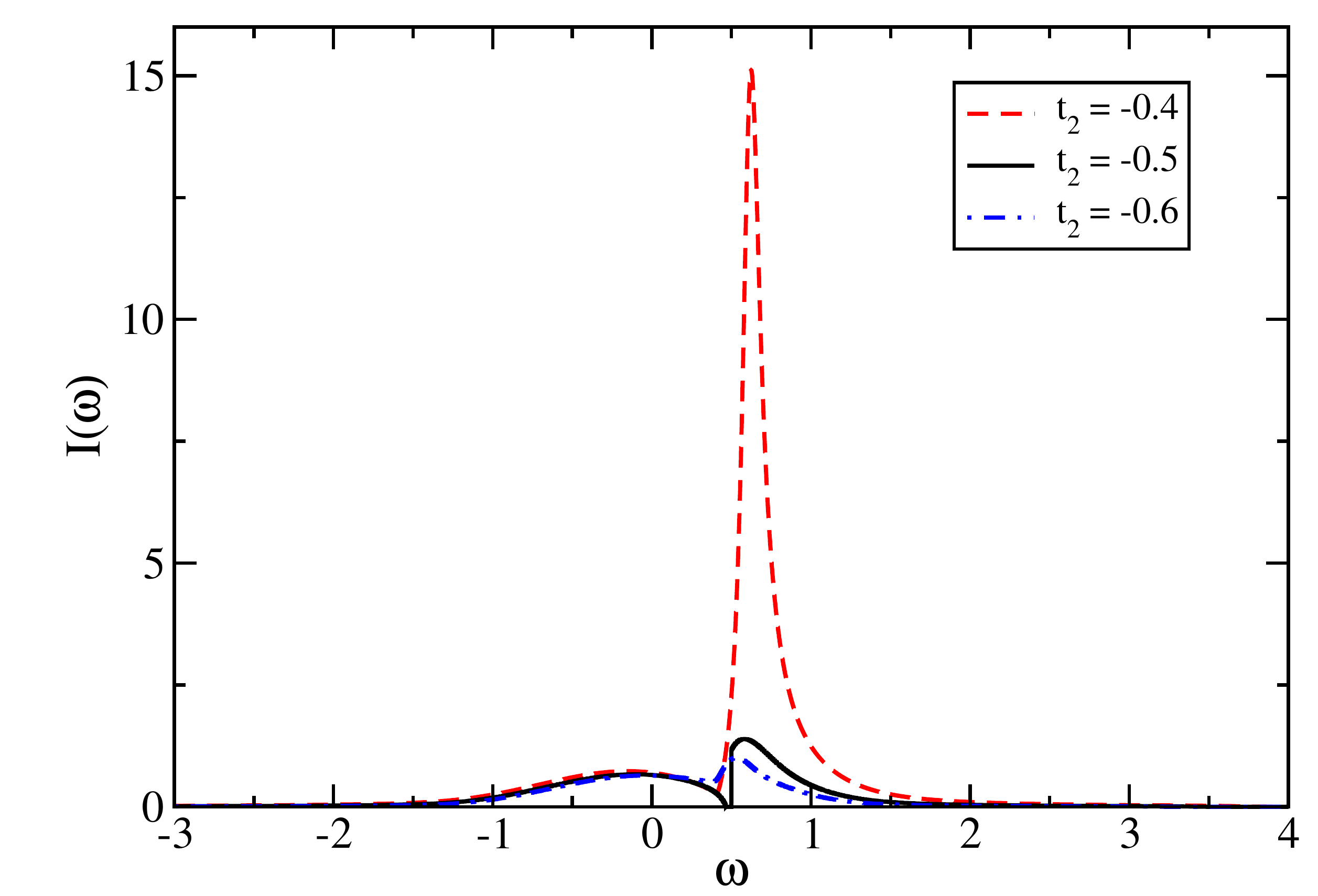}
\includegraphics[scale=0.19]{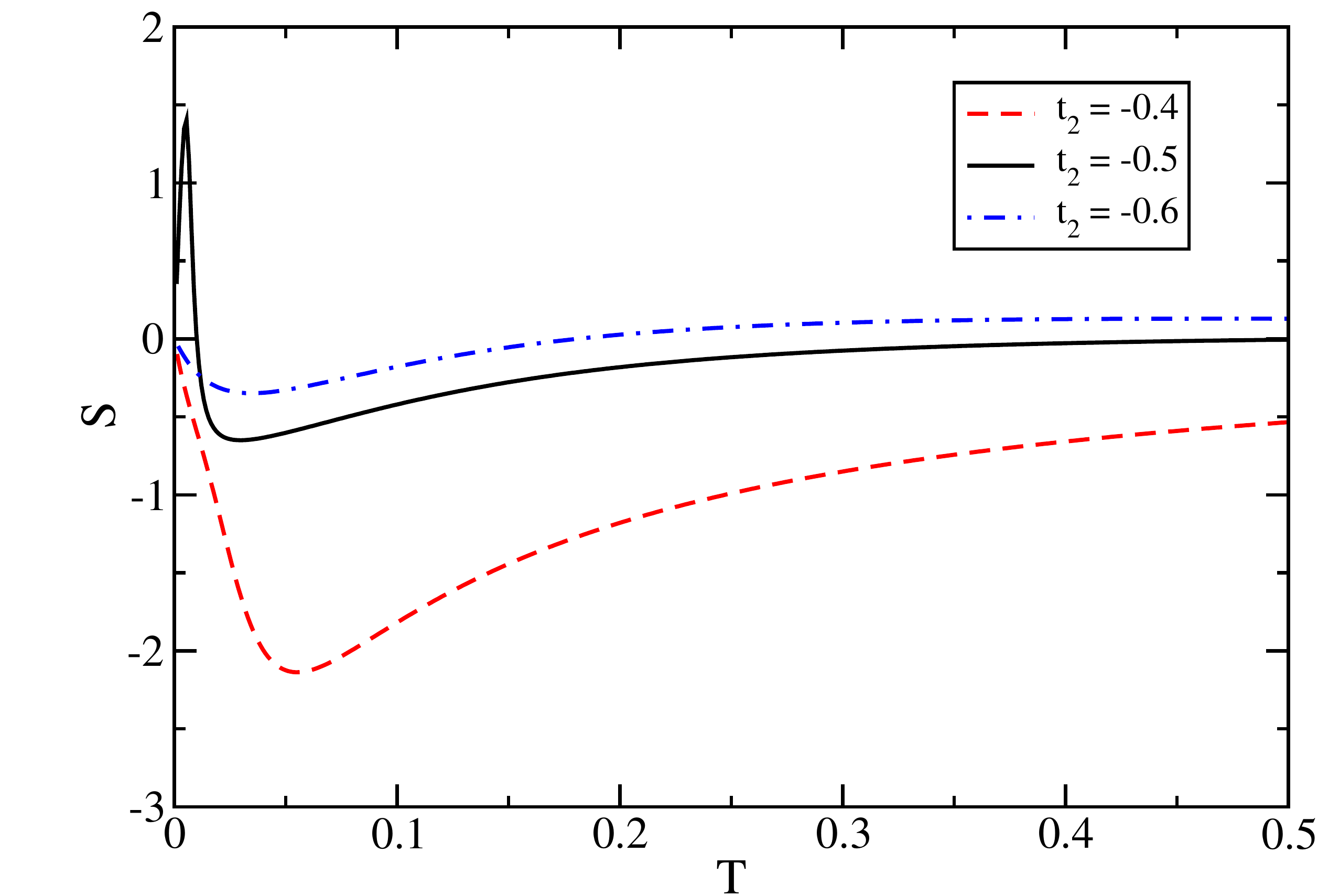}
\includegraphics[scale=0.19]{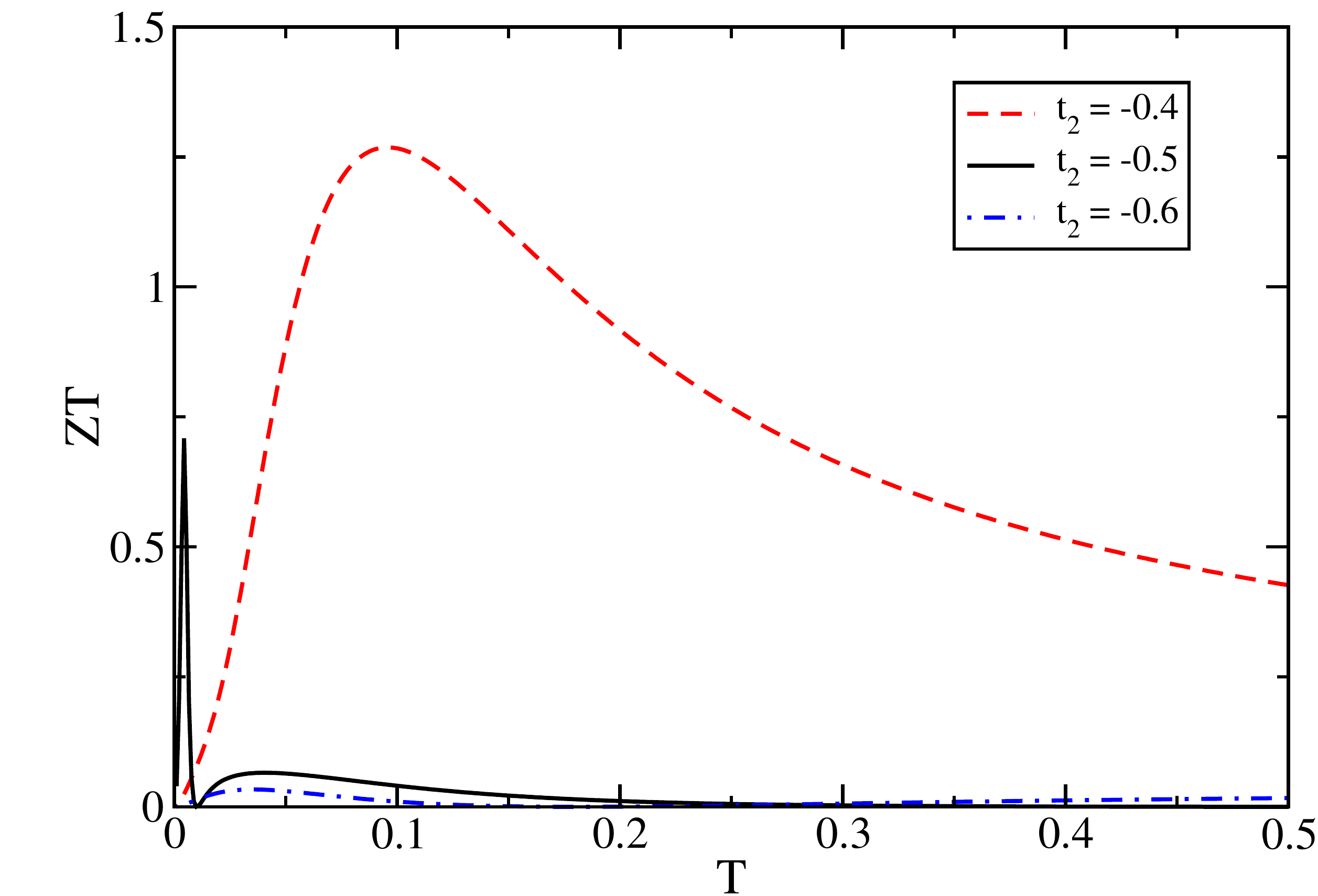}
\caption{(Color online) Density of states $A_d(\omega)$, transport function $I(\omega)$, and temperature dependences of the dc electric $\sigma_{\textrm{dc}}$ and thermal $\kappa_{\textrm{e}}$ conductivities, Seebeck coefficient $S$, and \textit{figure of merit} $ZT$ at half filling $n_f=n_d=1$ for $U=0.5$, $t_1=1$, and $t_2=-0.4$, $-0.5$, $-0.6$.}\label{fig:U05_t2_05}
\end{figure}

In the vicinity of another crossover point at $t_2=-0.5$ (no direct hopping between the sites
occupied by $f$-particles: $t^{++}=0$) the behaviour is altogether different (figure~\ref{fig:U05_t2_05}).
Exactly at the $t_2=-0.5$ value, the d.o.s. $A_d(\omega)$ contains two bands separated by a small gap, and the upper band possesses a singularity at the gap edge. It could be imagined that similar features should be observed for the transport function $I(\omega)$. However, the transport function displays only a finite abrupt change instead of a singularity above the gap. For the value of correlated hopping $t_2=-0.6$, the gap closes and the transport function is much smoother. On the other side of the crossover point, for the $t_2=-0.4$ value, the gap also closes and the shape of the d.o.s. does not change significantly in comparison with the case of $t_2=-0.6$ value, but now the transport function displays a strong enhancement with a narrow peak.
It should be noted that in many theoretical simulations, the calculation of two-particle quantities, including the transport function, is problematic, and sometimes an approximation $I(\omega)=\pi \Gamma A_d(\omega)$, replacing by one-particle quantity, is used instead (see, e.g.~\cite{tooski:055001}). Our results definitely show that such an approximation is not valid and produces altogether different transport functions for many cases.

Different shapes of the transport function in the vicinity of the $t_2=-0.5$ value manifest themselves in different transport properties. Exactly at the crossover point $t_2=-0.5$, the temperature dependences of the electric $\sigma_{\textrm{dc}}$ and thermal $\kappa_{\textrm{e}}$ conductivities as well as of the Seebeck coefficient are similar to the one in the doped small gap Mott insulator. For the $t_2=-0.6$ value, a metallic behaviour is observed, i.e., weak temperature dependence of the electric conductivity and small thermoelectric power. On the other side of the crossover point, an enhancement of all transport coefficients is observed which is more prominent for the electric conductivity and Seebeck coefficient than for the thermal conductivity and causes an increase of the thermoelectric \textit{figure of merit} $ZT$.

\begin{figure}[tbhp]
\centering
\includegraphics[scale=0.19]{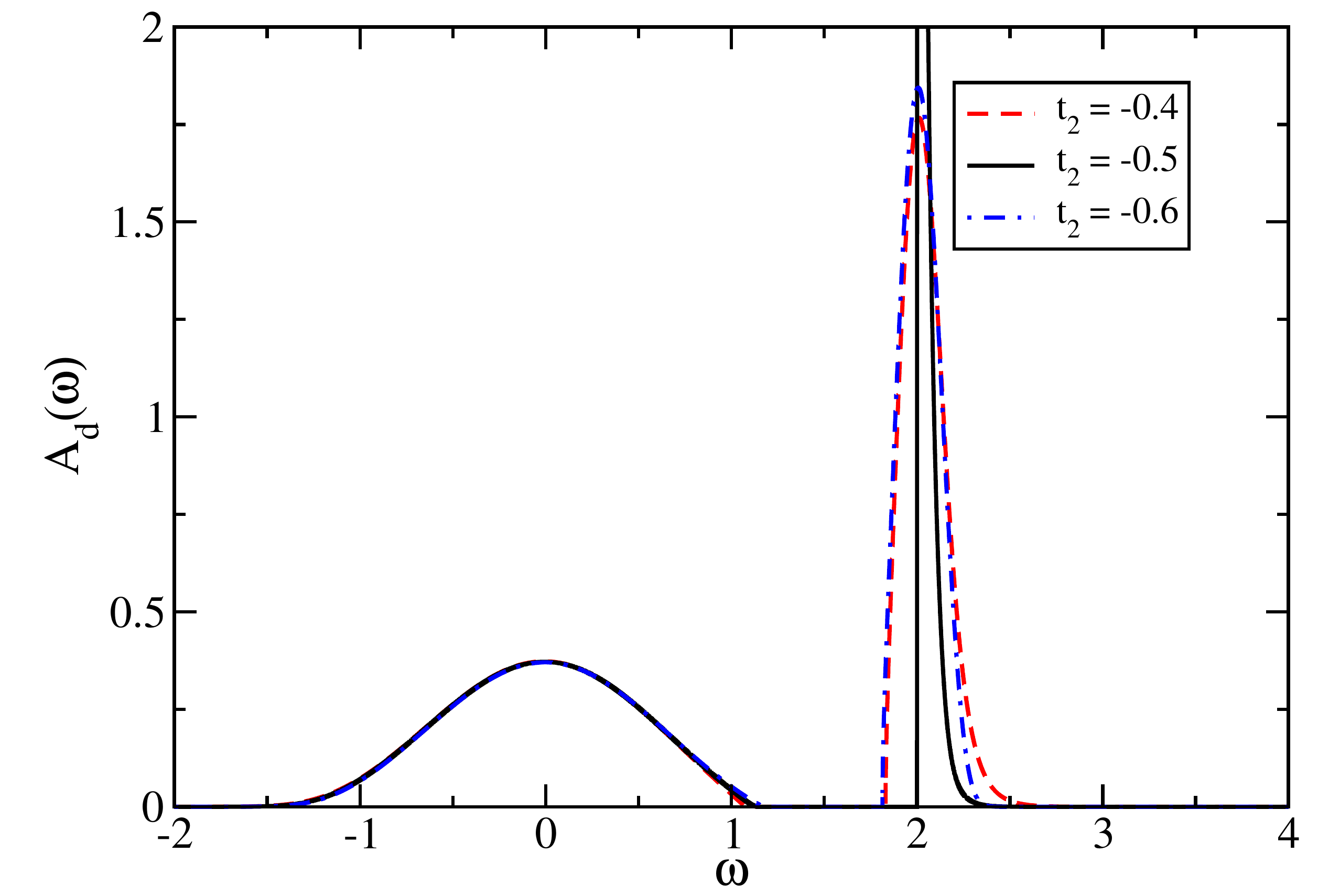}
\includegraphics[scale=0.19]{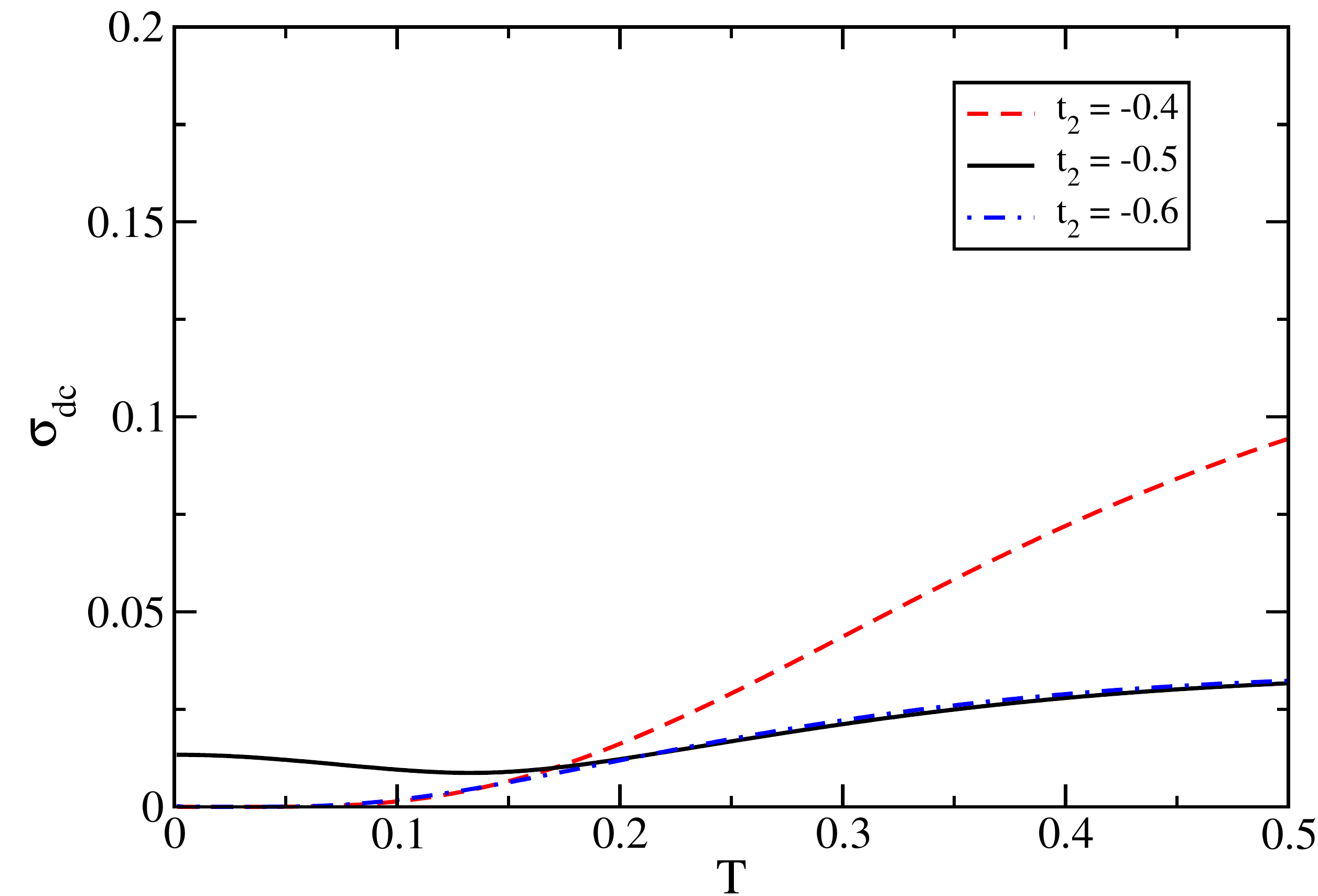}
\includegraphics[scale=0.19]{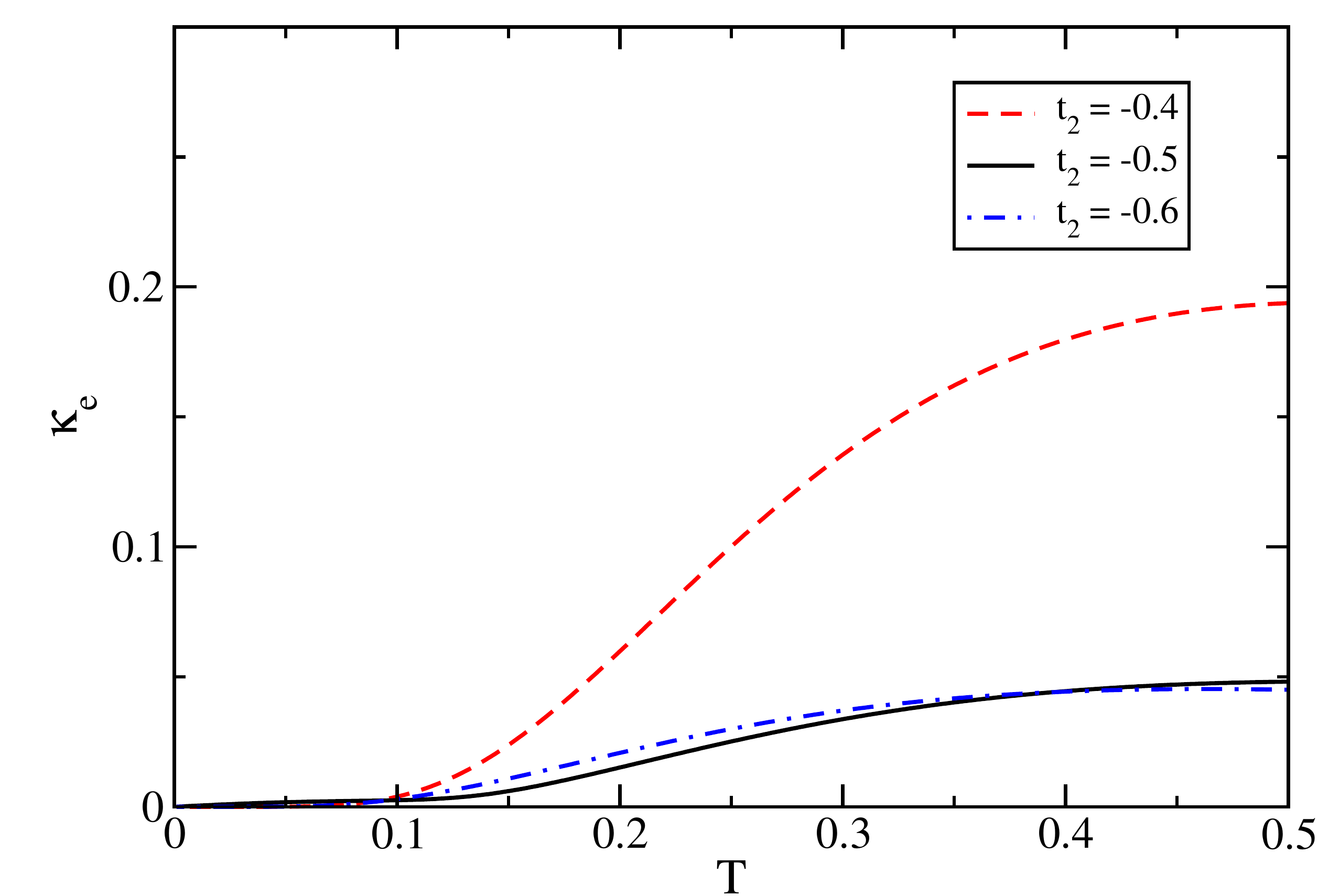}
\\
\includegraphics[scale=0.19]{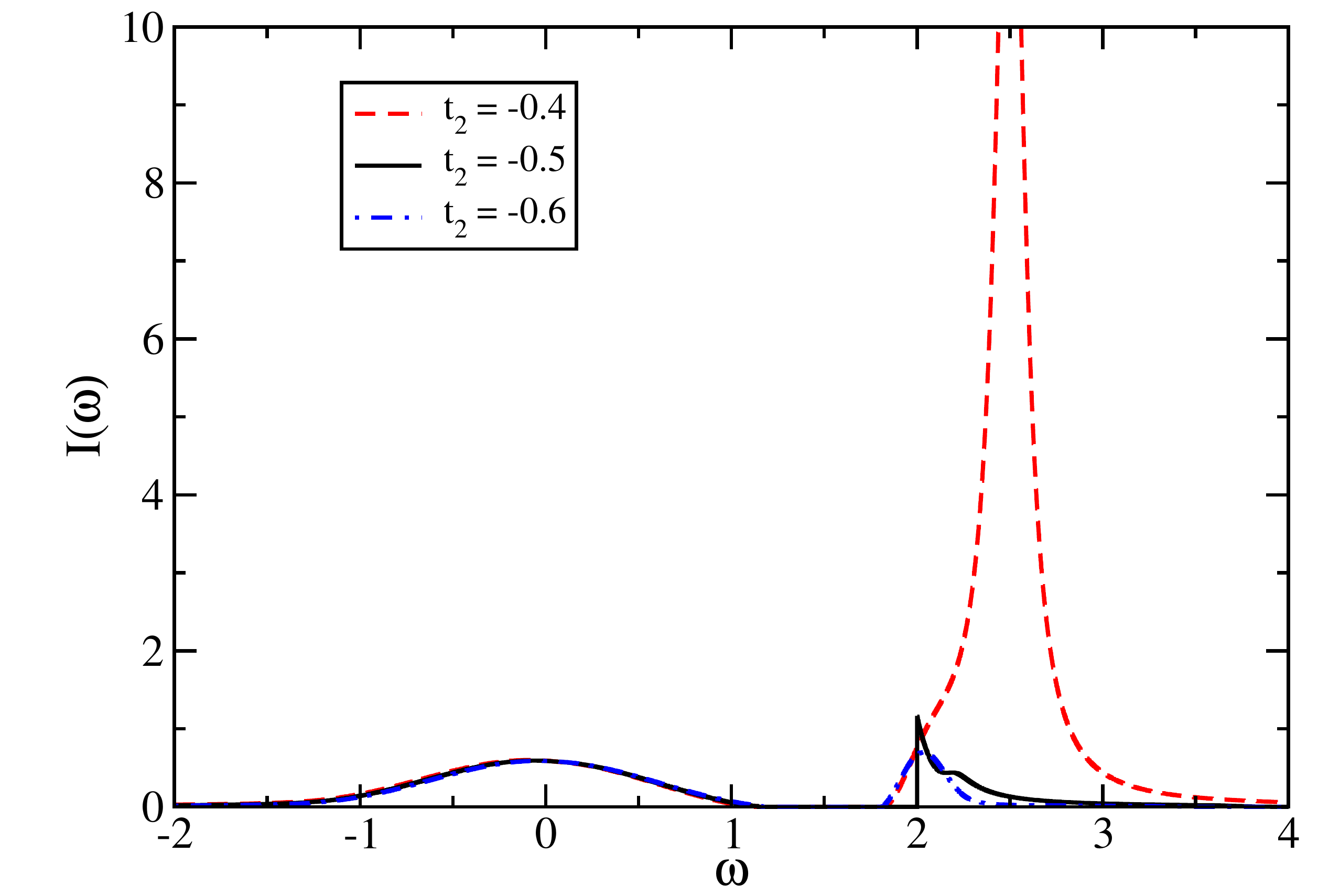}
\includegraphics[scale=0.19]{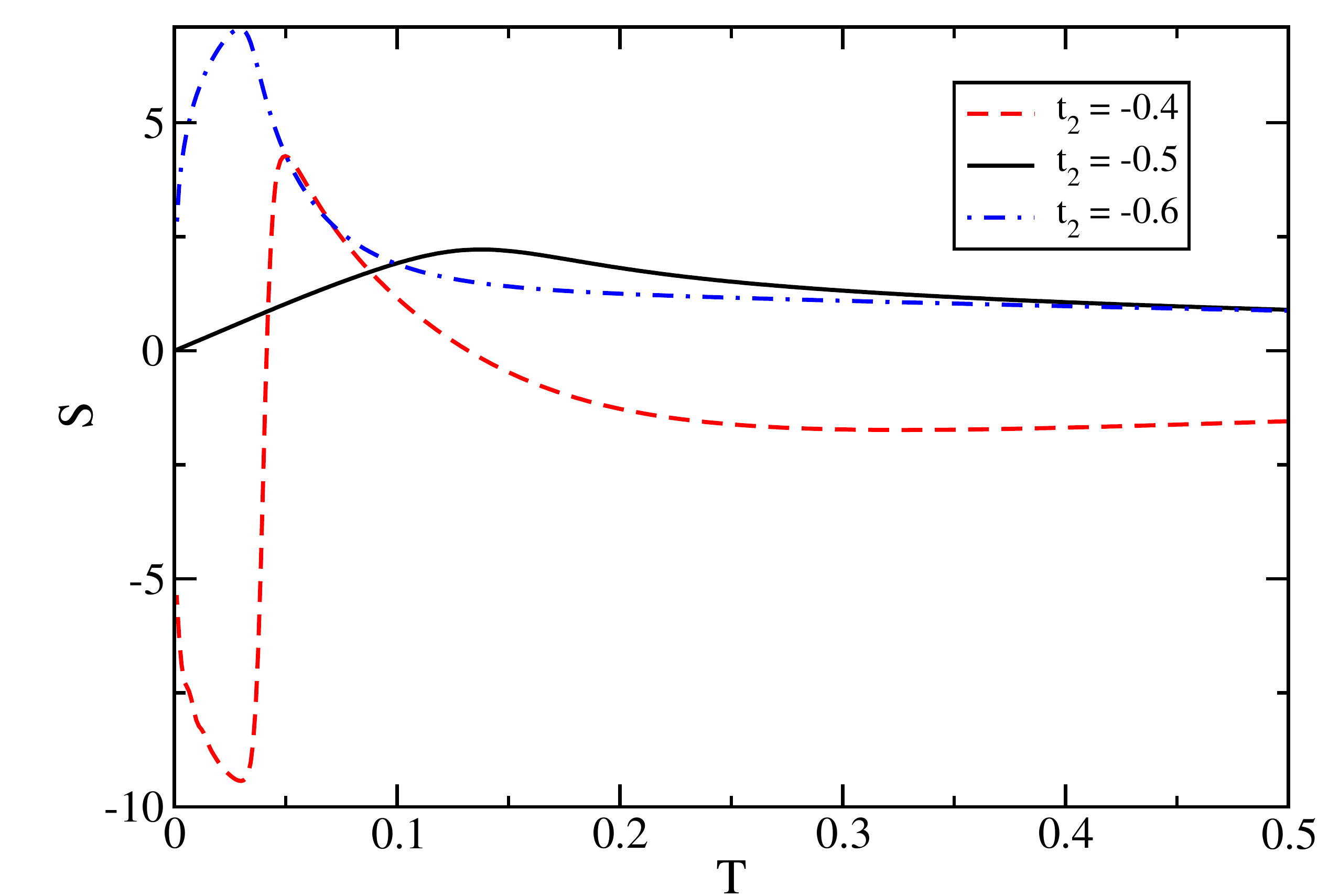}
\includegraphics[scale=0.19]{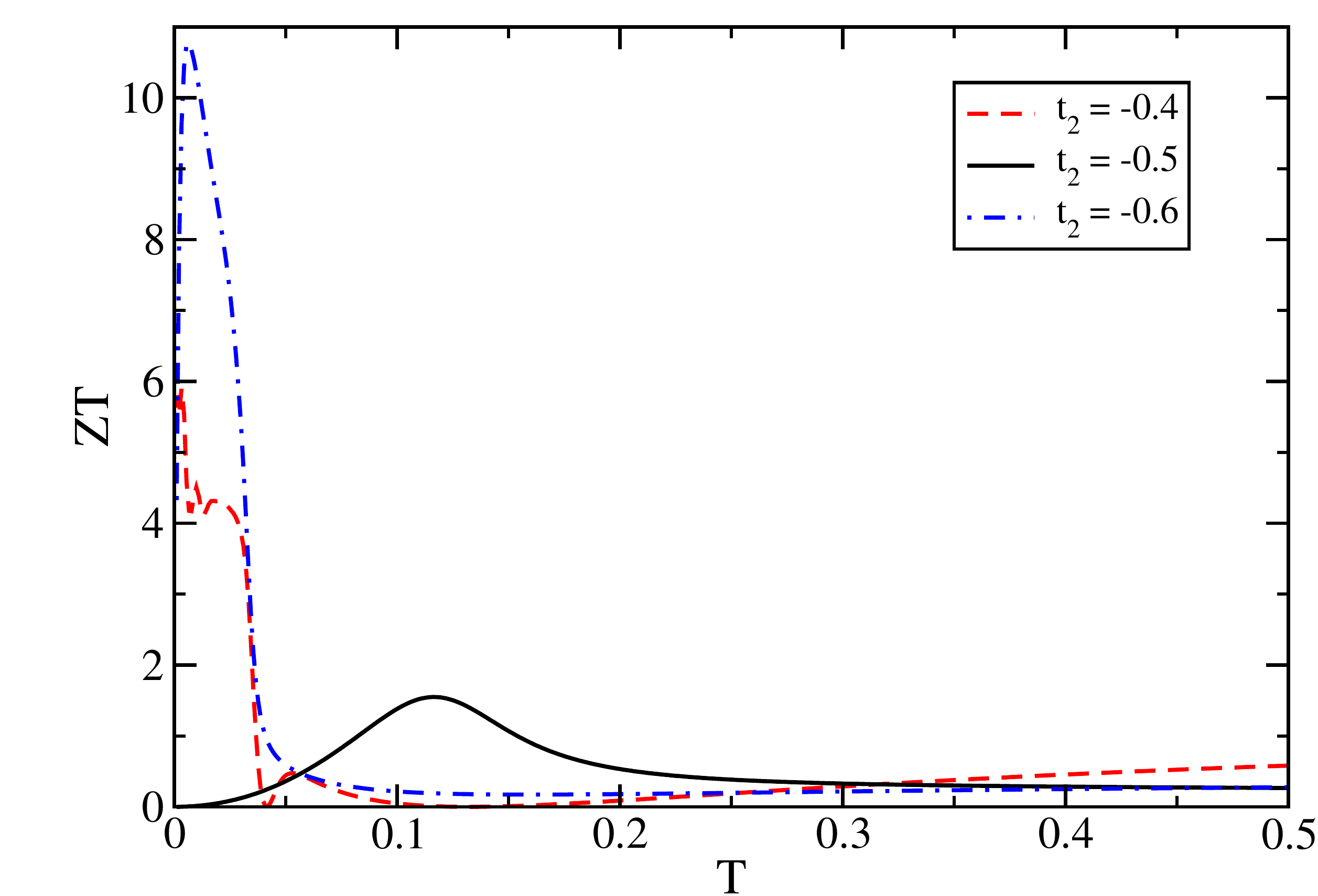}
\caption{(Color online) Same as in figure~\ref{fig:U05_t2_05} for $U=2$.}\label{fig:U20_t2_05}
\end{figure}

For large values of the Coulomb interaction $U=2$, the behaviour is quite different (figure~\ref{fig:U20_t2_05}).
Now, the d.o.s. always possesses the gap and the width of the upper band strongly reduces. Exactly at the crossover point $t_2=-0.5$, there is a singularity on the d.o.s. and there is only an abrupt change on the transport function similar to the small-$U$ case. For the $t_2=-0.6$ value, sharp features on the d.o.s. and transport function become smoother. On the other hand, for the $t_2=-0.4$ value, the d.o.s. is almost the same and the transport function develops a sharp peak. But now this peak is outside the Fermi window and less effects the transport properties. For the $t_2=-0.5$ value, the chemical potential is sticked at the upper edge of the lower band which results in a weak metallic behaviour of electric and thermal conductivities though the thermoelectric power behaves like in the doped Mott insulator. For other values of the correlated hopping, $t_2=-0.6$ and $-0.4$, the chemical potential shifts inside the gap, approaching the features of the transport function in the upper band, and the electric and thermal conductivities manifest the temperature dependences typical of the Mott insulator phase. Now, the Seebeck coefficient shows an enhancement and anomalous behaviour at low temperatures, when the extrema of the moments of Fermi window (figure~\ref{fig:Fw_n}) overlap with the features of the transport function following the change of the temperature.

\begin{figure}[!t]
\centering
\includegraphics[scale=0.185]{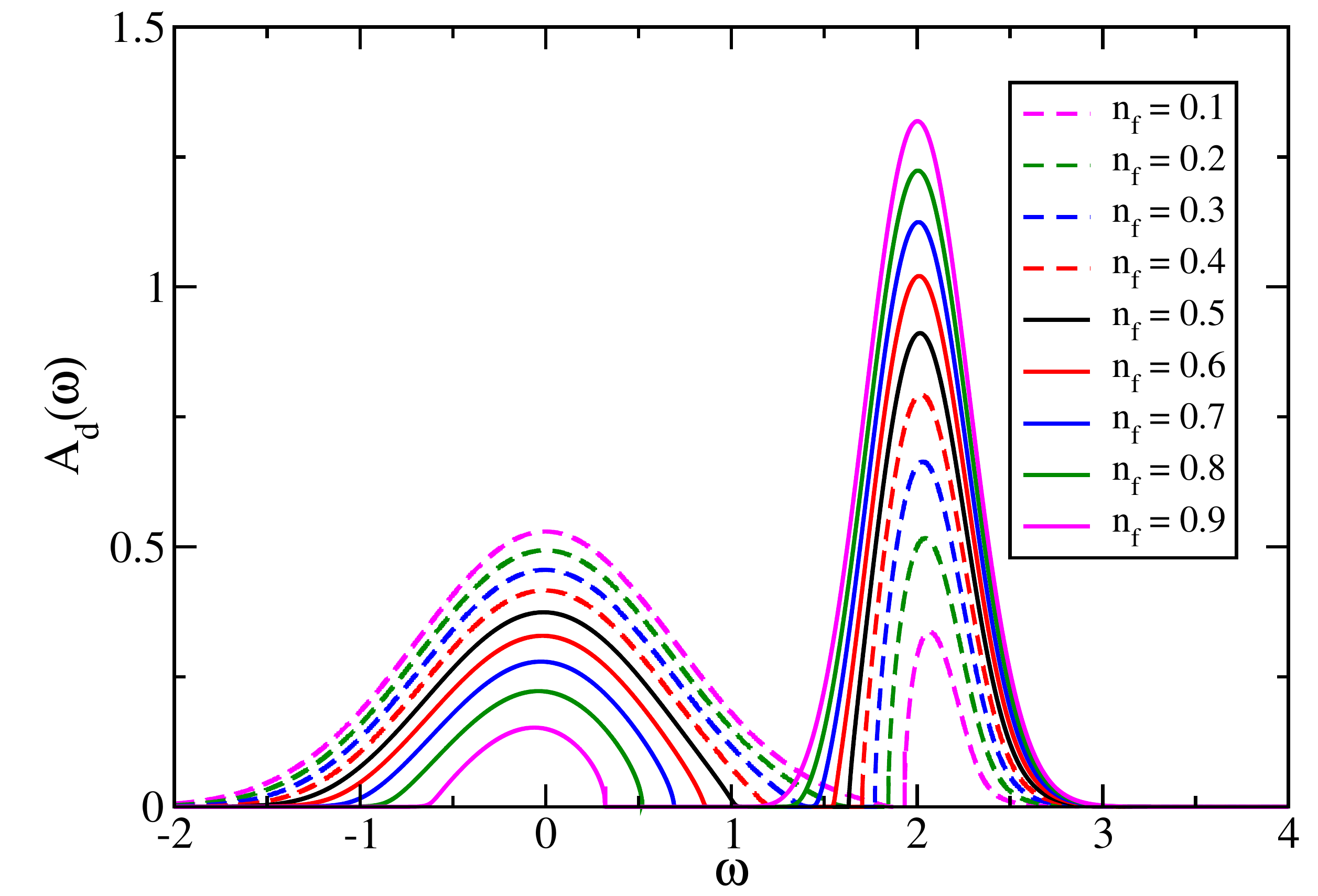}
\includegraphics[scale=0.185]{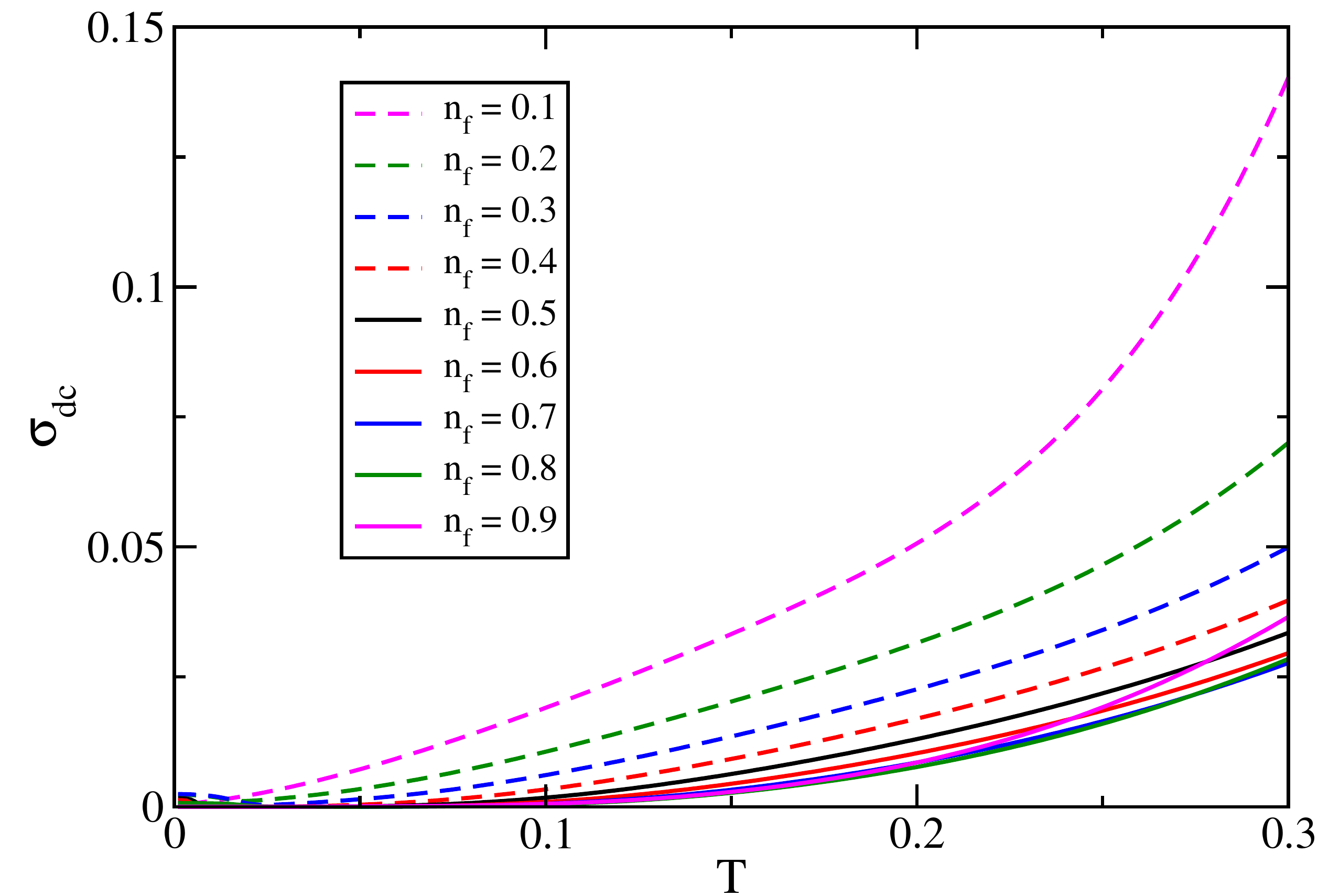}
\includegraphics[scale=0.185]{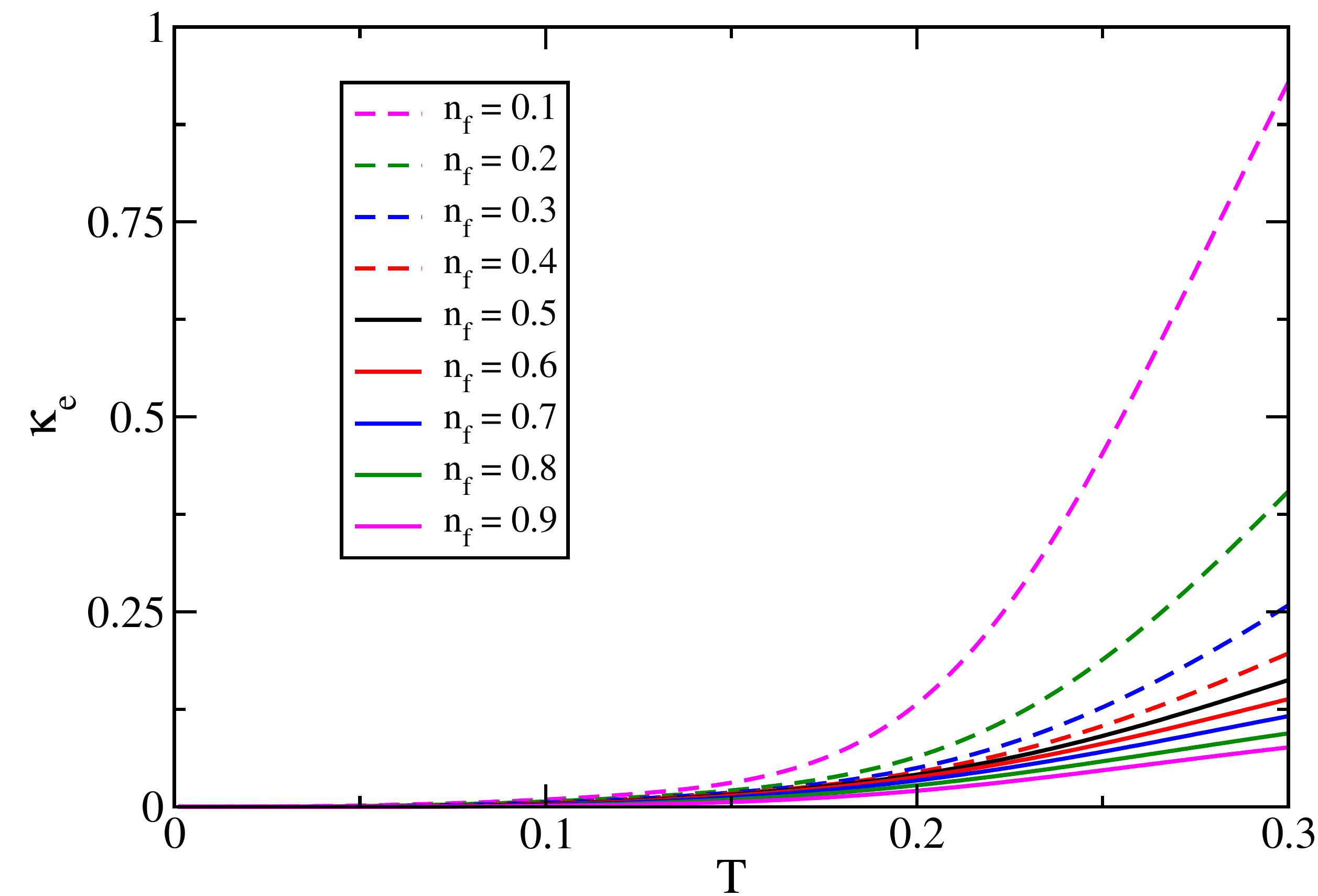}
\\
\includegraphics[scale=0.185]{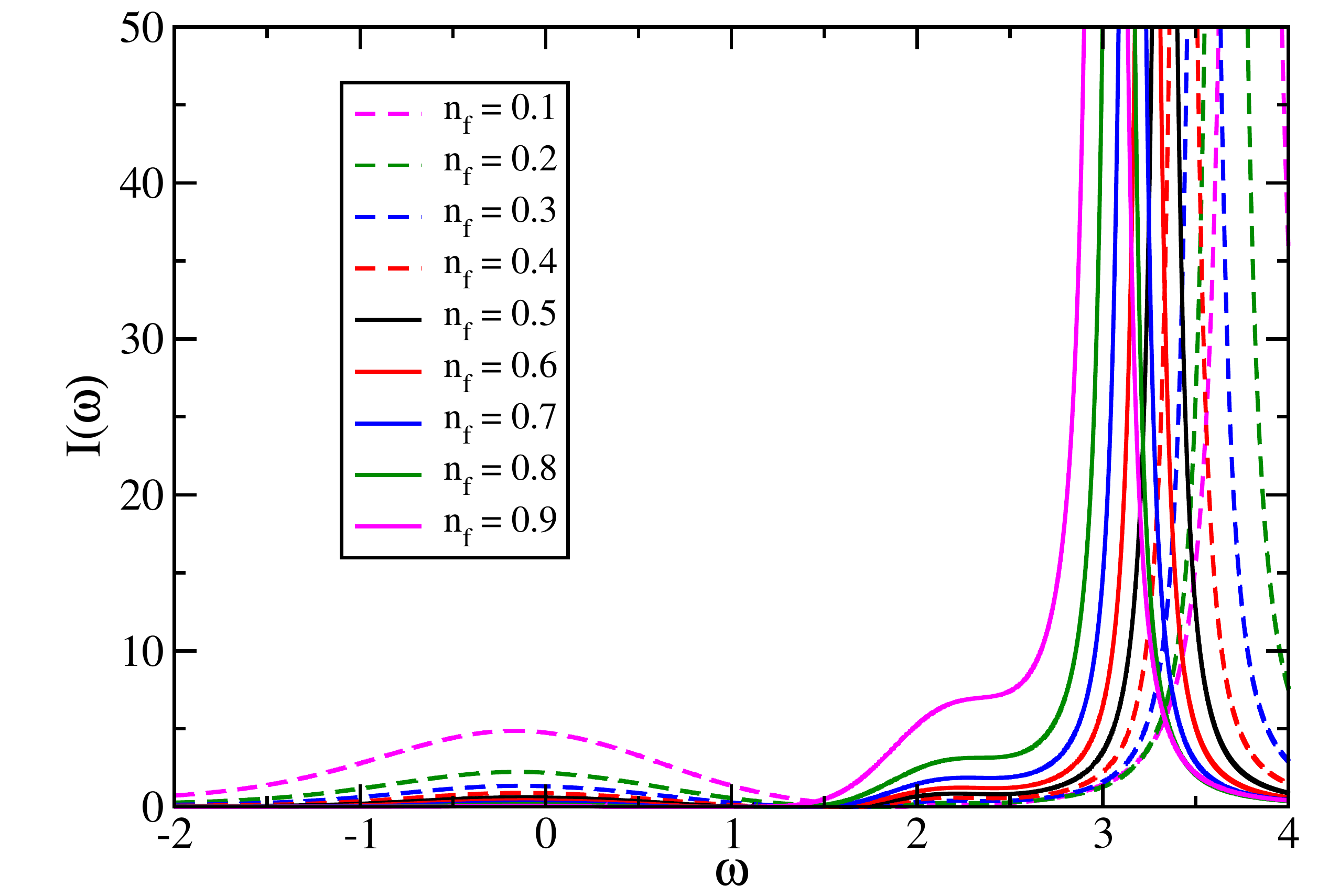}
\includegraphics[scale=0.185]{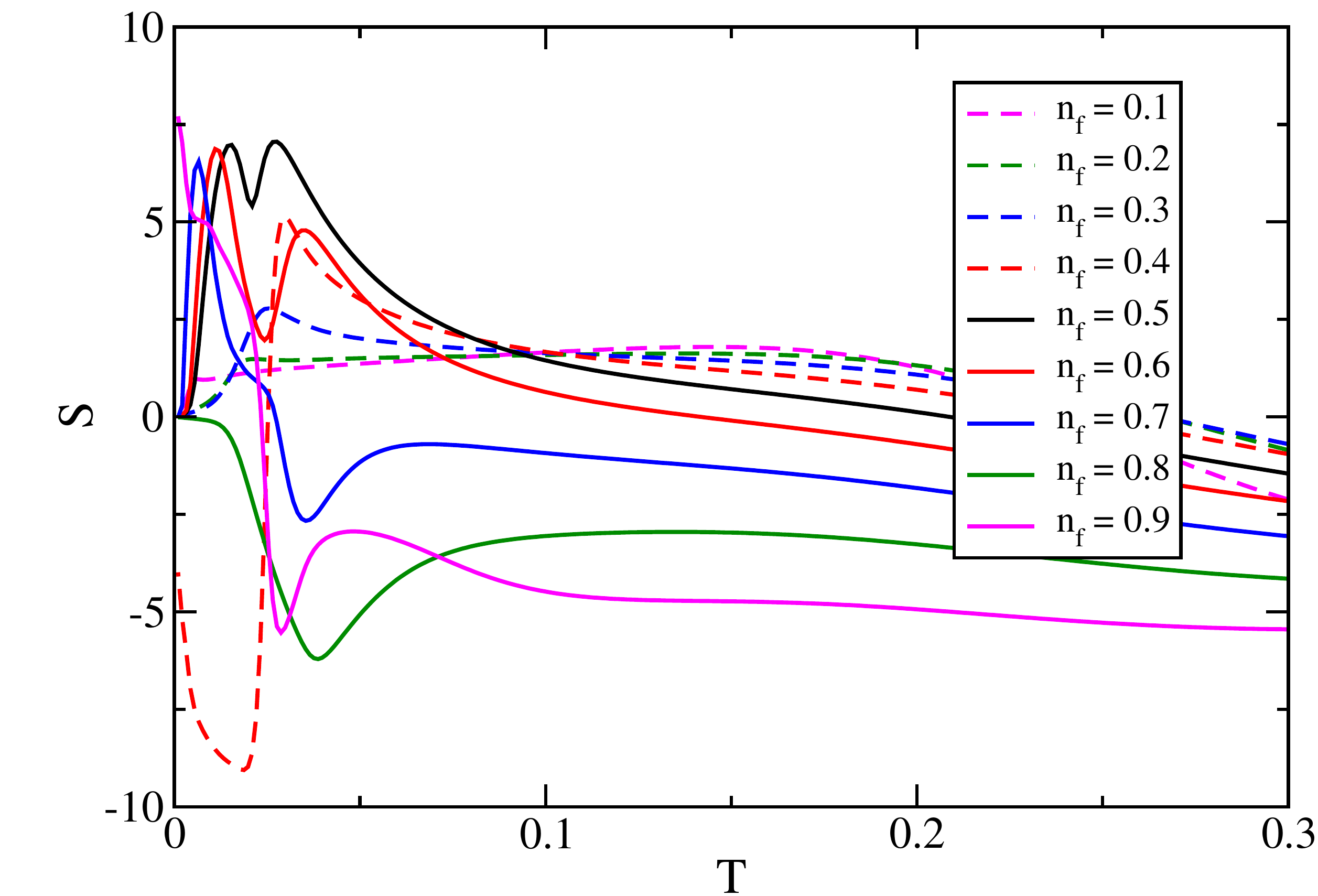}
\includegraphics[scale=0.185]{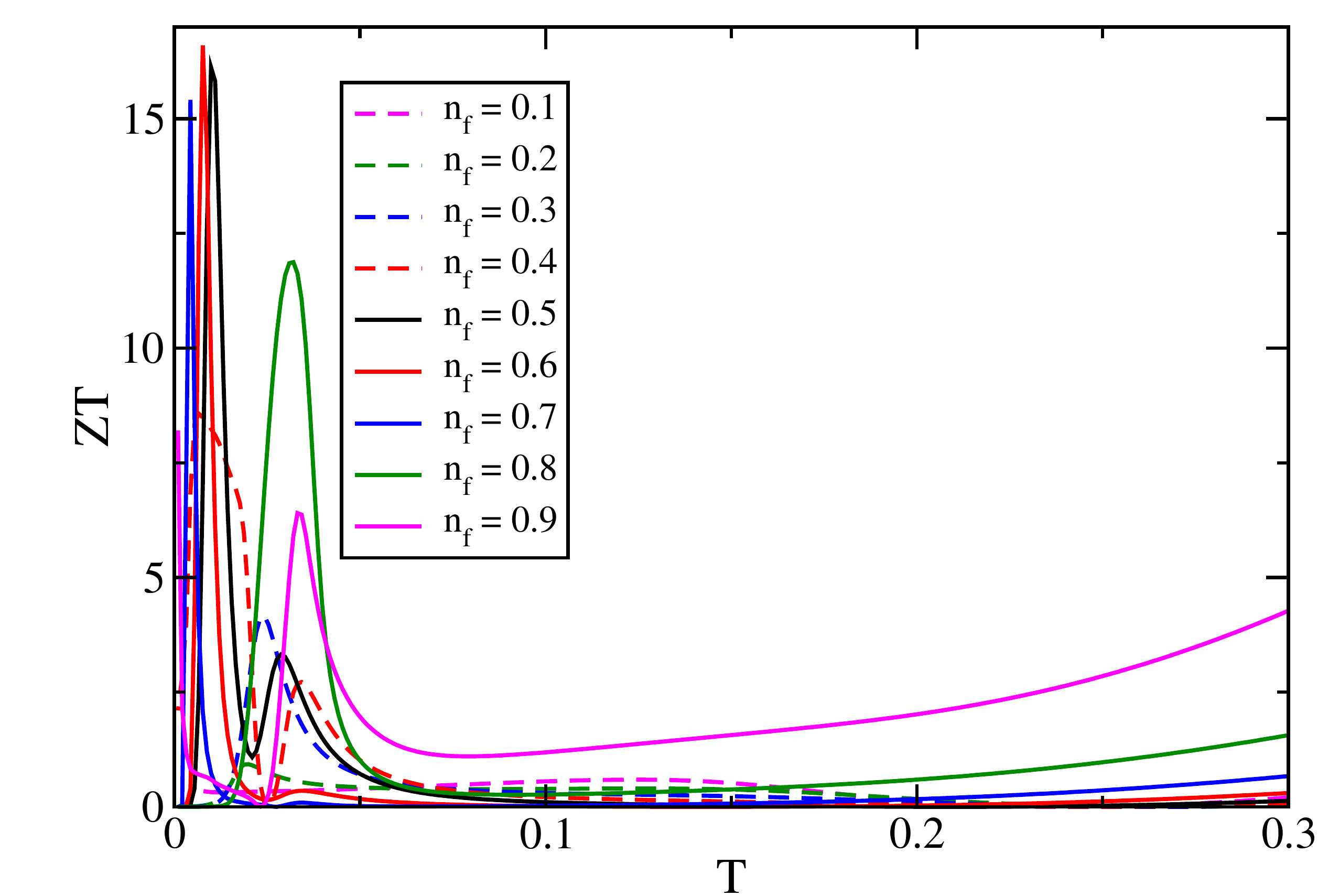}
\caption{(Color online) Density of states $A_d(\omega)$, transport function $I(\omega)$, and temperature dependences of the dc electric $\sigma_{\textrm{dc}}$ and thermal $\kappa_{\textrm{e}}$ conductivities, Seebeck coefficient $S$, and \textit{figure of merit} $ZT$ for $U=2$, $t_1=1$, $t_2=-0.3$, and different doping of $f$-states $n_f$ ($n_d+n_f=1$).}\label{fig:U20_t2_03}
\end{figure}

Such an enhancement of thermoelectric properties by correlated hopping is more prominent with doping. In figure~\ref{fig:U20_t2_03}
we present the results for the case of Mott insulator $U=2$ and for the intermediate value of
correlated hopping $t_2=-0.3$. The doping by $f$-particles over the half-filling $n_f\geqslant 0.5$ and
a simultaneous decrease of the $d$-states occupation ($n_d=1-n_f$) reduces the peak on a transport
function below the chemical potential, while the peak above it enhances and moves inside the Fermi window.
Due to the fancy profile of the transport function, different trajectories of the chemical potential
and Fermi window cover its features, with the change of temperature at different doping levels, at
different energies and produce anomalous temperature dependences of the Seebeck coefficient and
an enhancement of thermoelectric properties at low temperatures. On the other hand, a decrease of
the doping of $f$-states $n_f\leqslant 0.5$ makes the d.o.s. and transport function more symmetric
within the Fermi window and reduces the thermoelectric properties.

\vspace{-1mm}
\section{Conclusions}\label{sec:concl}

In this article we have investigated the effect of correlated hopping on the charge and heat transport and thermoelectric power in a correlated material described by the Falicov-Kimball model. Exact expressions for the transport function are derived for a homogeneous phase, which is assumed to be valid at low temperatures. We do not consider the possible phase transitions into the modulated phases~\cite{gajek:3473,wojtkiewicz:233103,wojtkiewicz:3467,cencarikova:42701}, which should be the subject of a separate investigation.

Depending on the value of correlated hopping, the crossover points which separate the regions with different shapes of the d.o.s. and transport function are clarified. The temperature dependences of the electrical and thermal conductivities and thermoelectric power are strongly effected by the presence of singularities and peaks on the transport function and by the temperature evolution of the chemical potential. The largest enhancement of thermoelectric properties is observed for the values of correlated hopping $-t_1/2<t_2<0$, when a direct hopping between the same many-body states at different sites reduces and an indirect hopping becomes important. It should be noted that the calculations of the correlated hopping amplitudes for different compounds, see e.g.~\cite{hubsch:196401,meir:196802}, provide absolute values close to this interval but disagree by its sign, which is the main factor according to our results.

Unfortunately, the use of the Gaussian d.o.s. produces non-physical results for the transport function outside the bands, which does not allow us to get reasonable values for the chemical potential at very low and very high temperatures and calls for additional investigations with another unperturbed
density of states.

\appendix

\section{Transport function in presence of correlated hopping}\label{sec:append}

After substitution of equation (\ref{chi_jj}) in (\ref{dc_cond}) we get an expression for the transport function in the form
\begin{equation}
I(\omega) = \frac{1}{2\pi} \sum_{\alpha\beta\alpha'\beta'} t^{\alpha\beta} t^{\alpha'\beta'} \frac{1}{N} \sum_{\bm{k}} \Im G_{\bm{k}}^{\beta\alpha'} (\omega) \Im G_{\bm{k}}^{\beta'\alpha}(\omega).
\end{equation}
The components of the lattice Green's function (\ref{G_k}) can be written as follows:
\begin{equation}\label{GkvsE}
G_{\bm{k}}^{\beta\alpha}(\omega) = \frac{A_{\beta\alpha}(\omega) - B_{\beta\alpha}\varepsilon_{\bm{k}}}{C(\omega) - D(\omega)\varepsilon_{\bm{k}} + P\varepsilon^2_{\bm{k}}}\,,
\end{equation}
where
\begin{align}
 A_{++}(\omega) &= Z^{--}(\omega), & A_{+-}(\omega) &= -Z^{+-}(\omega),
\nonumber\\
 A_{-+}(\omega) &= -Z^{-+}(\omega), & A_{--}(\omega) &= Z^{++}(\omega),
\end{align}
\begin{align}
 B_{++} &= t^{--}, & B_{+-} &= -t^{+-},
\nonumber\\
 B_{-+} &= -t^{-+}, & B_{--} &= t^{++},
\end{align}
and
\begin{align}
C(\omega) &= Z^{++}(\omega)Z^{--}(\omega) - Z^{+-}(\omega)Z^{-+}(\omega) = \det \mathbf{Z}(\omega),
\nonumber\\
D(\omega) &= Z^{++}(\omega) t^{--} + Z^{--}(\omega) t^{++} - Z^{+-}(\omega) t^{-+} - Z^{-+}(\omega) t^{+-},
\nonumber\\
P &= t^{++}t^{--} - t^{+-}t^{-+} = \det \begin{pmatrix}
                                         t^{++} & t^{+-} \\
                                         t^{-+} & t^{--}
                                        \end{pmatrix},
\end{align}
and we introduce a notation for the inverse irreducible part $\mathbf{Z}(\omega)=\bm{\Xi}^{-1}(\omega)$. The final analytic expression depends on the value of parameter $P$.

In the case of $P=0$, we have $\det \mathbf{t}_{\bm{k}}=0$, and the expression for the transport function is as follows:
\begin{align}
I(\omega) &= \frac{1}{4\pi} \sum_{\alpha\beta\alpha'\beta'} t^{\alpha\beta} t^{\alpha'\beta'}
\Biggl\{ \Re \left[\frac{A_{\beta\alpha'}(\omega) A_{\beta'\alpha}(\omega)}{D^2(\omega)}  F_{\infty}^{\prime}\left(\frac{C(\omega)}{D(\omega)}\right)\right]
\nonumber\\
& \qquad \qquad \qquad \qquad \qquad - \frac{\Re \left[ A_{\beta\alpha'}(\omega) A_{\beta'\alpha}(\omega)\right]}{|D(\omega)|^2} \Im F_{\infty}\left(\frac{C(\omega)}{D(\omega)}\right) \Biggm/ \Im \frac{C(\omega)}{D(\omega)}
\Biggr\},
\label{IwPeq0}
\end{align}
where
\begin{equation}
F_{\infty}(\zeta) = \int_{-\infty}^{+\infty} \rd \varepsilon \frac{\rho(\varepsilon)}{\zeta - \varepsilon}, \qquad F_{\infty}^{\prime}(\zeta) = \frac{\rd F_{\infty}(\zeta)}{\rd \zeta}
\end{equation}
are a Hilbert transform of the unperturbed density of states and its first derivative, respectively, and for the Gaussian d.o.s.
\begin{equation}
\rho(\varepsilon) = \frac{1}{W\sqrt{\pi}}\re^{-\varepsilon^2/W^2}
\end{equation}
we have
\begin{equation}
F_{\infty}^{\prime}(\zeta) = \frac{2}{W^2}\left[1 - \zeta F_{\infty}(\zeta)\right].
\end{equation}

In the general case of $P\neq0$, the transport function takes a more complicated form
\begin{align}
I(\omega) &= \frac{1}{8\pi P^2} \sum_{\alpha\beta\alpha'\beta'} t^{\alpha\beta} t^{\alpha'\beta'}
\left\{ 2\Re \left[\frac{\left[A_{\beta\alpha'}(\omega)-B_{\beta\alpha'}E_1(\omega)\right] \left[A_{\beta'\alpha}(\omega)-B_{\beta'\alpha}E_1(\omega)\right]}{\left[E_2(\omega)-E_1(\omega)\right]^2}  F_{\infty}^{\prime}\bigl(E_1(\omega)\bigr)\right]
\right.\nonumber\\
& \qquad \qquad \qquad \qquad \qquad \quad + 2\Re \left[\frac{\left[A_{\beta\alpha'}(\omega)-B_{\beta\alpha'}E_2(\omega)\right] \left[A_{\beta'\alpha}(\omega)-B_{\beta'\alpha}E_2(\omega)\right]}{\left[E_1(\omega)-E_2(\omega)\right]^2}  F_{\infty}^{\prime}\bigl(E_2(\omega)\bigr)\right]
\nonumber\\
& - \frac{1}{\Im E_1(\omega)} \Im \left[\frac{F_{\infty}\bigl(E_1(\omega)\bigr)}{\left[E_2(\omega)-E_1(\omega)\right]\left[E_2^*(\omega)-E_1(\omega)\right]} \biggl(\bigl[A_{\beta\alpha'}(\omega)-B_{\beta\alpha'}E_1(\omega)\bigr] \bigl[A_{\beta'\alpha}^*(\omega)-B_{\beta'\alpha}^*E_1(\omega)\bigr]
\right.\nonumber\\
& \qquad \qquad \qquad \qquad \qquad \qquad \qquad \qquad \qquad + \bigl[A_{\beta\alpha'}^*(\omega)-B_{\beta\alpha'}^*E_1(\omega)\bigr] \bigl[A_{\beta'\alpha}(\omega)-B_{\beta'\alpha}E_1(\omega)\bigr]\biggr)
\left.\vphantom{\frac{F_{\infty}^{\prime}\bigl(E_1(\omega)\bigr)}{\left[E_2(\omega)-E_1(\omega)\right]\left[E_2^*(\omega)-E_1(\omega)\right]}}\right]
\nonumber\\
& - \frac{1}{\Im E_2(\omega)} \Im \left[\frac{F_{\infty}\bigl(E_2(\omega)\bigr)}{\left[E_1(\omega)-E_2(\omega)\right]\left[E_1^*(\omega)-E_2(\omega)\right]} \biggl(\bigl[A_{\beta\alpha'}(\omega)-B_{\beta\alpha'}E_2(\omega)\bigr] \bigl[A_{\beta'\alpha}^*(\omega)-B_{\beta'\alpha}^*E_2(\omega)\bigr]
\right.\nonumber\\
& \qquad \qquad \qquad \qquad \qquad \qquad \qquad \qquad \qquad + \bigl[A_{\beta\alpha'}^*(\omega)-B_{\beta\alpha'}^*E_2(\omega)\bigr] \bigl[A_{\beta'\alpha}(\omega)-B_{\beta'\alpha}E_2(\omega)\bigr]\biggr)
\left.\left.\vphantom{\frac{F_{\infty}^{\prime}\bigl(E_1(\omega)\bigr)}{\left[E_2(\omega)-E_1(\omega)\right]\left[E_2^*(\omega)-E_1(\omega)\right]}}\right]
\right\},
\label{IwPneq0}
\end{align}
where $E_{1,2}(\omega)$ are roots of the denominator in equation (\ref{GkvsE})
\begin{equation}
E_{1,2}(\omega) = \frac{D(\omega)}{2P} \pm \sqrt{\left[\frac{D(\omega)}{2P}\right]^2 - \frac{C(\omega)}{P}}\,.
\end{equation}

For large energy values and for unperturbed Gaussian d.o.s., the imaginary parts in the last terms of equations (\ref{IwPeq0}) and (\ref{IwPneq0}) are small but their ratios are finite and produce a strong enhancement of the transport function and make its shape
altogether different from the one-particle d.o.s.

%\section*{Acknowledgments}

%\def\url#1{}
%\def\urlprefix{}
%\bibliographystyle{cmpj2}
%\bibliography{ashv,drsci}

\begin{thebibliography}{10}

\bibitem{mahan:42}
Mahan G., Sales B., Sharp J., Phys. Today, 1997, \textbf{50}, No.~3, 42--47; \bibdoi{10.1063/1.881752}.

\bibitem{costi:2519}
Costi T.A., Hewson A.C., Zlati\'c V., J. Phys.: Condens. Matter, 1994,
  \textbf{6}, No.~13, 2519; \bibdoi{10.1088/0953-8984/6/13/013}.

\bibitem{hubbard:238}
Hubbard J., Proc. R. Soc. London, Ser. A, 1963, \textbf{276}, No.~1365, 238--257; \bibdoi{10.1098/rspa.1963.0204}.

\bibitem{foglio:4554}
Foglio M.E., Falicov L.M., Phys. Rev. B, 1979, \textbf{20}, No.~11, 4554--4559;
  \bibdoi{10.1103/PhysRevB.20.4554}.

\bibitem{simon:7471}
Sim\'on M.E., Aligia A.A., Phys. Rev. B, 1993, \textbf{48}, No.~10, 7471--7477;
  \bibdoi{10.1103/PhysRevB.48.7471}.

\bibitem{blackman:2412}
Blackman J.A., Esterling D.M., Berk N.F., Phys. Rev. B, 1971, \textbf{4},
  No.~8, 2412--2428; \bibdoi{10.1103/PhysRevB.4.2412}.

\bibitem{hirsch:326}
Hirsch J.E., Physica C, 1989, \textbf{158}, No.~3, 326--336;
  \bibdoi{10.1016/0921-4534(89)90225-6}.

\bibitem{bulka:10303}
Bu\l{}ka B.R., Phys. Rev. B, 1998, \textbf{57}, No.~17, 10303--10306;
  \bibdoi{10.1103/PhysRevB.57.10303}.

\bibitem{didukh:7893}
Didukh L., Skorenkyy {\relax Yu}., Dovhopyaty {\relax Yu}., Hankevych V., Phys.
  Rev. B, 2000, \textbf{61}, No.~12, 7893--7908;\\
  \bibdoi{10.1103/PhysRevB.61.7893}.

\bibitem{kollar:045107}
Kollar M., Vollhardt D., Phys. Rev. B, 2001, \textbf{63}, No.~4, 045107; \bibdoi{10.1103/PhysRevB.63.045107}.

\bibitem{meir:196802}
Meir Y., Hirose K., Wingreen N.S., Phys. Rev. Lett., 2002, \textbf{89}, No.~19,
  196802; \bibdoi{10.1103/PhysRevLett.89.196802}.

\bibitem{hubsch:196401}
H\"ubsch A., Lin J.C., Pan J., Cox D.L., Phys. Rev. Lett., 2006, \textbf{96},
  No.~19, 196401; \bibdoi{10.1103/PhysRevLett.96.196401}.

\bibitem{tooski:055001}
Tooski S.B., Ram\v{s}ak A., Bu\l{}ka B.R., \v{Z}itko R., New J. Phys., 2014,
  \textbf{16}, No.~5, 055001;\\ \bibdoi{10.1088/1367-2630/16/5/055001}.

\bibitem{eckholt:093028}
Eckholt M., Garc\'\i{}a-Ripoll J.J., New J. Phys., 2009, \textbf{11}, No.~9,
  093028; \bibdoi{10.1088/1367-2630/11/9/093028}.

\bibitem{jurgensen:043623}
J\"urgensen O., Sengstock K., L\"uhmann D.-S., Phys. Rev. A, 2012, \textbf{86}, No.~4, 043623; \bibdoi{10.1103/PhysRevA.86.043623}.

\bibitem{liberto:013624}
Liberto M.D., Creffield C.E., Japaridze G.I., Smith C.M., Phys. Rev. A, 2014,
  \textbf{89}, 013624;\\ \bibdoi{10.1103/PhysRevA.89.013624}.

\bibitem{falicov:997}
Falicov L.M., Kimball J.C., Phys. Rev. Lett., 1969, \textbf{22}, No.~19,
  997--999; \bibdoi{10.1103/PhysRevLett.22.997}.

\bibitem{freericks:1333}
Freericks J.K., Zlati\'c V., Rev. Mod. Phys., 2003, \textbf{75}, No.~4,
  1333--1382; \bibdoi{10.1103/RevModPhys.75.1333}.

\bibitem{schiller:15660}
Schiller A., Phys. Rev. B, 1999, \textbf{60}, No.~23, 15660--15663;
  \bibdoi{10.1103/PhysRevB.60.15660}.

\bibitem{shvaika:075101}
Shvaika A.M., Phys. Rev. B, 2003, \textbf{67}, No.~7, 075101;
  \bibdoi{10.1103/PhysRevB.67.075101}.

\bibitem{shvaika:119901}
Shvaika A.M., Phys. Rev. B, 2010, \textbf{82}, No.~11, 119901;
  \bibdoi{10.1103/PhysRevB.82.119901}.

\bibitem{gajek:3473}
Gajek Z., Lema\'nski R., Acta Phys. Pol. B, 2001, \textbf{32}, No.~10, 3473--3476.

\bibitem{wojtkiewicz:233103}
Wojtkiewicz J., Lema\'nski R., Phys. Rev. B, 2001, \textbf{64}, No.~23, 233103; \bibdoi{10.1103/PhysRevB.64.233103}.

\bibitem{wojtkiewicz:3467}
Wojtkiewicz J., Lema\'nski R., Acta Phys. Pol. B, 2001, \textbf{32}, No.~10, 3467--3472.

\bibitem{cencarikova:42701}
\v{C}en\v{c}arikov\'a H., Farka\v{s}ovsk\'y P., Condens. Matter Phys., 2011, \textbf{14}, No.~4, 42701; \bibdoi{10.5488/CMP.14.42701}.

\bibitem{mahan:book}
Mahan G.D., Many-Particle Physics, 3rd Edn., Kluwer Academic/Plenum Publishers, New York,
   2000.

\bibitem{freericks:035133}
Freericks J.K., Zlati\'{c} V., Shvaika A.M., Phys. Rev. B, 2007, \textbf{75},
  No.~3, 035133; \bibdoi{10.1103/PhysRevB.75.035133}.

\bibitem{jonson:4223}
Jonson M., Mahan G.D., Phys. Rev. B, 1980, \textbf{21}, No.~10,
  4223--4229; \bibdoi{10.1103/PhysRevB.21.4223}.

\bibitem{jonson:9350}
Jonson M., Mahan G.D., Phys. Rev. B, 1990, \textbf{42}, No.~15,
  9350--9356; \bibdoi{10.1103/PhysRevB.42.9350}.

\bibitem{mahan:7436}
Mahan G.D., Sofo J.O., Proc. Natl. Acad. Sci. USA, 1996, \textbf{93}, No.~15,
  7436--7439; \bibdoi{10.1073/pnas.93.15.7436}.

\bibitem{freericks:195120}
Freericks J.K., Demchenko D.O., Joura A.V., Zlati\'{c} V., Phys. Rev. B, 2003,
  \textbf{68}, No.~19, 195120;\\ \bibdoi{10.1103/PhysRevB.68.195120}.

\bibitem{zlatic:155101}
Zlati\'{c} V., Boyd G.R., Freericks J.K., Phys. Rev. B, 2014, \textbf{89},
  No.~15, 155101; \bibdoi{10.1103/PhysRevB.89.155101}.

\bibitem{hoshino:1320}
Hoshino K., Niizeki K., J. Phys. Soc. Jpn., 1975, \textbf{38}, No.~5,
  1320--1327; \bibdoi{10.1143/JPSJ.38.1320}.

\end{thebibliography}

%
%% If you have problems with typesetting in ukrainian uncomment lines below.
%
%  \lastpage
%  \end{document}

%\clearpage

\ukrainianpart

\title{Вплив корельованого переносу на термоелектричні властивості: точний розв'язок для моделі Фалікова-Кімбала}
\author{А.М. Швайка}
\address{Iнститут фiзики конденсованих систем НАН України, вул. І.~Свєнцiцького, 1, 79011 Львiв, Україна}
%
%% якщо автор є один або автори є з однієї установи:
%
%  \author{1й Автор, 2й Автор, \ldots}
%  \address{Інститут\ldots}
%
%%

\makeukrtitle

\begin{abstract}
\tolerance=3000%
Досліджено вплив корельованого переносу на перенос заряду і тепла в моделі Фалікова-Кімбала. В рамках теорії динамічного середнього поля отримано точні розв'язки для електро- та теплопровідності і термо-е.р.с. Досліджено температурні залежності коефіцієнтів переносу для певних значень корельованого переносу, що відповідають суттєвій перебудові густини станів та транспортної функції. Встановлено випадки, які відповідають значному покращенню термоелектричних властивостей.
\keywords термо-е.р.с., корельований перенос, модель Фалікова-Кімбала
\end{abstract}

\end{document}